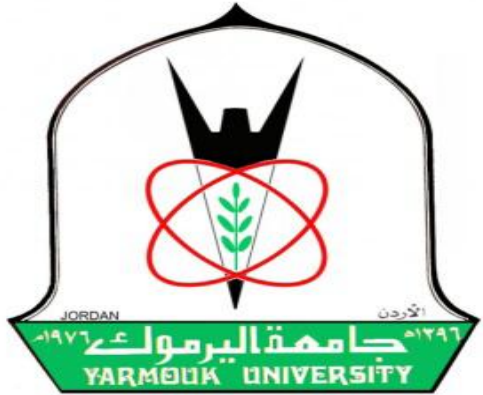

# Department of Telecommunications Engineering
# Faculty of Hijjawi for Engineering Technology
# Yarmouk University

# "Multi-layer Mechanism for Multicast Routing in Multi-hop Cognitive Radio Networks"

**Prepared by**
Mustafa Mahdi Ali

**Supervised by**
Dr. Haythem Bany Salameh

# "Multi-layer Mechanism for Multicast Routing in Multi-hop Cognitive Radio Networks"

By

Mustafa Mahdi Ali

Thesis Submitted in Partial Fulfillment of the Requirements for the Degree of
Master of Science in Wireless Communications Engineering

Submitted to

The Department of Telecommunications Engineering

Faculty of Hijjawi for Engineering Technology

Yarmouk University, Jordan

July, 2015

| **Committee Members** | **Signature and Date** |
|---|---|
| Dr. Haythem Bany Salameh (*Chairman*) | 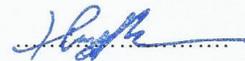 |
| Prof. Ahmad AL-shamali *(Member)* | 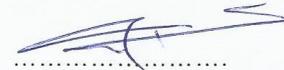 |
| Dr. Amer Magableh *(Member)* | 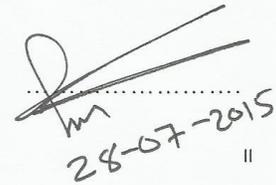 28-07-2015 |

# Dedication

I dedicate this work to **my mother**, **my father**, and **my country (Iraq)**.

# ACKNOWLEDGMENT

First, I am indebted in my work to almighty Allah, who has enabled me to complete my work on this thesis, and delight my way every moment on my life.

I would like to express my sincere gratitude to my supervisor Dr. Haytham Bany Salameh for the continuous support of my thesis work, for his encouragement, enthusiasm, and immense knowledge. His guidance helped me in all the time of research and writing of this thesis.

Besides my supervisor, I would like to express deep gratitude and thanks to the rest of my thesis committee members: Prof. Ahmad AL-shamali and Dr. Amer Magableh. I also wish to sincerely thank all the staff in the telecommunications engineering department at Faculty of Hijjawi for Engineering Technology at Yarmouk University.

Special loving thanks goes to my parents, brothers and friends for their unlimited love, support, encouragement, and for always doing everything for me when I need it the most.

Last but not the least; I would say thanks to all who have been supportive and well-wishers to me in this period of life.

All praise is to Allah

*Mustafa Mahdi Ali*

*July 2015*

# Table of Contents

# Table of Figures

# List of Tables

# Table of Abbreviations

| Symbol | Description |
|---|---|
| BER | Bit Error Rate |
| BS | Base Station |
| BW | Bandwidth |
| CR | Cognitive Radio |
| CRN | Cognitive Radio Network |
| DSA | Dynamic Spectrum Access |
| ETX | Expected Transmissions Count |
| FCC | Federal Communication Commission |
| IP | Internet Protocol |
| MA | Message Announcement |
| MANETs | Mobile Ad hoc Networks |
| MASA | Maximum Average Spectrum Availability |
| MDR | Maximum Data Rate |
| MIMO | Multiple-Input and Multiple-Output |
| MST | Minimum Spanning Tree |
| NP | Non-deterministic Polynomial-time |
| PDR | Packet Delivery Rate |
| $P_I$ | Idle Probability |
| POS | Probability of Success |
| PR | Primary Radio |
| PRNs | Primary Radio Networks |
| PU | Primary User |
| RF | Radio Frequency |
| RS | Random Selection |
| SDR | Software Defined Radio |
| SINR | Signal to Interference plus Noise Ratio |
| SPT | Shortest Path Tree |
| SU | Secondary User |

# Abstract

**Multi-layer Mechanism for Multicast Routing in Multi-hop Cognitive Radio Networks**

**By: Mustafa Mahdi Ali (2013976007)**

**Supervisor: Dr. Haytham Ahmad Bany Salameh**

Multicast routing is considered as one of the most important process in Mobile Ad hoc Networks (MANETs) to enable video sharing and data dissemination. Multicast routing can be used without any difficulties if a medium can be accessed by any transmitter at any given time. Unfortunately, challenges such as the unexpected behavior of primary users (PUs) and their access priority as well as network heterogeneity may appear in cognitive radio networks (CRNs). In this thesis, we propose a multi-layer multicast routing protocol for multi-hop mobile Ad Hoc CRN based on the concept of the minimum spanning tree (MST) and the shortest path tree (SPT). The proposed algorithm employs the probability of success (POS) metric in performing the channel assignment process.

Simulations experiments were conducted, to investigate the effectiveness of the proposed scheme, compare the performance of the proposed scheme with the-state-of-the-art schemes, and study the performance of the proposed scheme under various network conditions in terms of throughput and packet delivery rate (PDR).

Simulations results show that the SPT outperforms MST in term of throughput all the time.

# Chapter One: Introduction

Cognitive radio networks (CRNs) are smart networks that adapt their own parameters (such as operating frequency, modulation type, etc) depending on the surrounding environment such that the overall spectrum utilization is maximized while protecting the primary radio performance [1-8]. Cognitive radio (CR) has been adopted to tackle the spectrum underutilization problem caused by the used fixed spectrum allocation policy. Measurements by the Federal Communication Commission (FCC) reflect geographical and temporal variations in spectrum utilization ranging from 15 % to 85% [8].

CR exploits white and gray holes (i.e., available spectrum portions to CRs) to improve network throughput, connectivity, and security. CRs support dynamic and opportunistic spectrum access through the following actions [9]:-

(1)    Spectrum sensing: In spectrum sensing, the CR has to be aware of the surrounding radio frequency (RF) environment and detect any primary user (PU) activity; not to affect PU transmission. Specifically, changes in spectrum status due to PU activity and spectrum holes are identified in spectrum sensing process.

(2)    Spectrum management: The purpose of this action is to choose best available channel/s for CR transmissions.

(3)    Spectrum sharing: Spectrum sharing is based on dynamic spectrum access (DSA), where CRs access licensed and unlicensed spectrum portions opportunistically and dynamically and share them even among the same CRN users or between different CRNs.

(4)     Spectrum mobility: Whenever a PU is appeared, a CR should rapidly vacate the required spectrum by PU to another available portion of spectrum.

All these actions are achieved by using technologies such as software defined radio (SDR), adaptive filters, high digital signal processing, etc.

## 1.1  Literature Survey

Multicast routing protocols were proposed in [10-14] for mobile ad-hoc networks (MANETs), but none of them is suitable for CRNs because of the channel uncertainty and the dynamic spectrum allocation challenges in such networks. However, many routing protocols were proposed to support CRNs in [15-23], but none of these routing protocols enables data multicasting. In [24], the authors used different network parameters including retransmission, modulation, and scheduling proposed to formulate a cross layer optimization problem to multicast video in CRN. PU protection and fairness were taken in their consideration. For infrastructure CRNs, the authors proposed in [25] a multicast scheduling for multi-hop CRNs. This protocol uses power control and considers the interference to PU, fairness, and routers (relay) assignment. The link between any two nodes in a CRN depends not only on the transmission power and the distance between these nodes but also depends on the availability of at least a common channel between them which depends on PU activity (cannot be controlled by CRNs). This makes the construction of a multicast tree in CRNs (a challenging problem). Thus, the authors in [26] showed the relationship between PU traffic load and Constructing minimum energy multicast trees by transforming the problem of multicast into a directed Steiner tree problem. Simulations results showed that the simultaneous transmission is more convenient when the PU activity is low and the sequential transmission is more suitable when the activity of PUs is high. Besides that the authors also considered the CR consumed energy that used for sensing the available channels. In [27], the authors proposed a joint (routing, scheduling, power control, and channel assignment) distributed algorithm for multi-hop CRNs to improve network throughput. In their work the channel assignment was dynamically done in order to improve link capacity without causing interference to PU or other SUs while the bit error rate is guaranteed. For multicasting in CRNs, the authors in [28] proposed a scheme that depends on the network coding to enhance the throughput of the network. In [28] the interference limit (threshold) on PU and the signal to interference- plus- noise- ratio (SINR) were considered to construct minimum energy multicast problem which is a linear optimization problem (much easier than Non-deterministic Polynomial-time (NP) hard problems). In [29], the authors proposed a cross layer optimization scheme for video multicasting in CRN with a base station (BS). The main objective of their work is to guarantee video quality and CR users' fairness taking into account the different design factors (e.g., PU protection, scheduling, coding, and modulation). In [30], the multicast problem was studied for MIMO-based networks and two algorithms were proposed for robust multicast beamforming. In [31], the authors proposed power control in CRNs. They studied the effect of SU transmit power on the availability of spectrum opportunities at different SUs such that if a transmission power of the SU is low it can reach a small number of nodes. Thus more transmissions are needed. If the transmission power of the SUs is high, then it can reach larger number of nodes but in this case the source have to wait the best spectrum opportunity thus more sensing energy is needed. In [32], the authors proposed a multi-channel multicast and level channel assignment algorithm for wireless mesh networks to maximize throughput, reduce transmission delay, and minimize the number of relay nodes. In [33], the authors proposed multi-channel multi-radio routing schemes based on an efficient channel assignment algorithm. They proved that this algorithm is better than multi-channel multicast algorithm in terms of throughput and delay. In [34], the authors studied the multicast routing in multi-hop CRNs. By using a cross-layer approach via considering scheduling and routing, the authors support a set of multicast sessions through minimizing the required resources. This results in a mixed integer linear programming optimization problem. For multi-hop CRN, the authors in [35] proposed a novel routing metric, which jointly studied the effect of the PUs average channel availability time and the required transmission time to improve throughput. This algorithm tries to find the path between the source and a destination that has the maximum probability of success (POS). In [36], a distributed optimization algorithm for multi-hop CRNs was investigated by jointly considering the power control, routing, and scheduling to improve data rates for a set of user sessions. The authors in [37] investigated the use of relaying to improve spectrum utilization. Specifically, they proposed the geometric condition under which a (SU) could transmit over a single channel. If the destination is out of the transmission range of the (SU) transmitter, multi-hop relying will be used. In addition, two multi-hop routing (nearest-neighbor routing and farthest-neighbor routing) are proposed. The authors in [38] proposed a distributed on-demand multicast routing and a channel allocation algorithm for mesh CRNs. The purpose of the work in [38] is to minimize the access delay. The authors in [39] used the probability of success (POS) metric to perform multicast in single-hop CRNs. The video source transmits video packets to all destinations across a unified channel. In [40], the authors proposed expected transmission count metric (ETX) as a routing protocol for multi-hop wireless networks. The ETX metric finds all the paths that give high throughput and reduces the expected number of retransmissions.

In the networks, the multicast means the source send packets to multiple nodes in the network (i.e., the source sends the same packets to multiple nodes in the network). Multicast may use multi-source instead of single source. The path from the source to each destination in the network may be single hop or multi-hop. The multicast reduces communication cost, improve channel efficiency, provide effective use of energy and bandwidth, minimize the sender and router processing, and minimize delivery delay [41].

Unicast is a type of communication where the data is sent from one source to one receiver. In the unicast, there is only one sender, and one receiver. If some device needs to send a message to multiple devices, it will have to send multiple unicast messages, each message addressed to a specific device. So, the sender has to send a separate message to each destination, and to do that it has to know the exact IP address of each destination [42].

Another type of transmission is broadcast which means the transmission of the message from the single source to all destinations in a specific network. So, all destinations received the same message at the same time by using single IP address [43].

## 1.2 Motivation

The multicast is very important for various applications such as video conferencing, data disseminations, disaster relief, and military purposes. Multicast saves spectrum resource, reduces the communication cost, and improves channel efficiency. However, we aim at designing an efficient multicast routing protocol that improves network throughput with high packet delivery rate (PDR) for CRNs. The proposed protocol accounts for the unique feature of CRNs environment.

## 1.3 Contribution

The significance of this work is outlined in performing multicast for multi-hop CRNs. This is done by converting the original network topology to a SPT and/or MST in multi-layer manner. Transmission from one layer to another layer is performed using the POS-based channel assignment scheme that combines the effect of both link quality and average spectrum availability time such that the overall network throughput is improved.

## 1.4 Thesis Outlines

The rest of this thesis is organized as follows. In Chapter two, the proposed multicasting protocol is introduced. In Chapter three, simulation results and discussion are presented. Finally, in Chapter four, conclusion remarks and future work are given.

# Chapter Two: The Proposed Multicasting Protocol

## 2.1  System Model

We consider a CRN that coexists with several PRNs in the same geographical area, where there is one transmitter (source) attempts to deliver multicast messages to *Nr* destinations over a set of available channels C as shown in Figure 2.1. The status of each channel is modeled as a two states Markov model, alternating between busy and idle. Busy state represents that the channel is occupied by a PU, so, this channel cannot be used by the SUs. The idle state represents that the channel is not used by the PUs, so, this channel can opportunistically be used by the SUs. We assume infrastructureless ad hoc multi-hop CRN. We assume that a common control channel is available to coordinate the transmissions in the CRN. Table 2.1 summarizes the main notations used in this chapter.

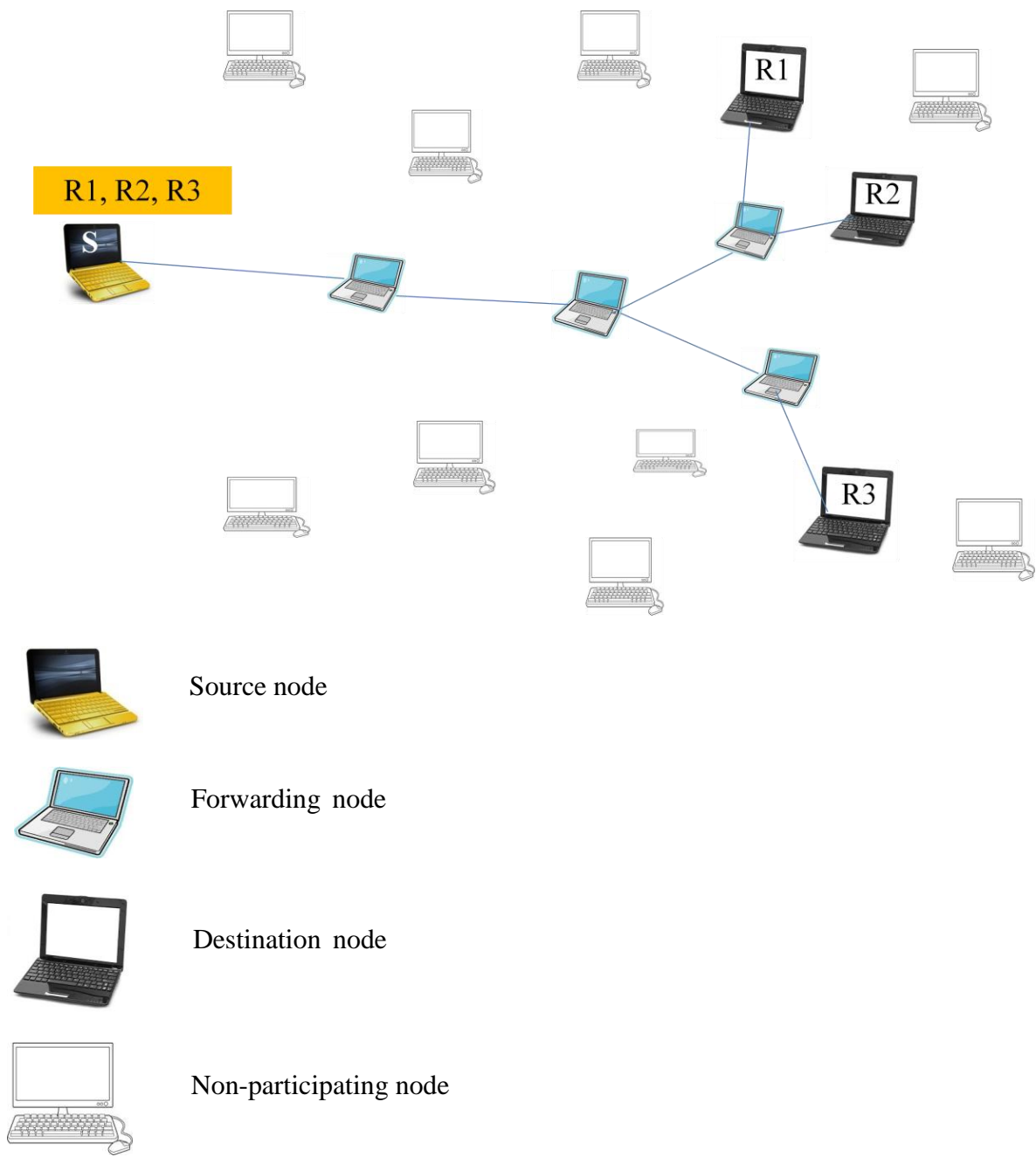

**Fig 2.1 Network model**

**Table 2.1 Summary of notations used in the algorithm**

| Parameter | Description |
|---|---|
| $p_t$ | Cognitive radio transmission power |
| Nr | Total number of destinations |
| N | Total number of nodes |
| $P_{suc(j)}^{i-k}$ | Probability of success (POS) between nodes $i$ and $k$ over channel $j$ |
| $Tr_{(j)}^{(i-k)}$ | Transmission rate between nodes $i$ and $k$ over channel $j$ |
| $Tr_{(j)}^{(i-k)}$ | Required transmission time to transmit a packet between nodes $i$ and $k$ over channel $j$ |
| $\mu_j$ | Average availability time of channel $j$ |
| $Pr_{(j)}^{(i-k)}$ | Power received between nodes $i$ and $k$ over channel $j$ |
| $\xi_{(j)}^{(i-k)}$ | Channel power gain between nodes $i$ and $k$ over channel $j$ |
| $D$ | Data packet size |
| $N_0$ | Thermal power spectral density |
| $Bw$ | Channel bandwidth |
| $n$ | Path loss exponent |
| M | Total number of channels for PRNs (every channel may be idle or busy) |
| C | Available channels for the CRN |
| $d$ | The distance between any two nodes in the network topology |
| $\lambda$ | Wave length |

## 2.2 Problem Definition

Given the aforementioned network model, we aim at designing a routing protocol for video multicasting such that the throughput is improved. At first, we construct MST and SPT for the original topology. Multi-layer route could be resulted; from the upper layer to the lower layer. The source computes the POS for all destinations/other relays in each sub-tree over all available channels. After that, the transmitter at each layer decides the group of receivers that should participate in the session.

### 2.2.1 Multicast Routing Using Probability of Success

In this thesis, we consider a multi-layer on-demand multicast routing protocol over multi-hop CRNs. Here, a video streaming node (source node) multicasts video packets to the receiver nodes over single session (to save spectrum) across multi-layer. The main problem of multicasting in CRNs is to find a set of common channels between the source and the destinations to send the video packets from the source to the destinations. Selecting the common channel for multicasting is a challenging issue, such that the network performance is improved. Channel availability and channel quality have a huge impact on network performance in CRNs. The scheme that chooses the channel with maximum average spectrum availability (MASA) does not take into consideration the quality of that channel because channel availability does not reflect the channel quality. On the other hand, the scheme that chooses the channel with maximum data rate (MDR) does not consider the availability time of that channel. The POS metric jointly considers the effect of both channel quality and its availability time [7].

The authors in [7] presented a closed-form expression for the POS between any two nodes $i$ and $k$ over channel $j \in C$ based on a stochastic model of PUs activities under a Rayleigh fading channel model as in equation (1): -

$$P_{suc(j)}^{i-k} = exp\left(\frac{-Tr_{(j)}^{(i-k)}}{\mu_j}\right) \quad (1)$$

where: $P_{suc(j)}^{i-k}$ is the POS between any two nodes $i$ and $k$ over channel $j$, $\mu_j$ is the average spectrum availability time for channel $j$, and $Tr_{(j)}^{(i-k)}$ is the required transmission time to send a packet from node $i$ to $k$ over channel j, and is calculated as in (2). From (1), it is clear that the probability of successful transmission between any two nodes $i$ and $k$ over channel j can be improved by choosing a channel with minimum required transmission time (higher data rate) and maximum average availability time [7].

$$Tr_{(j)}^{(i-k)} = \frac{D}{R_{(j)}^{(i-k)}} \quad (2)$$

where $D$ is the packet size (in bits) and $R_{(j)}^{(i-k)}$ is the data rate between nodes $i$ and $k$ over channel j, and can be calculated as [7]: -

$$R_{(j)}^{(i-k)} = (Bw) \, log_2\left(1 + \frac{Pr_{(j)}^{(i-k)}}{Bw * N_0}\right) \quad (3)$$

where $N_0$ represents the thermal power spectral density in (Watt /Hz), and $Bw$ is the channel bandwidth for channel $j$. $Pr_{(j)}^{(i-k)}$ represents the received power from transmitter $i$ to receiver $k$, which is given by [7]:-

$$Pr_{(j)}^{(i-k)} = \frac{p_t}{d^n}(\frac{\lambda}{4\pi})^2 (\xi_{(j)}^{(i-k)}) \tag{4}$$

where $p_t$ is the cognitive radio transmission power, $d$ is the distance between any two nodes, $n$ is the path loss exponent, and $\xi_{(j)}^{(i-k)}$ is the channel power gain between nodes $i$ and $k$ over channel $j$. For Rayleigh fading, $\xi_{(j)}^{(i-k)}$ is exponentially distributed with mean 1 [7].

The cost of using POS as a channel assignment scheme is to know the average available time of the channel and to determine the quality of that channel. While the cost of using maximum average spectrum availability scheme or maximum data rate scheme requires only knowing the average availability time of the channel or determining the channel quality, respectively.

### 2.2.2 Shortest Path Tree (SPT) and Minimum Spanning Tree (MST)

The SPT finds the loop free path with minimum distance (cost) from the source (root) to each vertex (node) in the topology. For any connected and undirected graph with a single source (node a) as shown in Figure 2.2, the SPT can be found using Dijkstra's algorithm (see Figure 2.3). Note that the SPT assumes an existence of a source (root) node in order to build the tree, so that if the source node is changed, the whole tree should be changed consequently (it is not valid for all sources). The total running time for Dijkstra's algorithm is $(m + n \log n)$, where $m$ is the number of edges (links) that connect the nodes and $n$ is the number of vertices [44], [45].

The MST is a spanning tree whose edges (links) sum to minimum distance (cost). In other words, a MST is a tree formed from a subset of the edges that includes every vertex in the topology and the total distance (cost) of all the edges is as low as possible (see Figure 2.4).

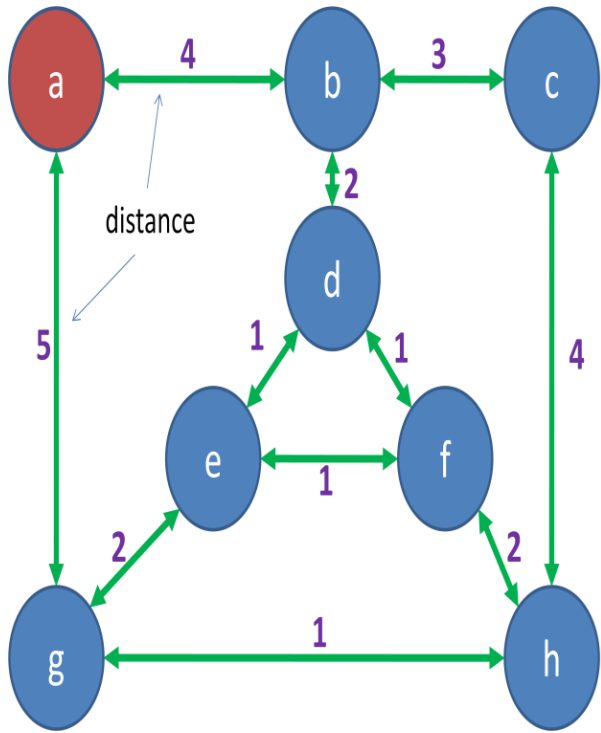
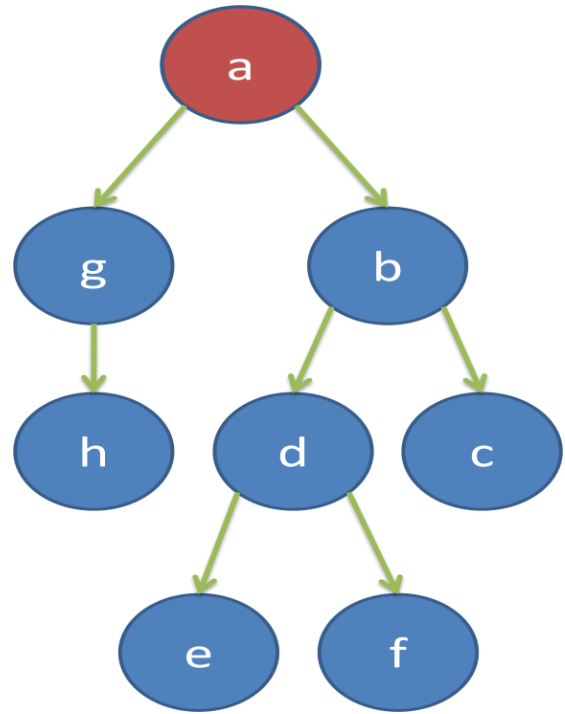

**Fig. 2.2 connected and undirected topology with $n = 8$ and $m = 11$.**

**Fig. 2.3 The resulting shortest path tree.**

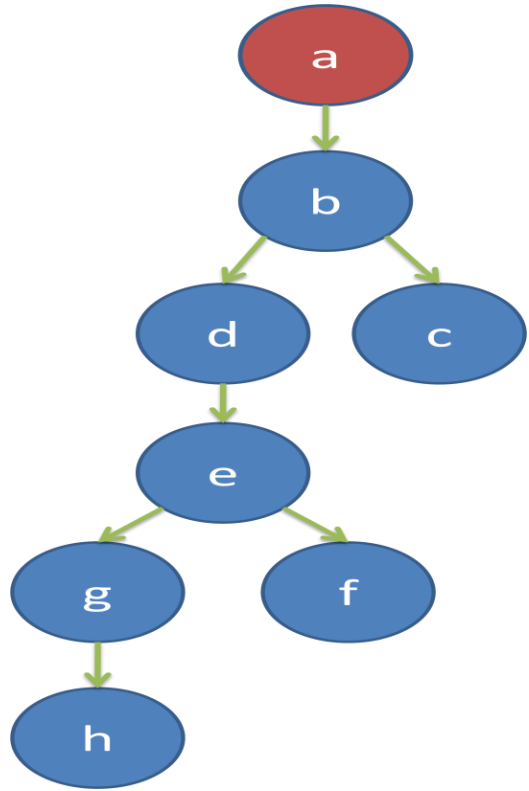

**Fig 2.4 The resulting minimum spanning tree with minimum distance equals to 14.**

The MST can be found using Kruskal's algorithm and the total running time of this algorithm is $(m \log m)$, where $m$ is the number of edges [44], [45]. The MST is valid for any node in the topology is considered as a source. Thus, there is no need to re-build the tree when the source is changed. The MST for Figure 2.2 is shown in Figure 2.4 with total minimum distance (weight) is 14. For the SPT and MST the number of edges always equal to $(n-1)$. For the topology in Figure 2.2 with $n = 8$ and $m = 11$, the resulting SPT and MST have the number of edges is equals to 7. It is worthy to mention that for MST the source (root) node always need more number of hops to reach all nodes than SPT for the same topology. The source in Figure 2.4 need five hops to reach the last node in the MST, while the source in SPT needs only three hops to reach the last node in that tree as shown in Figure 2.3.

### 2.2.3 Problem Statement

For a given random CRN topology with limited available number of channels, the aim is to maximize the overall network throughput and enhancing the packet delivery rate in the CRN. Here, we assume a single secondary source that tries to multicast video packets to a limited number of secondary destinations, assuming a multi-hop environment. However, finding the path from the source to each destination in a multi-hop CRN is a quite challenging. The problem can be summarized as transforming the random network topology into MST or SPT to find the loop free path from the source to each destination in the CRN across relay nodes (routers). In addition, we investigate the performance of the two trees under different channel assignment schemes in order to find the scheme that achieves the maximum possible network throughput.

## 2.3 The Proposed Solution

The main idea of the scheme is to transmit information from a given source to multi destinations in a multi-hop CRN using MST and/or SPT such that the network throughput is improved. Initially, the scheme converts the network topology (undirected graph) to a tree rooted at the source and spanning all the nodes. Because of the multi-hop property in the CRN, a multi-layer transmission is needed to send information packets form the source to all destinations in the tree. The following steps summarize the channel assignment process using POS metric:

> **Step 1:** At the beginning, the source (or relay) i send a Message Announcement (MA) that contains Source ID, Group ID, and its available channel list to its destinations.
>
> **Step 2:** When a CR node k receives the MA, it computes the POS for each channel $j \in C$ using equation 1 then the node k sends an acknowledgment packet back to the source i.
>
> **Step 3:** The source (or relay) i waits for a predetermined time-out to receive all acknowledgment packets from all participating nodes.
>
> **Step 4:** After the time-out, the source (or relay) finds the minimum POS for each channel $j \in C$ and specify the channel that have maximum POS of the minimum.
>
> **Step 5:** The source selects a unified channel from the available channels that have the maximum POS of the minimum by using the rule {POS max = Max {(Min) POS} $\forall j \in C$}.

**A simplified POS algorithm.**

## 2.3.1 Numerical Example Using The SPT

Suppose that we have the network topology with single source (node 1) as shown in Figure 2.5. Assume that the CRN parameters are given in Table 2.2.

**Table 2.2 Parameters' values of the numerical example**

| Parameter | Value |
|---|---|
| Network area | 200m×200m |
| Total number of nodes (N) | 15 |
| Total number of destinations (Nr) | 5 nodes (node 6 - node 10) |
| Total number of primary channels (M) | 6 |
| Average availability time ($\mu_j$) | [10 20 30 40 50 60] ms |
| Idle probability ($P_I$) | 0.7 |
| Transmission power ($p_t$) | 0.1 W |
| Packet size ($D$) | 4KB |

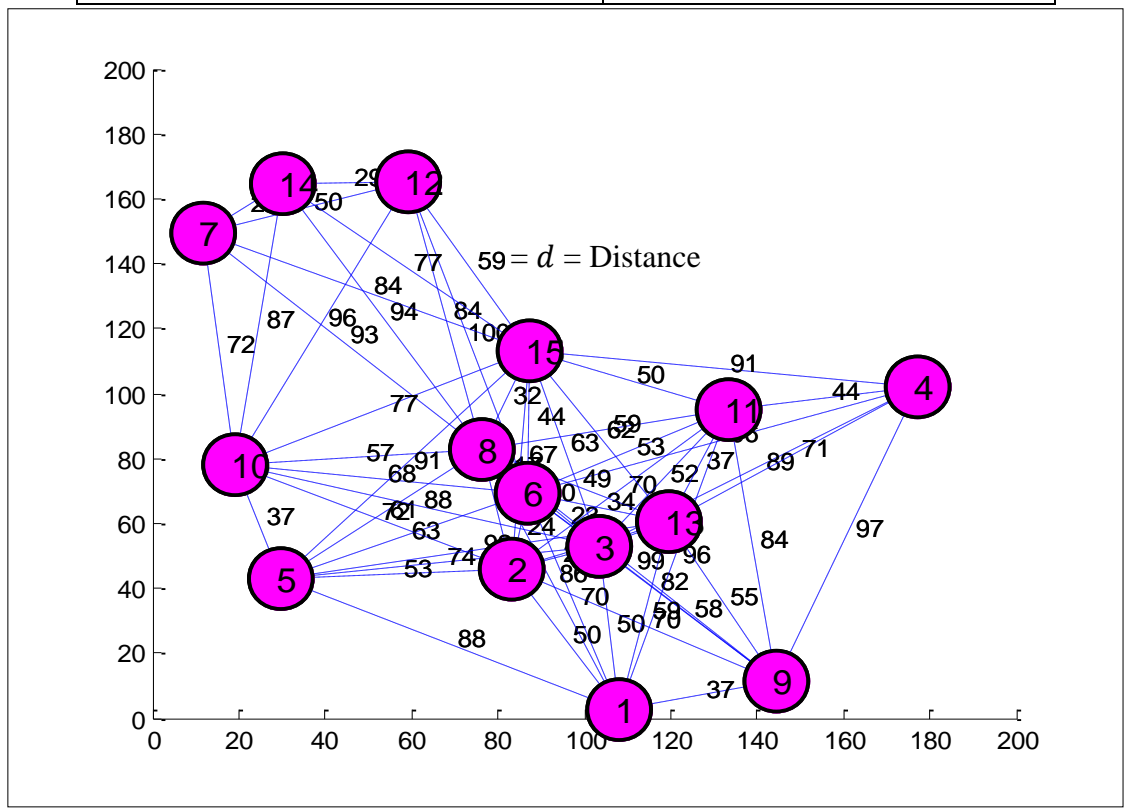

**Fig. 2.5 Network topology with 15 nodes and a single source (node 1).**

Given the above network parameters and topology, the algorithm generates the associated SPT (see Figure 2.6), which is rooted at the source (node 1) and spanning all the nodes in the network topology. After that the source finds a path to each destination in the tree as shown in Figure 2.6, i.e.,

Path to node 6:       1 ⟶ 6           (1 hop)

Path to node 7:       1 ⟶ 8 ⟶ 7      (2 hops)

Path to node 8:       1 ⟶ 8           (1 hop)

Path to node 9:       1 ⟶ 9           (1 hop)

Path to node 10:      1 ⟶ 2 ⟶ 10     (2 hops)

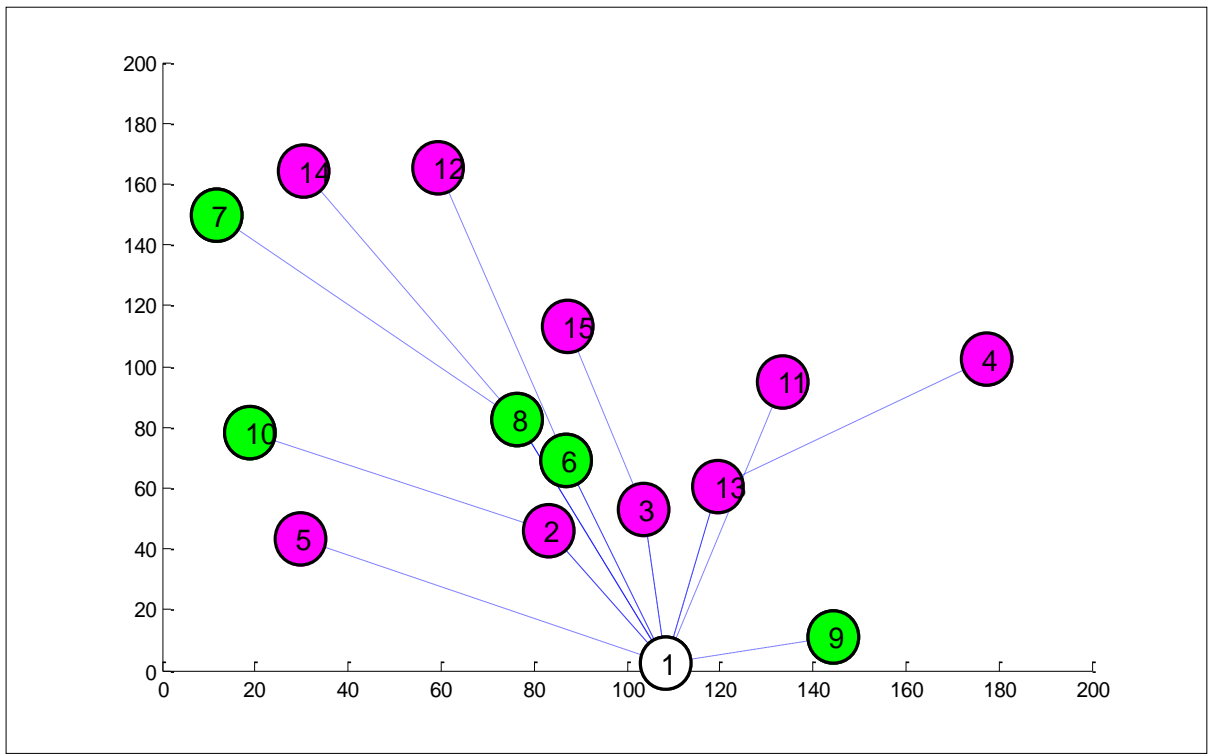

**Fig. 2.6 The resulting SPT for the given network topology.**

When the loop free path is known from the source to each destination in the network, the POS between the source and its receivers for all sub-trees over all links are found by using the equations 1-4. The algorithm divides the tree into layers to do the channel assignment process using POS metric as follows:

**The First Layer Transmission: -** represent the transmission from the source (node 1) to nodes (6, 8, 9, and 2) with unavailable (busy) channels 2 and 3. At first, the POS must be found between the source node and other nodes (6, 8, 9, and 2) over all links. After finding the minimum POS over all available channels (i.e., minimum POS for CH1 is 0.2903, minimum POS for CH4 is 0.7716, minimum POS for CH5 is 0.8869, and minimum POS for CH6 is 0.796), we find the maximum of the minimum POS (Max of {0.2903, 0.7716, 0.8869, 0.796} = 0.8869) through all channels and links to assign that channel for the current transmission as indicated in Table 2.3. Accordingly channel 5 will be selected to complete the first layer transmission.

Table 2.3 POS between node 1 and nodes (6, 8, 9, and 2) at different channels

| CH \ Nodes | CH1 | CH2 | CH3 | CH4 | CH5 | CH6 |
|---|---|---|---|---|---|---|
| **1-6** | 0.534 | 0 | 0 | 0.7716 | 0.8895 | 0.9073 |
| **1-8** | 0.2903 | 0 | 0 | 0.8222 | 0.89 | 0.8691 |
| **1-9** | 0.6563 | 0 | 0 | 0.9207 | 0.9037 | 0.936 |
| **1-2** | 0.6658 | 0 | 0 | 0.9071 | 0.8869 | 0.796 |

**The Second Layer Transmission: -** which is includes two sub trees with two new sources 2 and 8. Note that node 2 is a relay node in the network (it is not a destination). This node contributes in delivering packets from the source to the different destinations.

**Sub Tree 1: -** in sub tree 1, the transmission from the node 2 to the destination node 10 with busy channels 1 and 2 is considered a unicast transmission. In this case the source (node 2) chooses the best available channel that has the highest POS (i.e., channel 6) for the current transmission as shown in Table 2.4.

**Table 2.4 POS between node 2 and 10**

| CH / Nodes | CH1 | CH2 | CH3 | CH4 | CH5 | CH6 |
|---|---|---|---|---|---|---|
| **2-10** | 0 | 0 | 0.842 | 0.8048 | 0.7958 | 0.91 |

**Sub Tree 2: -** in this sub tree, the transmission is performed from node 8 to destination node 7 with busy channels 2, 5, and 6. Recall that node 14 is neither a destination nor a relay node. So, the transmission between nodes 8 and 7 is unicast. The best available channel with the highest POS (i.e., channel 4) is chosen from the available channels for current transmission as indicated in Table 2.5.

**Table 2.5 POS between node 8 and 7**

| CH / Nodes | CH1 | CH2 | CH3 | CH4 | CH5 | CH6 |
|---|---|---|---|---|---|---|
| **8-7** | 0.1939 | 0 | 0.768 | 0.8093 | 0 | 0 |

By comparing the required transmission time $Tr_{(j)}^{(i-k)}$ of the selected channels with the actual available time of those channels, only destinations 6, 9, and 10 will successfully receive the multicast packets with throughputs $V_k$ are as follow:

$V_6 = D / Tr_{(5)}^{(1-6)} = (4)(8)(1024) / 0.0059 = 5.5539$ Mbps

$V_7 = V_8 = 0$ (the packets have not been received because the required transmission time to transmit a packet of the selected channel is more than the actual available time of that channel)

$V_9 = D / Tr_{(5)}^{(1-9)} = (4)(8)(1024) / 0.0051 = 6.4251$ Mbps

$V_{10} = D / (Tr_{(5)}^{(1-2)} + Tr_{(6)}^{(2-10)}) = (4)(8)(1024) / (0.006 + 0.0057) = 2.8$ Mbps

The total and average throughput is 14.7 Mbps and 2.9 Mbps, respectively.

Since only three out of five destinations have received packets, the packet delivery rate (PDR) is:

$3/5 = 0.6 = 60\%$.

# Chapter Three: Simulation Results and Discussion

In this chapter, we conduct simulations experiments to evaluate the performance of the proposed scheme using MATLAB [46]. We compare the proposed scheme (i.e., POS) with the maximum average spectrum availability time (MASA), maximum data rate (MDR) and random selection (RS) schemes.

## 3.1 Simulation Setup

First, we generate a random topology with one CR source, N CR nodes, and Nr CR destinations within (200 ×200) area that coexist with several PRNs in the same geographical area. We consider a Rayleigh fading channel model with path-loss exponent (n) = 4 (i.e., indoor environment) to describe the gain between any two communicating nodes in the network [7], [39]. We set the bandwidth and the thermal-noise power spectral density to BW = 1 MHz, and $N_0=10^{-18}$ W/Hz for all channels. The data packet size and transmission power is set to D = 4 KB and $P_t$ = 0.1 Watt, respectively. The total number of primary channels (M) is 20 with average spectrum availability time ($\mu_j$) that ranges from 2 ms to 70 ms. The average availability time for each PR channel of PRNs is modeled as an Idle/Busy Markov model with average ON and OFF duration of $\mu_j$ and $\lambda_j$, respectively. The idle probability for each channel j is given by $P_I = \mu_j / (\mu_j + \lambda_j)$. We consider three cases for idle probability (i.e., $P_I$ = 0.9 (low

PR traffic load), $P_I = 0.5$ (moderate PR traffic load), and $P_I = 0.1$ (high PR traffic load)). We suppose there are guard band channels and exclusive channel occupancy.

## 3.2 Performance Evaluation for Shortest Path Tree (SPT)

According to the SPT method, the proposed algorithm finds the minimum cost (i.e., distance) from the source to the each destination in the network. The shortest path tree is implemented in Matlab using Djikestra algorithm [46]. In this section, we conduct simulations to evaluate the performance of the proposed scheme with the state-of-the-art schemes. Then, we evaluate the performance of the proposed scheme under different network parameters. In this set of experiment, we consider the total number of nodes (N) is 40 and the total number of destinations (Nr) is 16.

### 3.2.1 Performance Evaluation under Low Traffic Load ($P_I = 0.9$)

The simulation results provided in this subsection show the performance evaluation of the proposed scheme in terms of throughput and packet delivery rate. The results also show the improvement gained of using the proposed scheme compared to other schemes.

**3.2.1.1 Throughput and Packet Delivery Rate Performance versus Channel Bandwidth**

Figures 3.1 and 3.2 respectively show the throughput and the packet delivery rate (PDR) as a function of the channel bandwidth (BW). Note that as the channel BW increases the throughput and packet delivery rate increase. This is because the channel capacity (i.e., data rate) is proportional to the channel BW. However, the POS scheme outperforms MASA,

MDR, and RS schemes in term of the packet delivery rate by up to 12.5%, 85%, and 133%, respectively. Also, the POS scheme outperforms MASA, MDR, and RS schemes in term of throughput by 10%, 45%, and 66%, respectively.

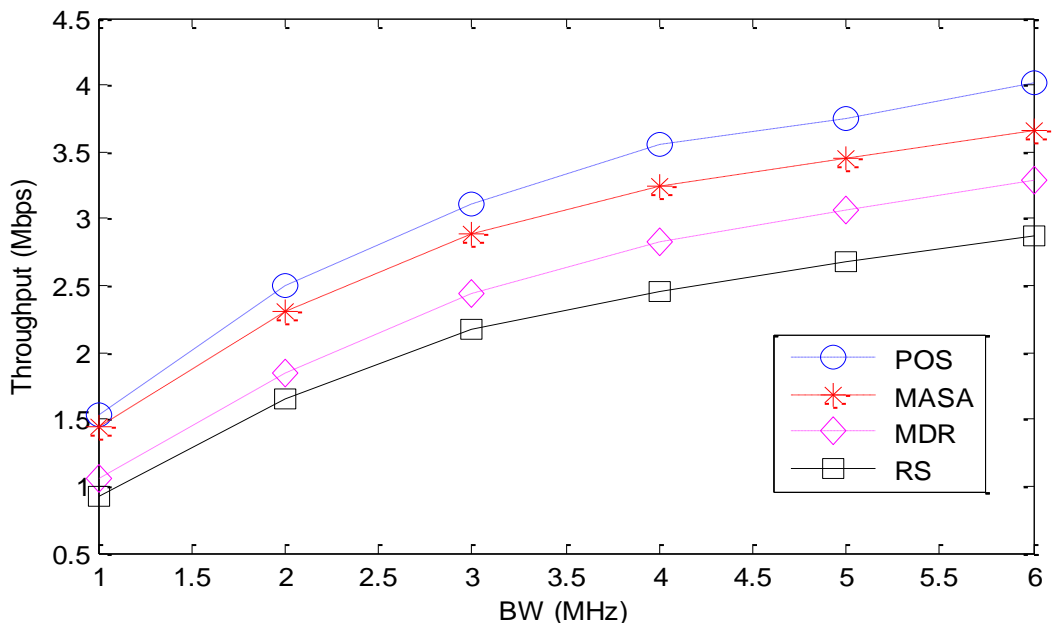

**Fig. 3.1 Throughput vs. channel bandwidth under $P_I = 0.9$.**

(N = 40, Nr = 16, M = 20, $P_t$ = 0.1W, D = 4KB)

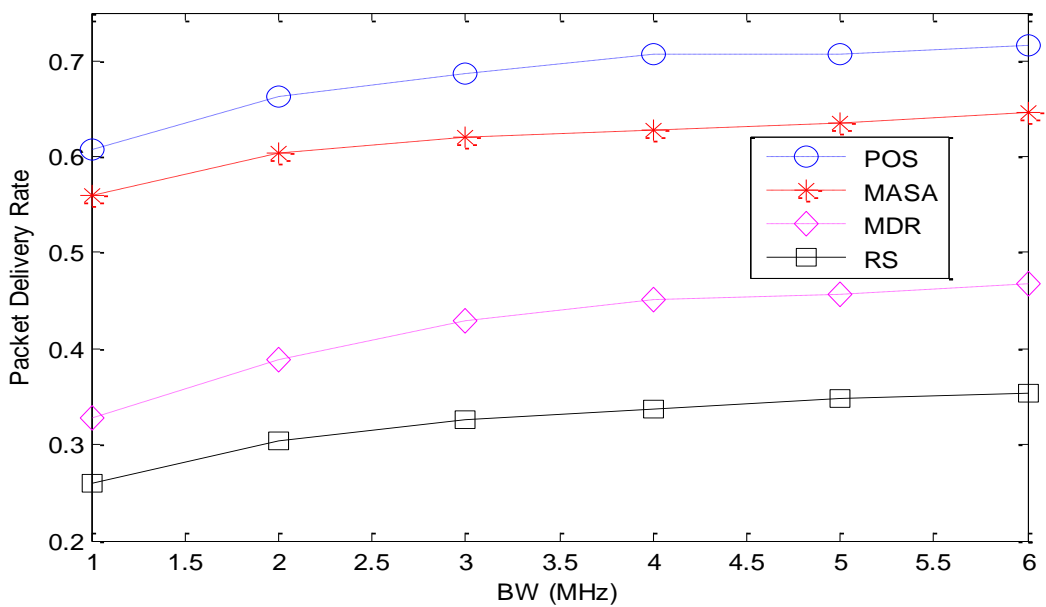

**Fig. 3.2 PDR vs. channel bandwidth under $P_I = 0.9$.**

(N = 40, Nr = 16, M = 20, $P_t$ = 0.1W, D = 4KB)

### 3.2.1.2 Throughput and Packet Delivery Rate Performance versus The packet size

Figures 3.3 and 3.4 plot the throughput and the packet delivery rate (PDR) for all the schemes versus the packet size (D).

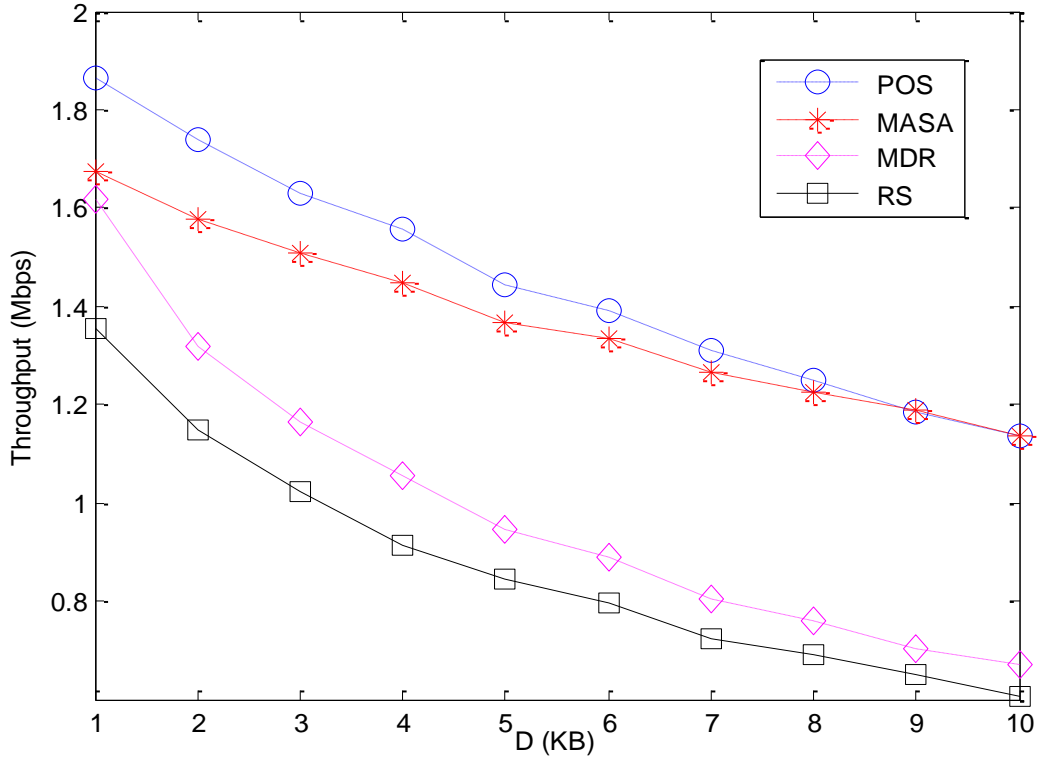

**Fig. 3.3 Throughput vs. the packet size under $P_I = 0.9$.**

(N = 40, Nr = 16, M = 20, $P_t$ = 0.1W, BW = 1MHz)

Figure 3.3 shows the significant improvement gain in term of throughput. Figure 3.4 shows the improvement gain in term of the packet delivery rate. The proposed scheme outperforms MASA, MDR, and RS schemes in term of the packet delivery rate by up to 9.3%, 110%, and 162%, respectively. Also, the proposed scheme outperforms MASA, MDR, and RS schemes in term of the throughput by up to 11.28 %, 69.5 %, and 87 %, respectively.

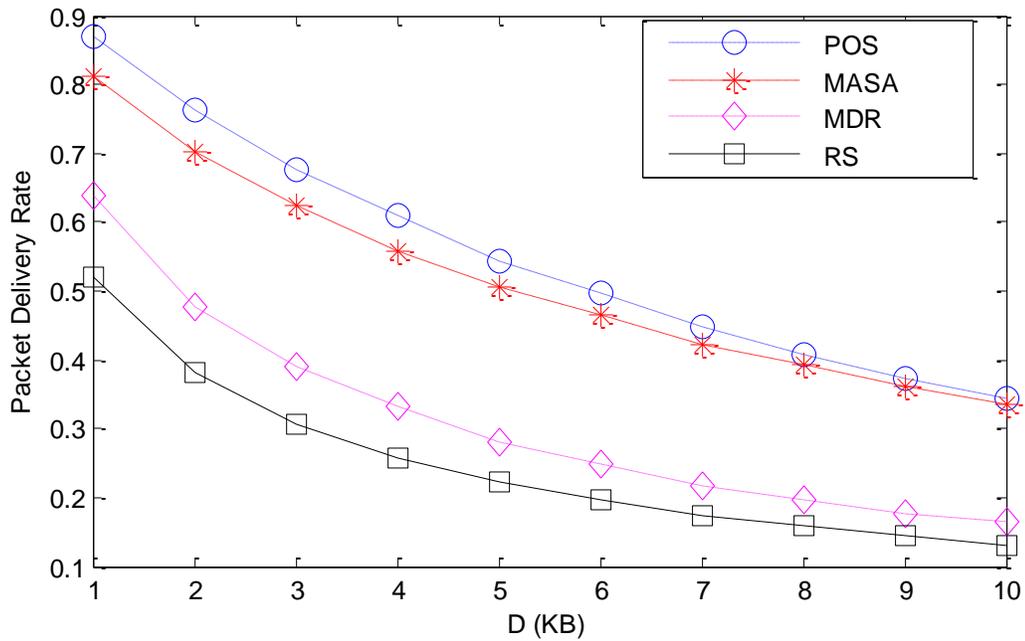

**Fig. 3.4 PDR vs. the packet size under $P_I = 0.9$.**

(N = 40, Nr = 16, M = 20, $P_t$ = 0.1W, BW = 1MHz)

Note that as the packet size increases, the PDR and throughput decreases. This can be explained using equation (2) as follows: - as the packet size increases (and consequently increased the required transmission time), the required channel availability time increases which in turn makes finding the appropriate channel harder.

**3.2.1.3 Throughput and Packet Delivery Rate Performance versus Number of Primary Channels**

Increasing the number of primary channels improve network throughput and PDR because increasing the probability of available channels for CRNs as shown in Figures 3.5 and 3.6. As can be noted here increasing the number of channels will enhance throughput and PDR. The proposed scheme outperforms MASA, MDR, and RS schemes in term of the packet delivery rate by up to 9.4%, 86%, and 137%, respectively. Also, the proposed scheme outperforms

MASA, MDR, and RS schemes in term of the throughput by up to 12.8%, 47.8%, and 67.6%, respectively.

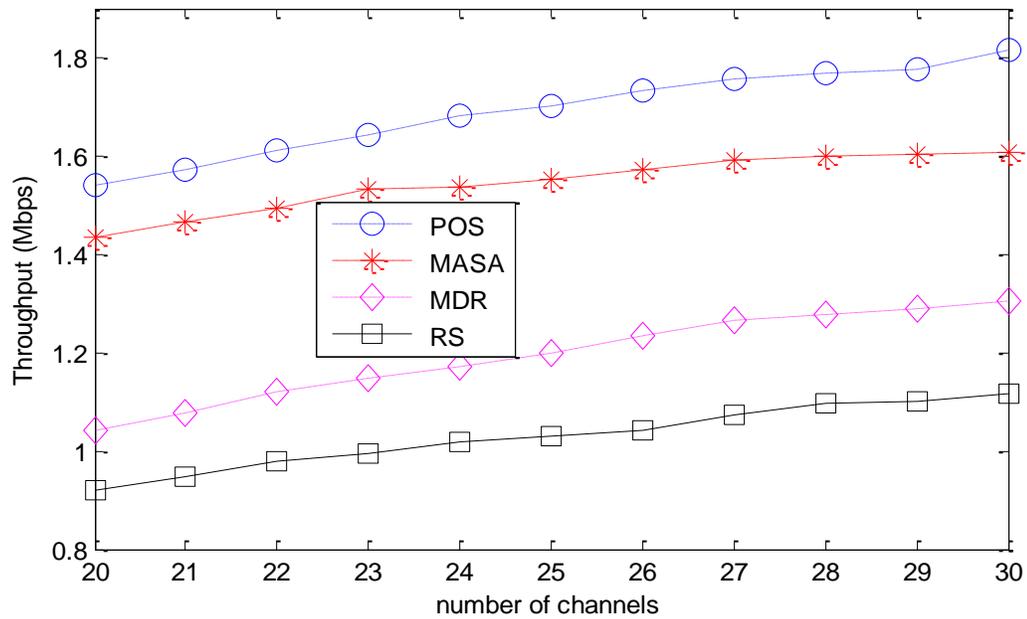

**Fig. 3.5 Throughput vs. number of primary channels under $P_I$ = 0.9.**

(N = 40, Nr = 16, $P_t$ = 0.1W, BW = 1MHz, D = 4KB)

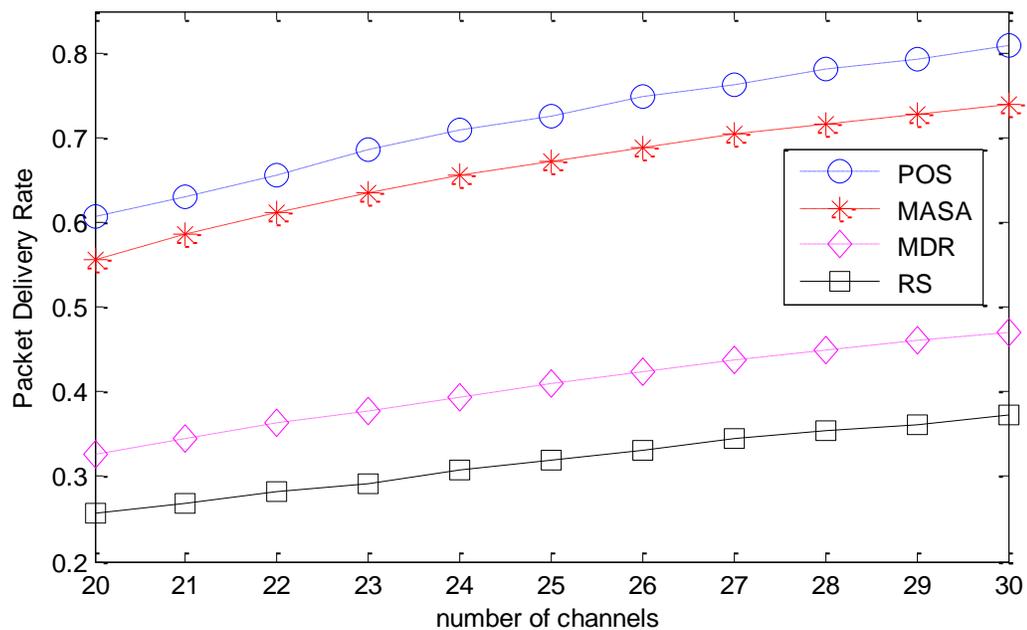

**Fig. 3.6 PDR vs. number of primary channels under $P_I$ = 0.9**

(N = 40, Nr = 16, $P_t$ = 0.1W, BW = 1MHz, D = 4KB)

## 3.2.1.4 Throughput and Packet Delivery Rate Performance versus The Transmission Power

It is well known that increasing the transmission power can significantly improve the achieved data rate, decrease the required transmission time, and result in improving network performance (see Figures 3.7 and 3.8). However, the maximum transmission power for CR users are very limited, compared to the PU transmission power [47]. One watt is too high for such applications because this will increase interference significantly. The proposed scheme outperforms MASA, MDR, and RS schemes in term of the packet delivery rate by up to 8.6%, 85%, and 133%, respectively. Also, the proposed scheme outperforms MASA, MDR, and RS schemes in term of the throughput by up to 10%, 48%, and 68.6%, respectively.

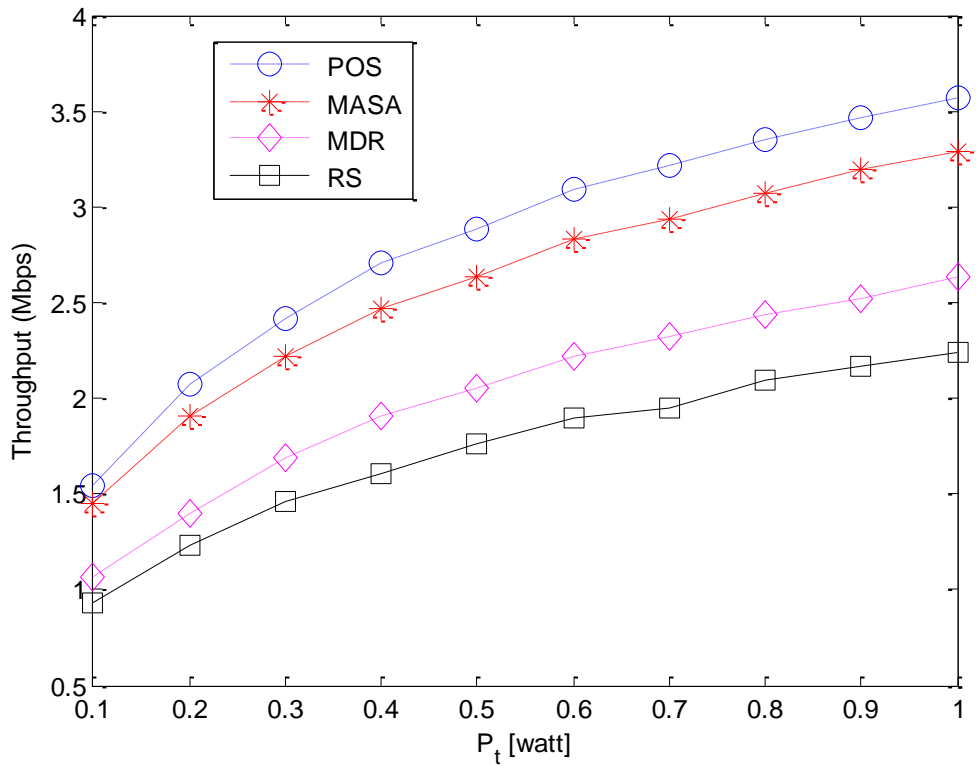

**Fig. 3.7 Throughput vs. the transmission power under $P_I$ = 0.9.**

(N = 40, Nr = 16, M = 20, BW = 1MHz, D = 4KB)

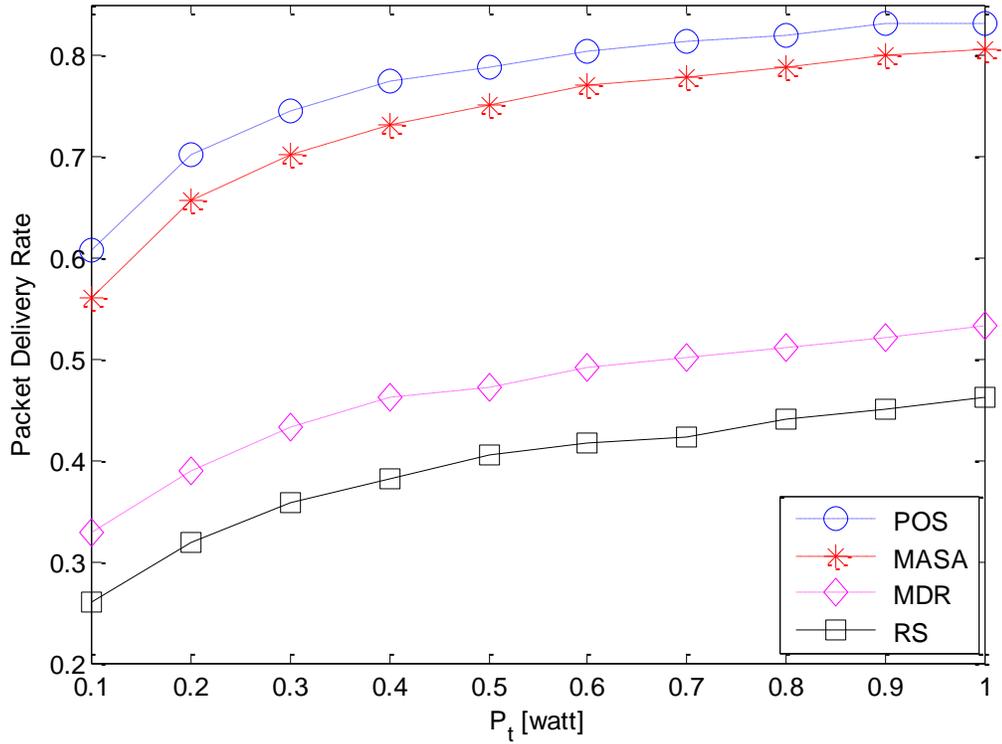

**Fig. 3.8 PDR vs. the transmission power under $P_I = 0.9$.**

(N = 40, Nr = 16, M = 20, BW = 1MHz, D = 4KB)

### 3.2.2 Performance Evaluation under Moderate Traffic Load ($P_I = 0.5$)

The PUs traffic load significantly affects the overall CRN performance. In this subsection, we study the performance of the proposed scheme compared to the other schemes in terms of throughput and PDR under different network parameters at moderate traffic load ($P_I = 0.5$).

**3.2.2.1 Throughput and Packet Delivery Rate Performance versus Channel Bandwidth**

As we mentioned before, as the channel bandwidth increases the network performance is improved. The same number of channels with higher bandwidth yields more available spectrum, and thus the channel capacity is increased as shown in Figures 3.9 and 3.10.

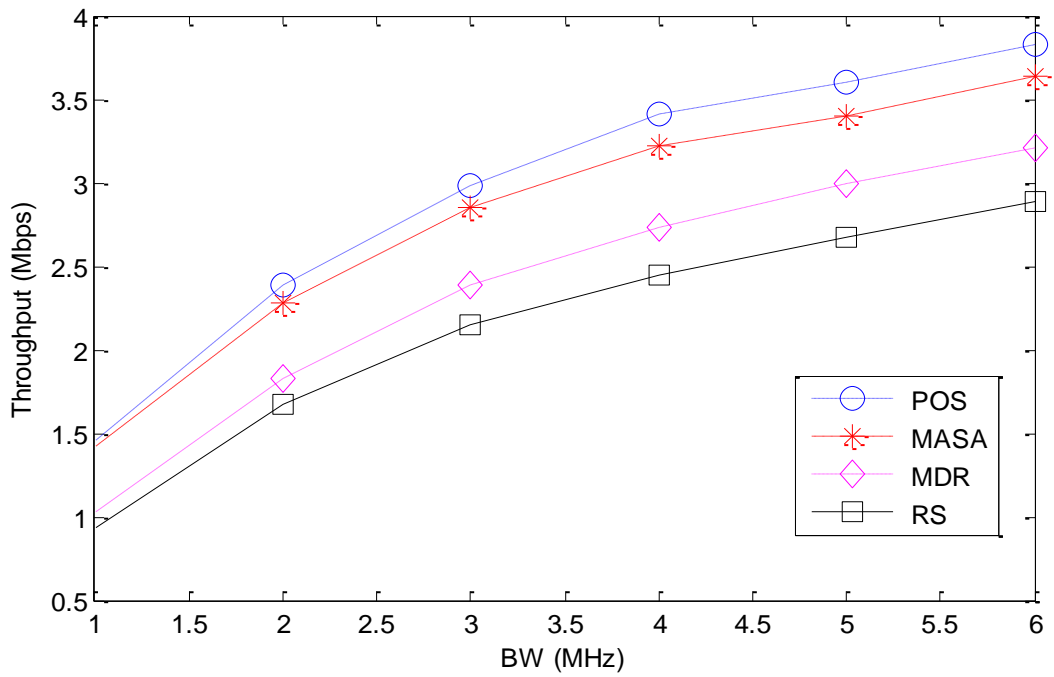

**Fig. 3.9 Throughput vs. channel bandwidth under $P_I = 0.5$.**

($N = 40$, $Nr = 16$, $M = 20$, $P_t = 0.1W$, $D = 4KB$)

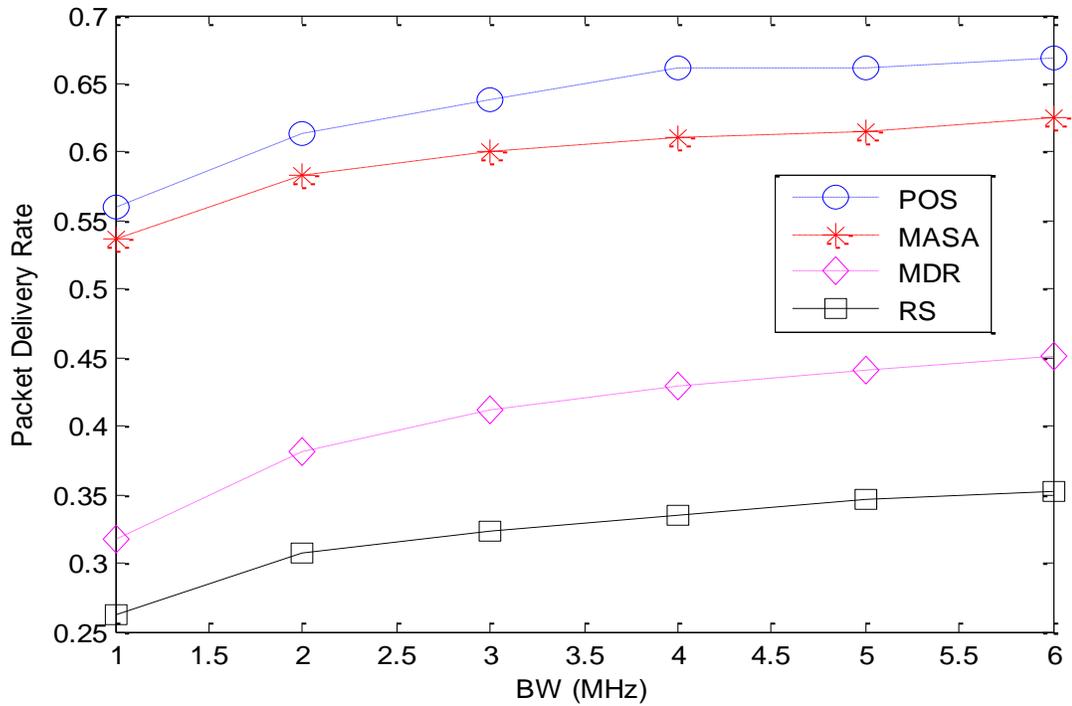

**Fig. 3.10 PDR vs. channel bandwidth under $P_I = 0.5$.**

($N = 40$, $Nr = 16$, $M = 20$, $P_t = 0.1W$, $D = 4KB$)

Specifically the proposed scheme outperforms MASA, MDR, and RS schemes in term of throughput by up to 6.1%, 41.6%, and 55%, respectively. Also, the proposed scheme outperforms MASA, MDR, and RS schemes in term of the packet delivery rate by up to 8.3%, 76.7%, and 113%, respectively.

**3.2.2.2 Throughput and Packet Delivery Rate Performance versus The packet size**

Figures 3.11 and 3.12 show the performance evaluation of the proposed scheme compared to the other schemes in terms of throughput and PDR versus the packet size (D in KB). We note that increasing the packet size degrades network performance. Increasing the packet size increases the required transmission time and hence the probability of appearing an PU meanwhile CR transmission is increased. Specifically the proposed scheme outperforms MASA, MDR, and RS schemes in term of throughput by up to 7%, 64%, and 76%, respectively. Also, the proposed scheme outperforms MASA, MDR, and RS schemes in term of the packet delivery rate by up to 5.6%, 97%, and 138.5%, respectively.

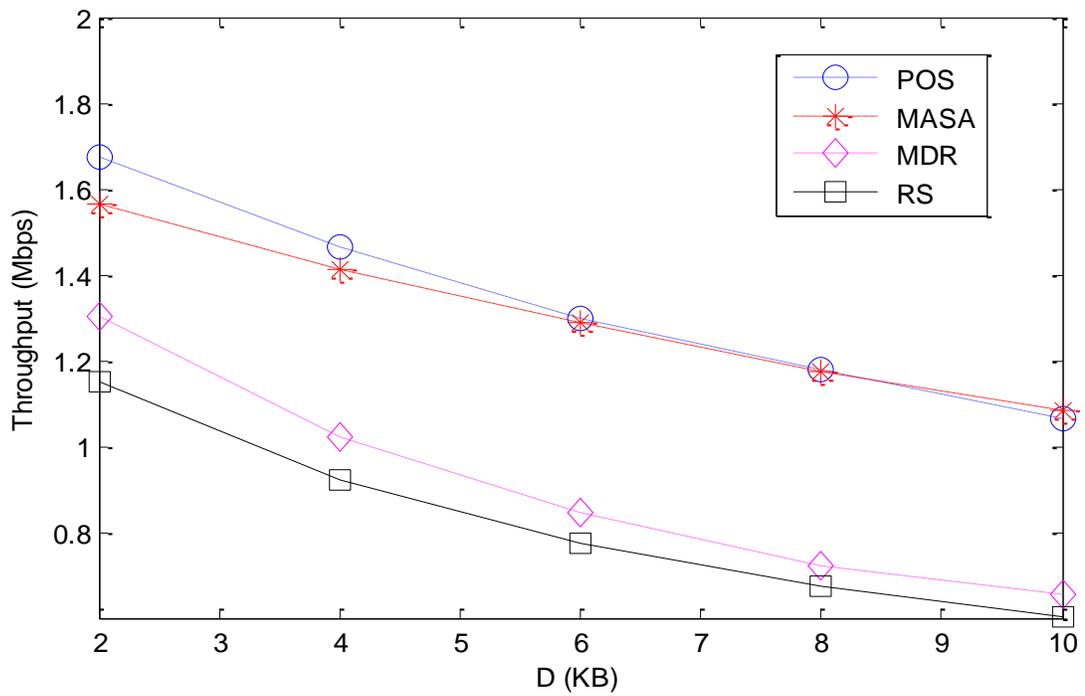

**Fig. 3.11 Throughput vs. the packet size under $P_I = 0.5$.**

($N = 40$, $Nr = 16$, $M = 20$, $P_t = 0.1W$, $BW = 1MHz$)

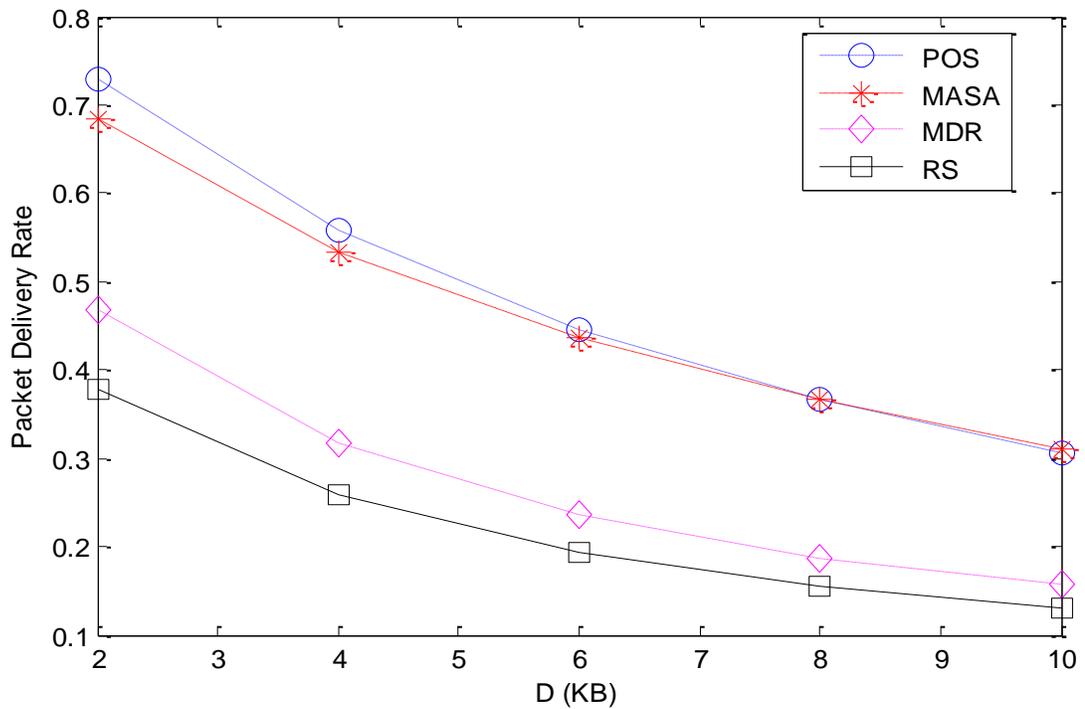

**Fig. 3.12 PDR vs. the packet size under $P_I = 0.5$.**

($N = 40$, $Nr = 16$, $M = 20$, $P_t = 0.1W$, $BW = 1MHz$)

### 3.2.2.3 Throughput and Packet Delivery Rate Performance versus Number of Primary Channels

Figures 3.13 and 3.14 respectively, show the throughput and PDR performance versus number of primary channels (M). Obviously, increasing the channel bandwidth or the number of channels will improve network performance. Specifically, increasing the number of channels will increase the number of idle channels to be utilized by SUs. When idle probability is equal to 0.5 that means on average half of primary channels is can be utilized by SUs. However, the proposed scheme outperforms MASA, MDR, and RS schemes in term of throughput by up to 10%, 43%, and 60%, respectively. Also, the proposed scheme outperforms MASA, MDR, and RS schemes in term of the packet delivery rate by up to 7.6%, 77%, and 121%, respectively.

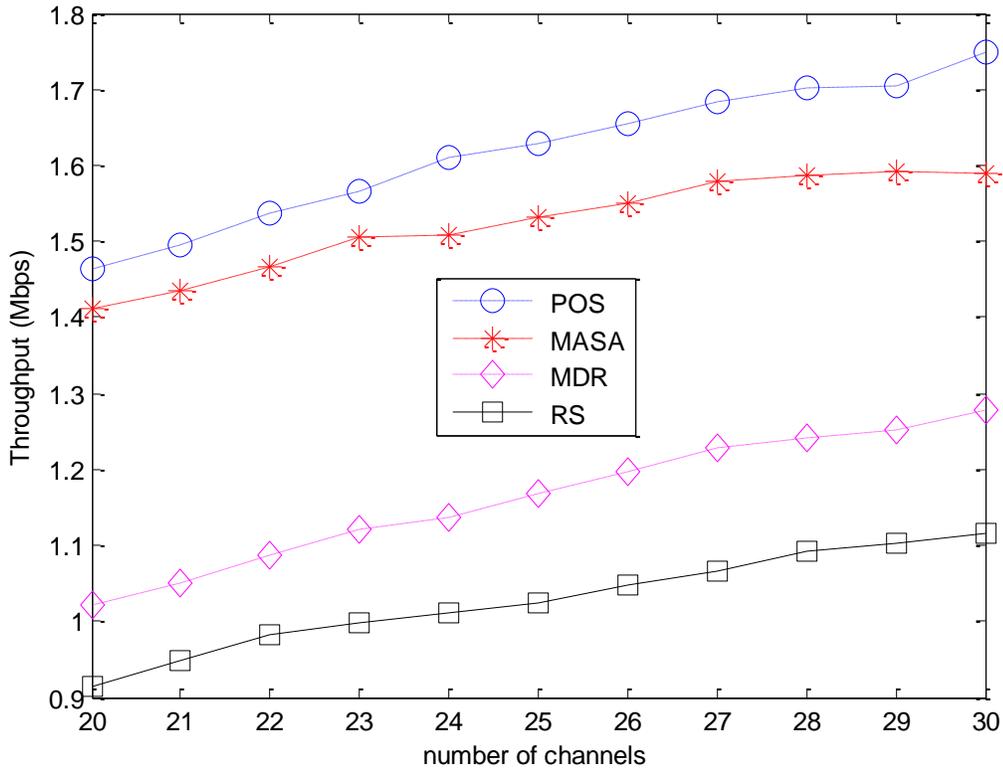

**Fig. 3.13 Throughput vs. number of primary channels under $P_I = 0.5$.**

(N = 40, Nr = 16, $P_t$ = 0.1W, BW = 1MHz, D = 4KB)

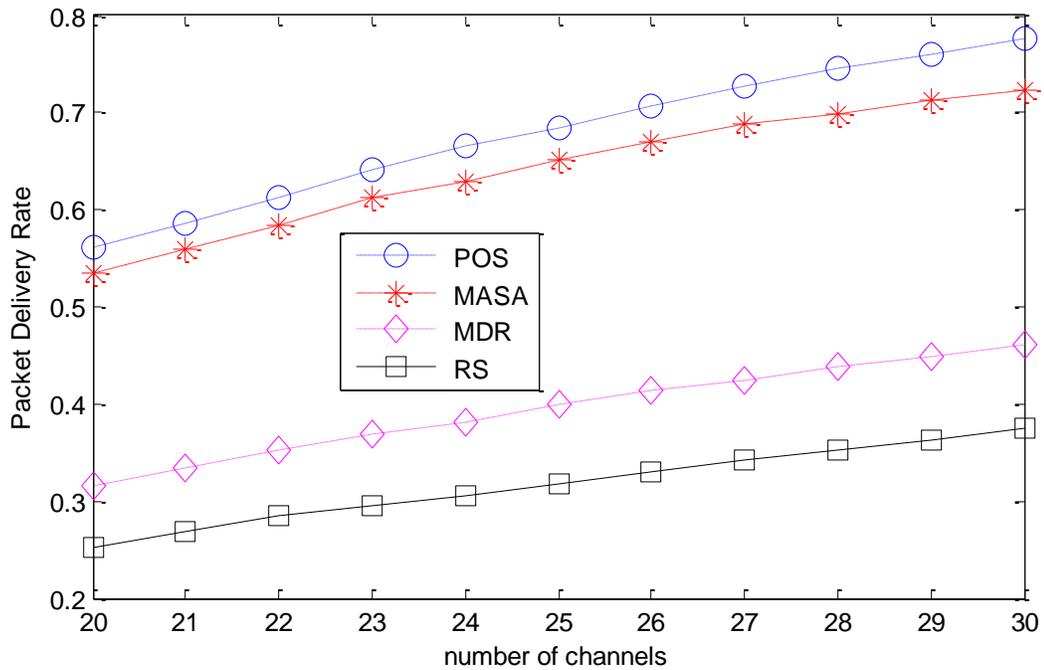

**Fig. 3.14 PDR vs. number of primary channels under $P_I = 0.5$.**

($N = 40$, $Nr = 16$, $P_t = 0.1W$, $BW = 1MHz$, $D = 4KB$)

**3.2.2.4 Throughput and Packet Delivery Rate Performance versus The Transmission Power**

In Figures 3.15 and 3.16, we investigate the CRN performance versus the CR transmission power. An improvement gain increases in CRN performance as the transmission power increases. However, one watt is too high for such applications because this will increase interference significantly. The proposed scheme outperforms MASA, MDR, and RS schemes in term of throughput by up to 7%, 41.7%, and 62%, respectively. Also, the proposed scheme outperforms MASA, MDR, and RS schemes in term of the packet delivery rate by up to 4.5%, 76.7%, and 113%, respectively.

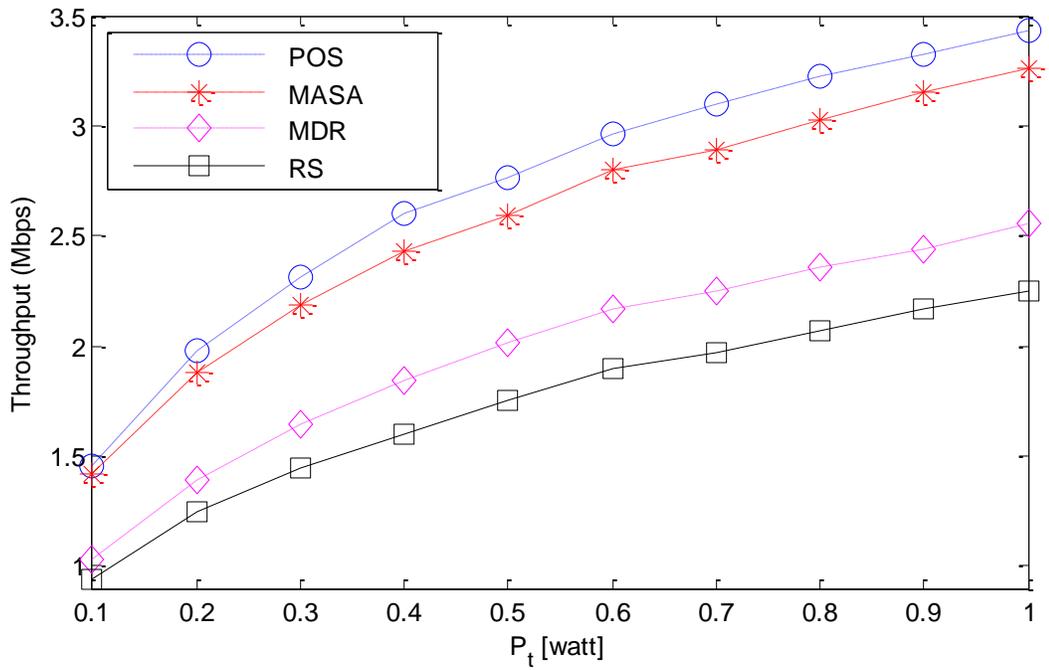

**Fig. 3.15 Throughput vs. the transmission power under $P_I = 0.5$.**

(N = 40, Nr = 16, M = 20, BW = 1MHz, D = 4KB)

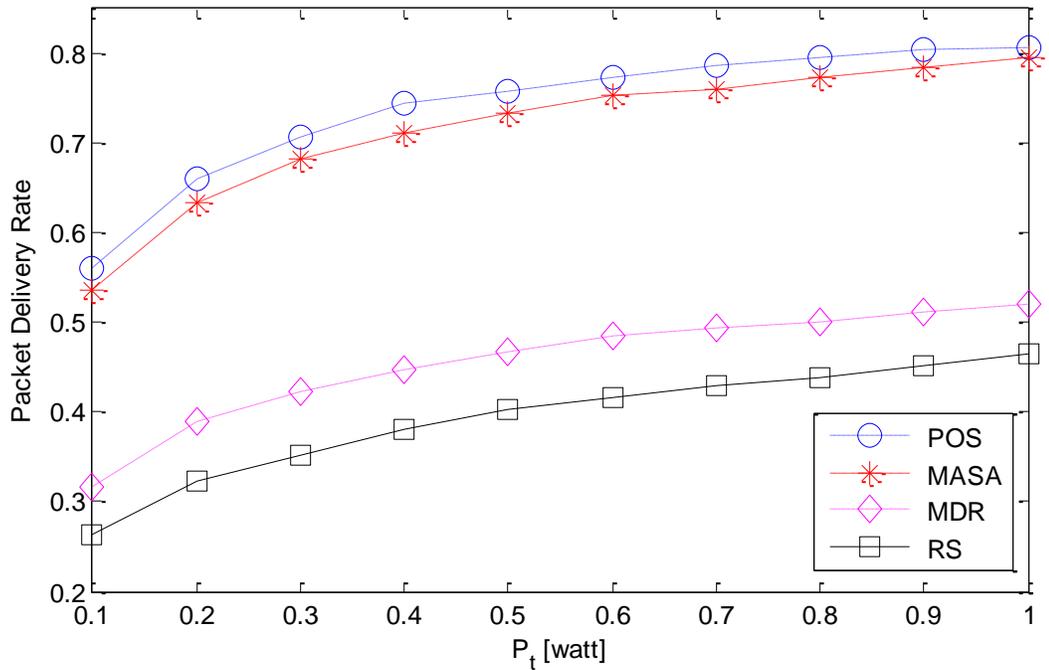

**Fig. 3.16 PDR vs. the transmission power under $P_I = 0.5$.**

(N = 40, Nr = 16, M = 20, BW = 1MHz, D = 4KB)

## 3.2.3 Performance Evaluation under High Traffic Load ($P_I = 0.1$)

With high PU traffic load, the channels are busy most of the times. Specifically, the number of available channels for the CRN will be very small. Since the number of available channels is very low, the process of channel assignment for SU is limited to a small set and sometimes to a single choice. As a result, we may notice that the performance of POS and MASA schemes is almost the same and outperform MDR and RS schemes in all cases. In other words, the improvement gain of the proposed scheme compared to the other schemes when the idle probability is equal to 0.9 is higher than when the idle probability is equal to 0.5. At the end of this Section, we will observe the same result when the idle probability changes from 0.5 to 0.1. This is because when the idle probability is high, more channels are available for CRN. This results in more choices for the channel assignment. However, when we have more available channels, the proposed scheme shows better performance over other schemes because of, the better channel assignment process.

### 3.2.3.1 Throughput and Packet Delivery Rate Performance versus Channel Bandwidth

Figures 3.17 and 3.18 show the performance evaluation of the proposed scheme using SPT compared to the other schemes. Recall that increasing channel bandwidth (BW) improves CRN performance. The proposed scheme performance is comparable to the MASA and they outperform the other schemes.

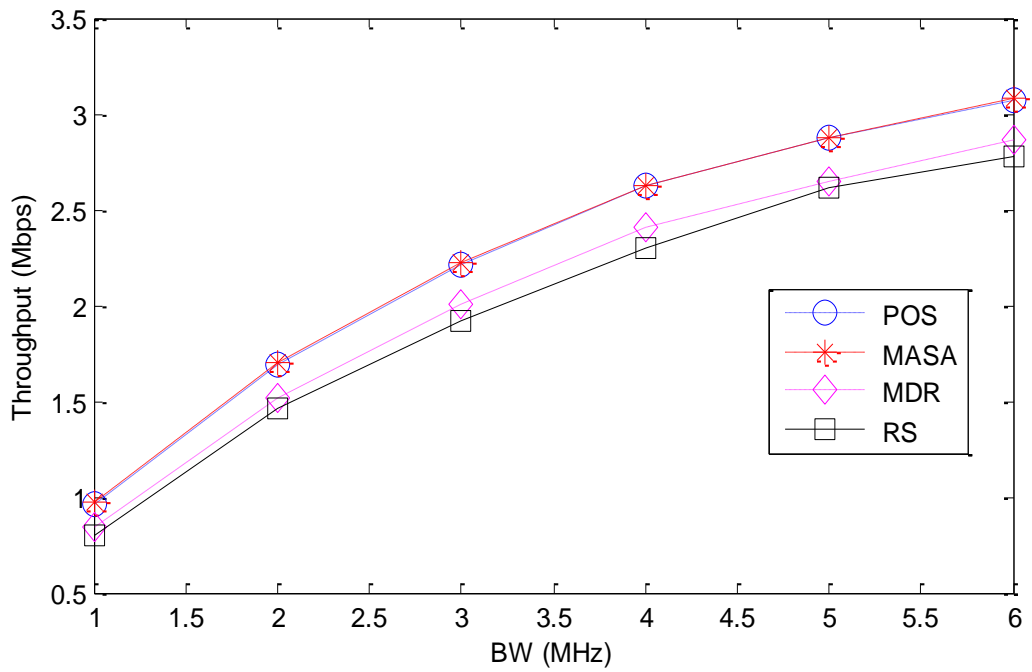

**Fig. 3.17 Throughput vs. channel bandwidth under $P_I = 0.1$.**

(N = 40, Nr = 16, M = 20, $P_t$ = 0.1W, D = 4KB)

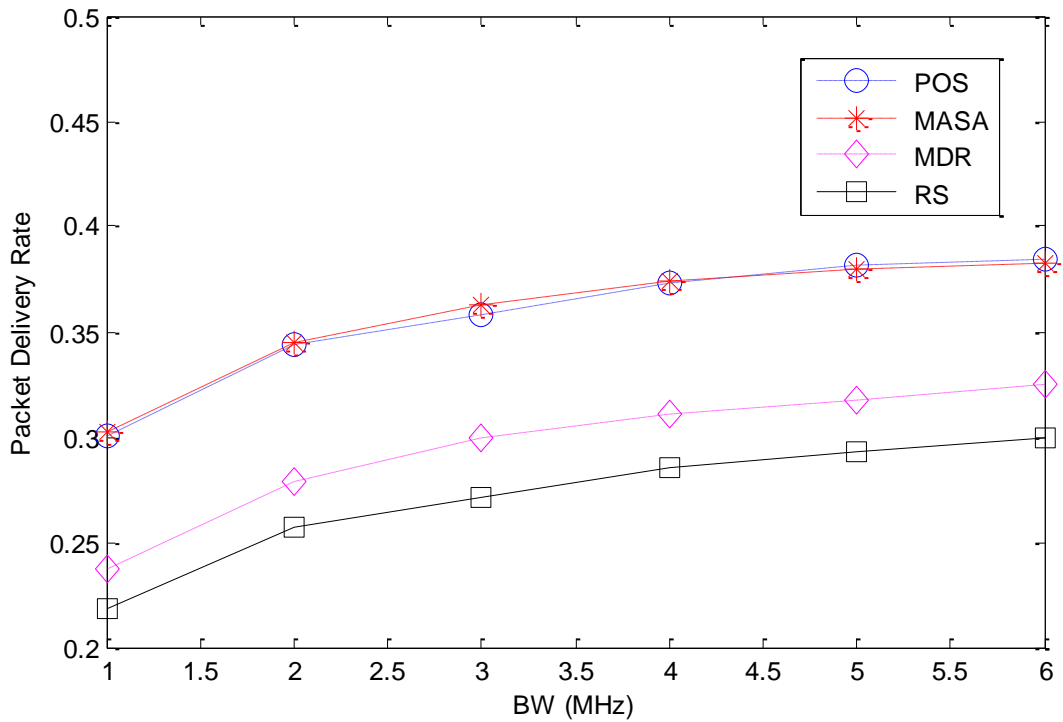

**Fig. 3.18 PDR vs. channel bandwidth under $P_I = 0.1$.**

(N = 40, Nr = 16, M = 20, $P_t$ = 0.1W, D = 4KB)

### 3.2.3.2 Throughput and Packet Delivery Rate Performance versus The Packet Size

As we discussed before, increasing the packet size results in degradation the CRN performance (see Figures 3.19 and 3.20). However, the performance of the proposed scheme is comparable to the performance of the MASA scheme, since in the case of high PU activity the channels are barely available. So, the most significant factor for successful SU transmissions is finding a suitable available channel with average available time that is greater than the required time for the SU transmission.

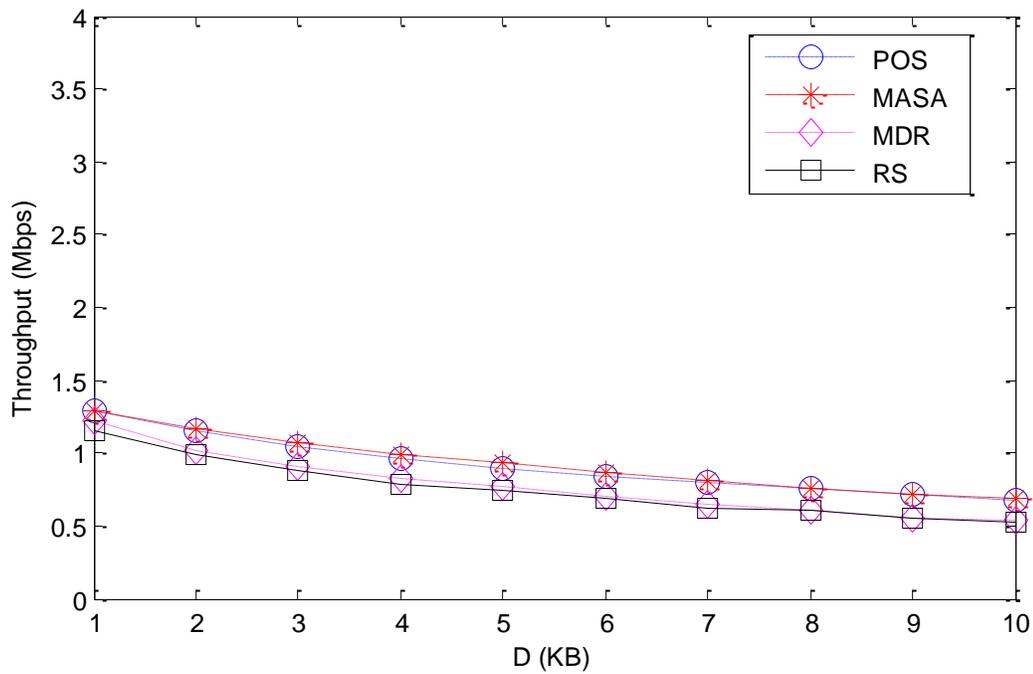

**Fig. 3.19 Throughput vs. the packet size under $P_I = 0.1$.**

(N = 40, Nr = 16, M = 20, $P_t$ = 0.1W, BW = 1MHz)

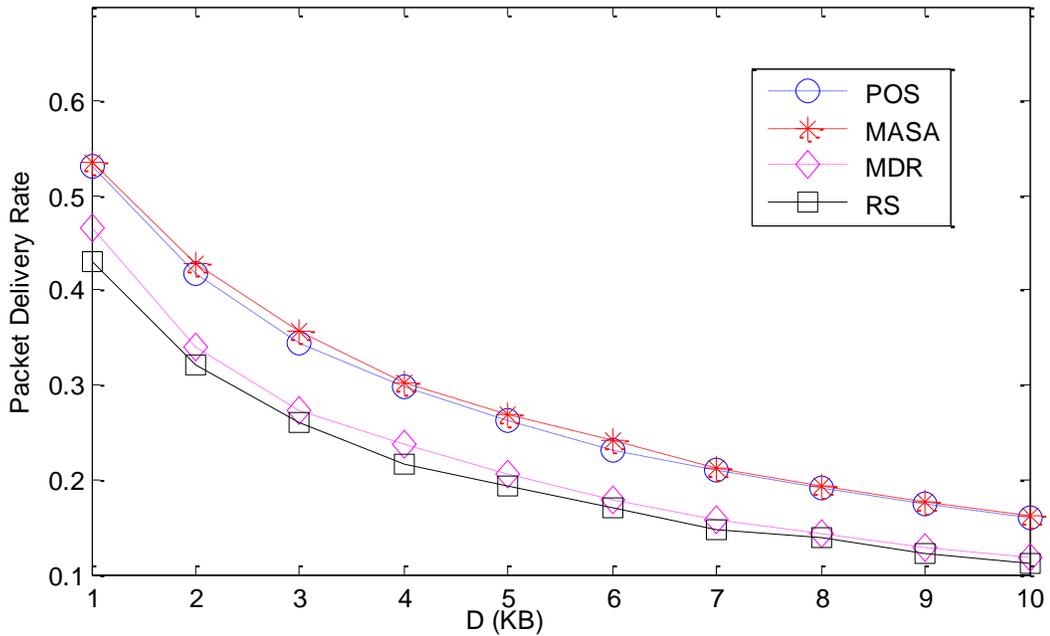

**Fig. 3.20 PDR vs. the packet size under $P_I = 0.1$.**

(N = 40, Nr = 16, M = 20, $P_t$ = 0.1W, BW = 1MHz)

### 3.2.3.3 Throughput and Packet Delivery Rate Performance versus Number of Primary Channels

The CRN performance is improved when increasing the number of channels. For idle probability equal to 0.1, the number of available channels for the CRN on average is 2-3. However, the performance of the proposed scheme outperforms both the RS and MDR schemes, but it is comparable to the MASA scheme (see Figures 3.21 and 3.22) under high PU activity (e.g., $P_I = 0.1$).

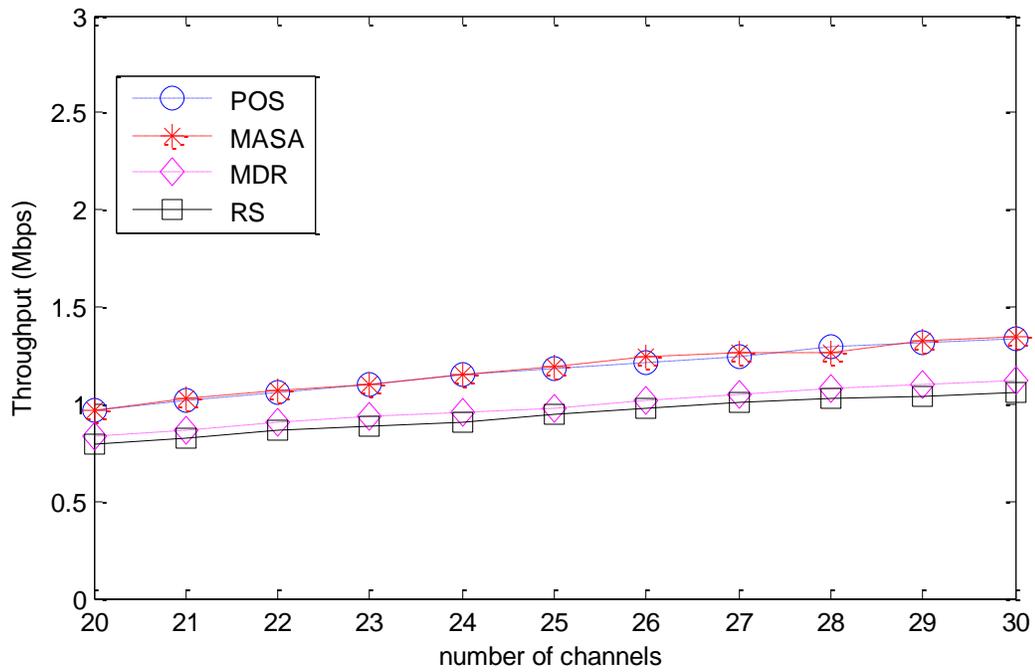

**Fig. 3.21 Throughput vs. number of primary channels under $P_I = 0.1$.**

(N = 40, Nr = 16, $P_t$ = 0.1W, BW = 1MHz, D = 4KB)

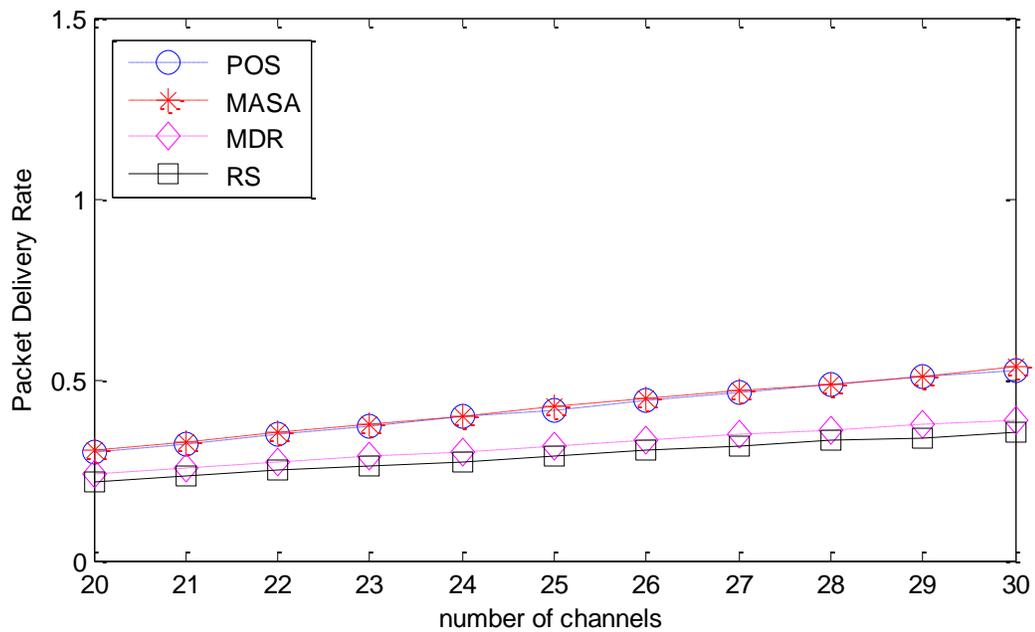

**Fig. 3.22 PDR vs. number of primary channels under $P_I = 0.1$.**

(N = 40, Nr = 16, $P_t$ = 0.1W, BW = 1MHz, D = 4KB)

**3.2.3.4 Throughput and Packet Delivery Rate Performance versus The Transmission Power**

As we observed before, increasing the transmission power improves the CRN performance. However, one watt is too high for such applications because this will increase the interference significantly. Specifically, the performance of the proposed scheme is comparable to the MASA scheme and they outperform the other schemes (see Figures 3.23 and 3.24).

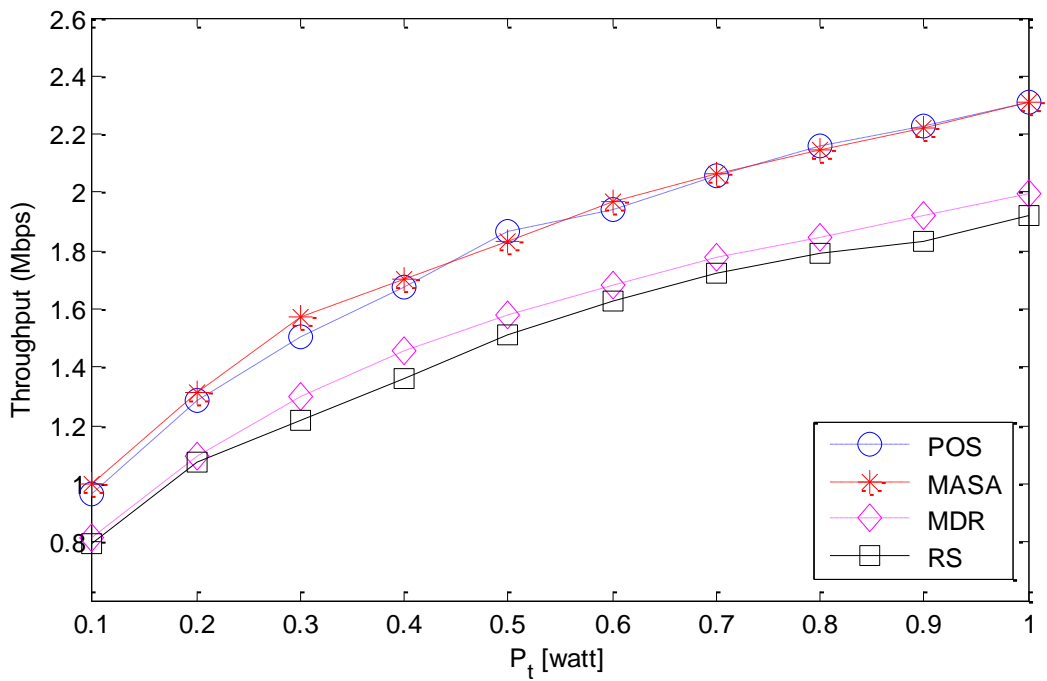

**Fig. 3.23 Throughput vs. the transmission power under $P_I = 0.1$.**

(N = 40, Nr = 16, M = 20, BW = 1MHz, D = 4KB)

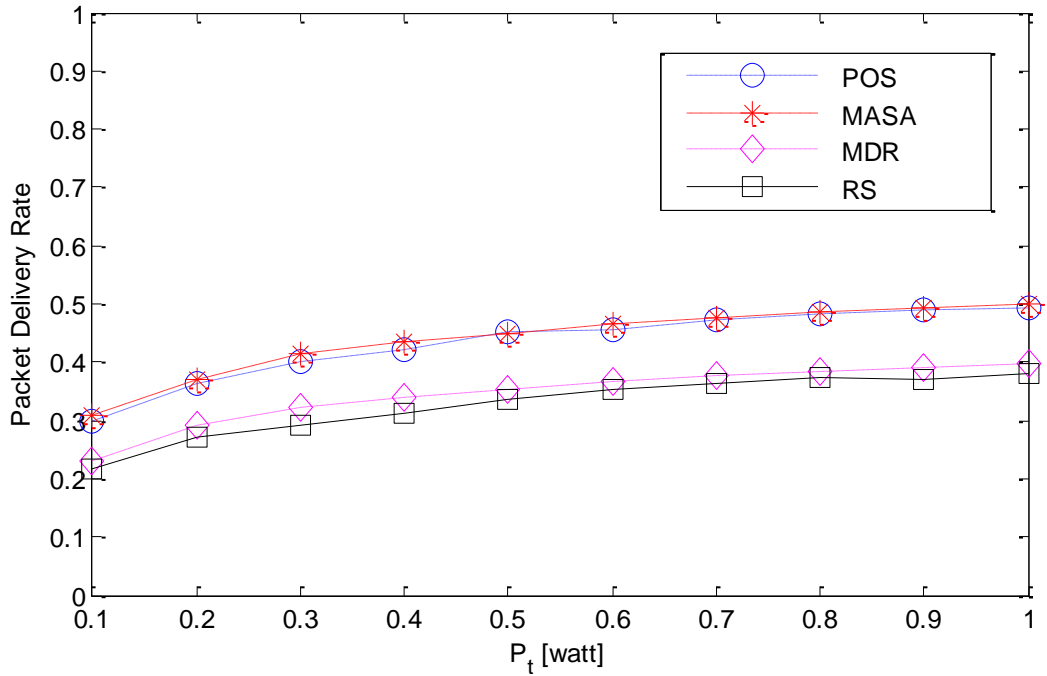

**Fig. 3.24 PDR vs. the transmission power under $P_I = 0.1$.**
(N = 40, Nr = 16, M = 20, BW = 1MHz, D = 4KB)

### 3.2.4 Impact of PUs Traffic Load

The performance of the proposed scheme outperforms the other schemes under low-to-moderate PU activity. Whereas, its performance is comparable to that of the MASA scheme under high PU activity (e.g., $P_I \leq 0.3$), meanwhile it outperforms MDR and RS schemes (see Figures 3.25 and 3.26). The proposed scheme outperforms MASA, MDR, and RS schemes in term of throughput by up to 7%, 47%, and 66.5%, respectively. Also, the proposed scheme outperforms MASA, MDR, and RS schemes in term of packet delivery rate by up to 8.6%, 85.5%, and 135%, respectively.

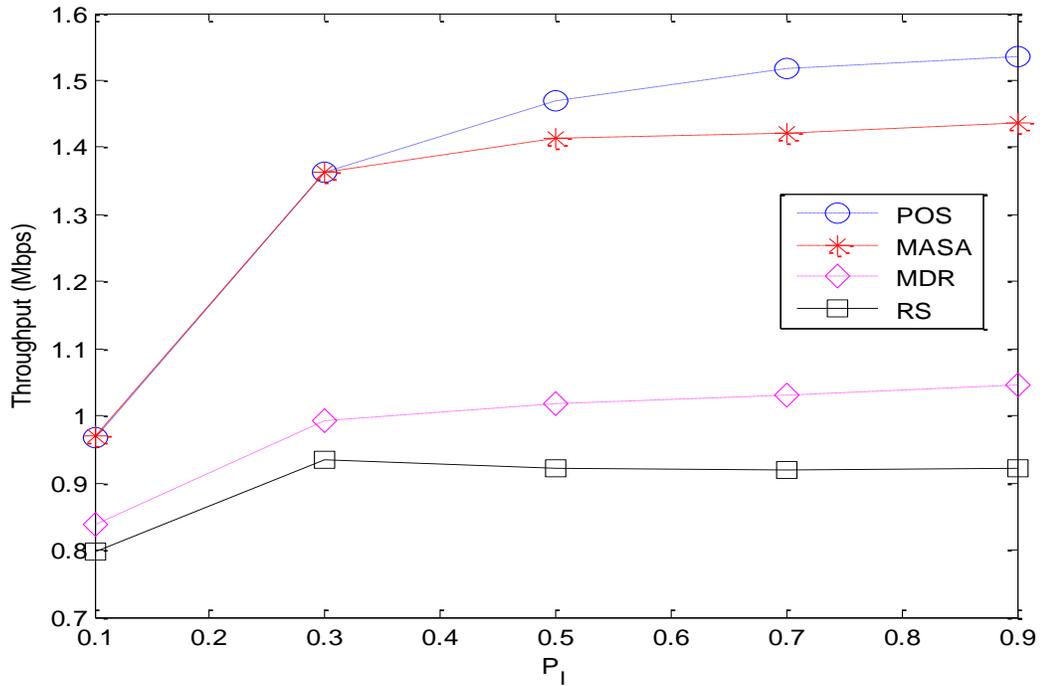

**Fig. 3.25 Throughput vs. idle probability.**

(N = 40, Nr = 16, M = 20, $P_t$ = 0.1W, BW = 1MHz, D = 4KB)

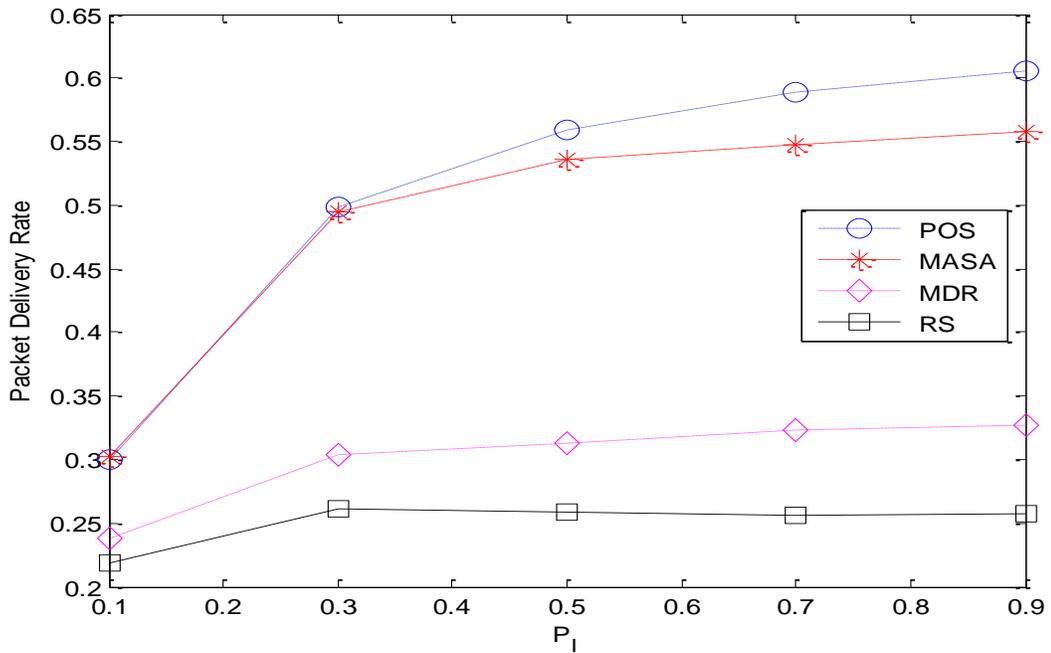

**Fig. 3.26 PDR vs. idle probability.**

(N = 40, Nr = 16, M = 20, $P_t$ = 0.1W, BW = 1MHz, D = 4KB)

## 3.3 Performance Evaluation for Minimum Spanning Tree (MST)

The MST finds the paths that connect all nodes within the topology with minimum cost or weight (i.e. distance). Kruskal's algorithm is used to implement the MST algorithm in Matlab [46]. This section provides the simulation results of the MST. All results provided in this section were evaluated with the total number of nodes (N) is 20, the total number of destinations (Nr) is 11, and one source node. The proposed scheme is also compared to other schemes under different network conditions.

### 3.3.1 Performance Evaluation under Low Traffic Load ($P_I = 0.9$)

Under low traffic loads, more spectrum resources are available to be utilized in better manner than high traffic loads. The resource (i.e., idle channels) is basically depending on the idle probability, lower traffic load results in higher network performance. However, higher traffic load reduces resources such that the performance is degraded and resources are not satisfying the transmissions demand. In other words, under high traffic load, most of the PU channels are utilized.

**3.3.1.1 Throughput and Packet Delivery Rate Performance versus Channel Bandwidth**

As discussed before, increasing the channel bandwidth is resulting in a better network performance. Figures 3.27 and 3.28 show the performance of the proposed scheme and other schemes versus the channel bandwidth. The proposed scheme outperforms MASA, MDR, and RS schemes in term of throughput by up to 42.6%, 52%, and 122.6%, respectively. Also, the proposed scheme outperforms MASA, MDR, and RS schemes in term of packet delivery rate by up to 18%, 142%, and 236%, respectively.

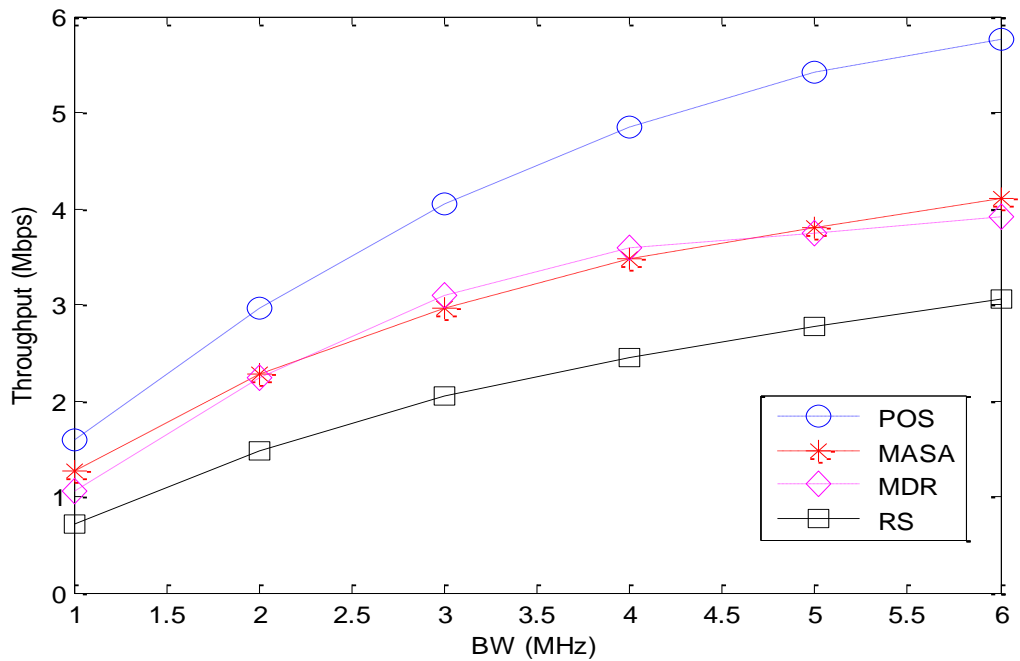

**Fig. 3.27 Throughput vs. channel bandwidth under $P_I = 0.9$.**

(N = 20, Nr = 11, M = 20, $P_t$ = 0.1W, D = 4KB)

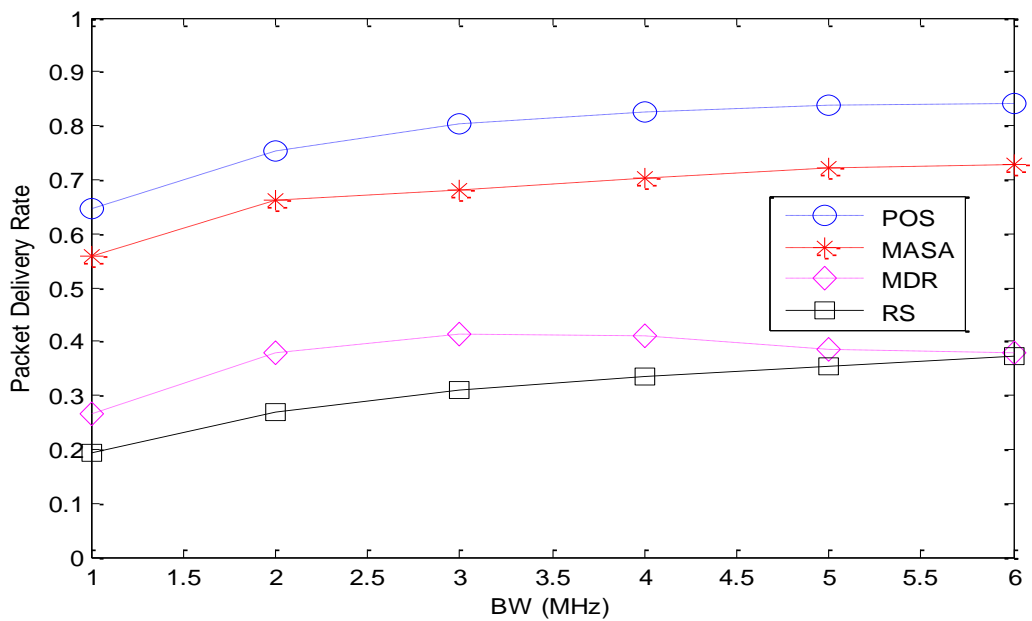

**Fig. 3.28 PDR vs. channel bandwidth under $P_I = 0.9$.**

(N = 20, Nr = 11, M = 20, $P_t$ = 0.1W, D = 4KB)

### 3.3.1.2 Throughput and Packet Delivery Rate Performance versus The Packet Size

As observed earlier in this Chapter, due to the increase in the transmission time when increasing the packet size, the probability of successful transmission is decreased. Specifically, each available channel remains available for a limited period of time. Each SU should finish its transmission within that time in order to successfully send its packet. However, increasing the packet size is resulting in increasing the SU transmission time. Hence, the SU will require more time than the channel available time and, hence the packet will be dropped (see Figures 3.29 and 3.30).

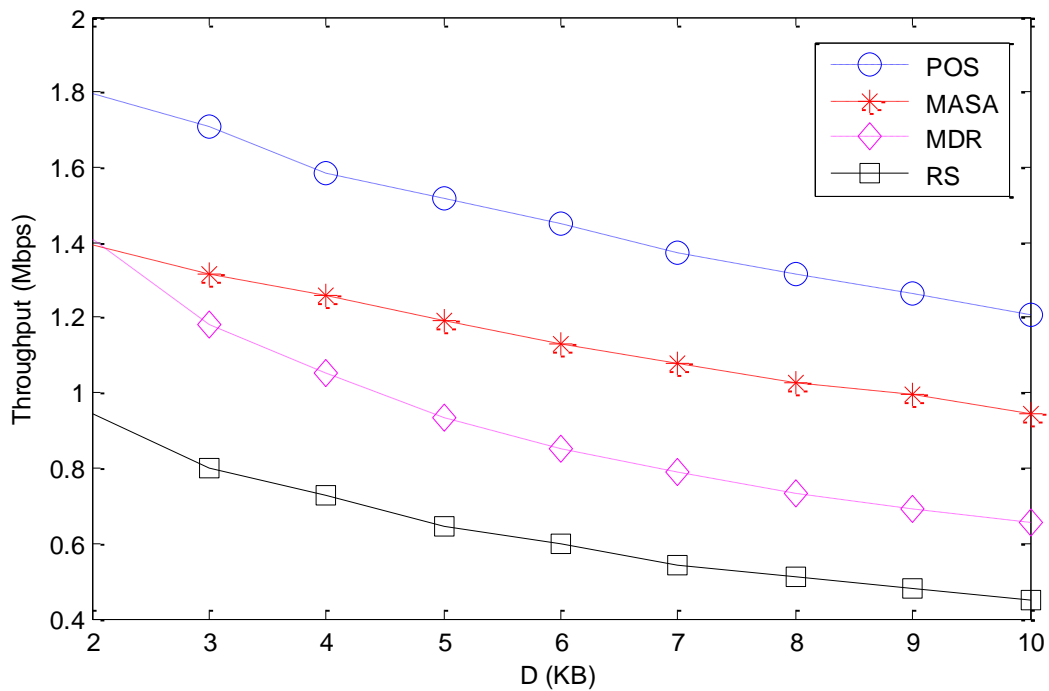

Fig. 3.29 Throughput vs. the packet size under $P_I = 0.9$.

($N = 20$, $Nr = 11$, $M = 20$, $P_t = 0.1W$, $BW = 1MHz$)

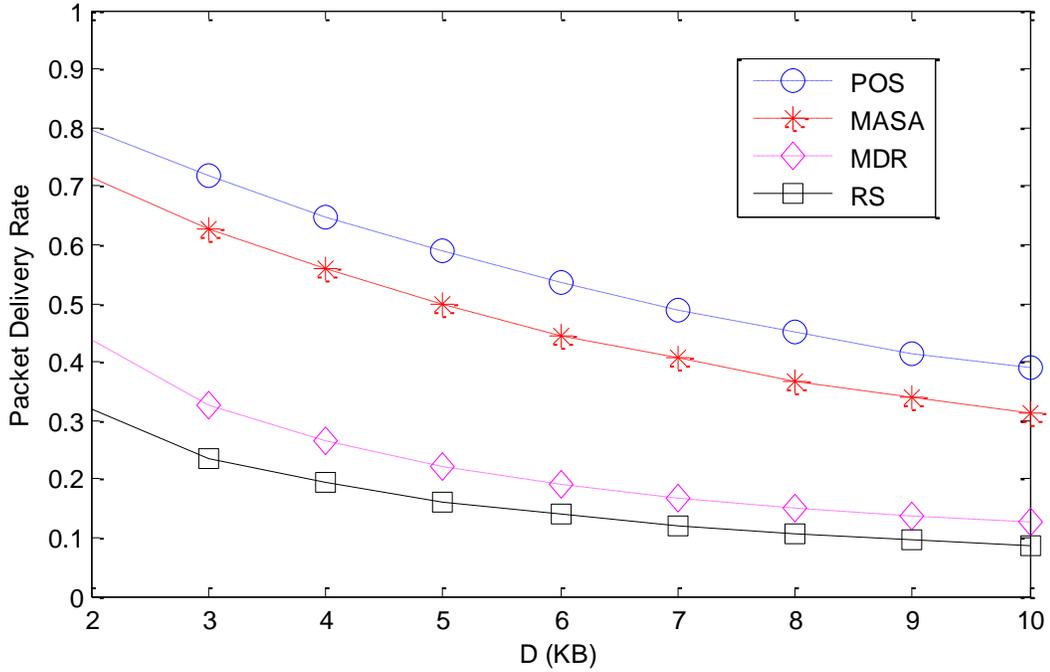

**Fig. 3.30 PDR vs. the packet size under $P_I = 0.9$.**

(N = 20, Nr = 11, M = 20, $P_t$ = 0.1W, BW = 1MHz)

Specifically, the proposed scheme outperforms MASA, MDR, and RS schemes in term of throughput by up to 29.4%, 84.6%, and 169%, respectively. Also, the proposed scheme outperforms MASA, MDR, and RS schemes in term of packet delivery rate by up to 24.3%, 209%, and 343%, respectively.

### 3.3.1.3 Throughput and Packet Delivery Rate Performance versus Number of Primary Channels

As shown in Figures 3.31 and 3.32, increasing number of primary channels (M) enhances the network performance. Specifically, the proposed scheme outperforms MASA, MDR, and RS schemes in term of throughput by up to 30%, 52.5%, and 121.7%, respectively. Also, the proposed scheme outperforms MASA, MDR, and RS schemes in term of packet delivery rate by up to 16%, 143.5%, and 238%, respectively.

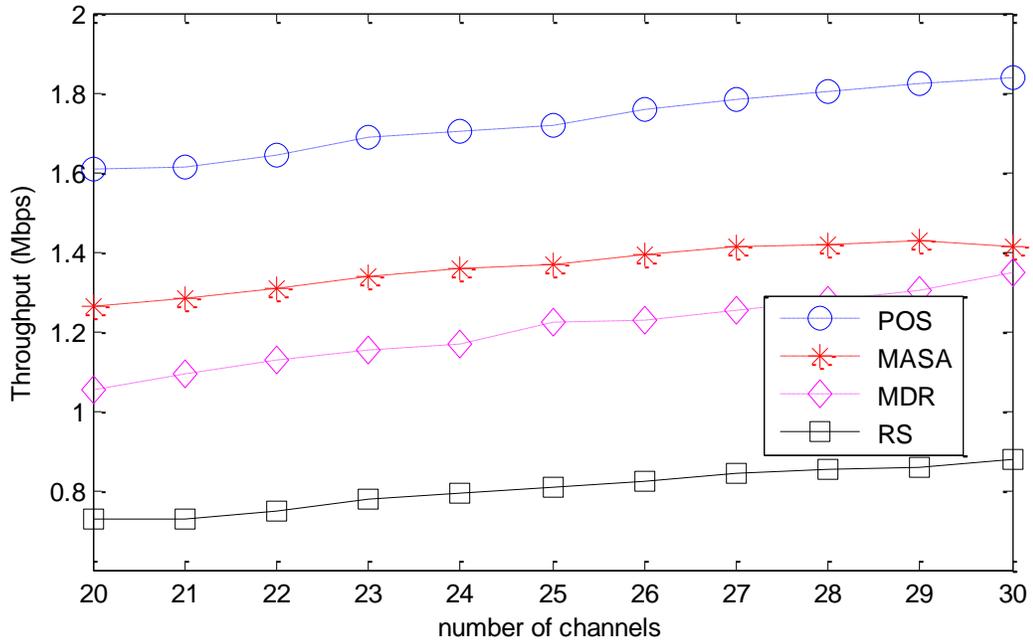

**Fig. 3.31 Throughput vs. number of primary channels under $P_I = 0.9$.**

($N = 20$, $N_r = 11$, $P_t = 0.1W$, $BW = 1MHz$, $D = 4KB$)

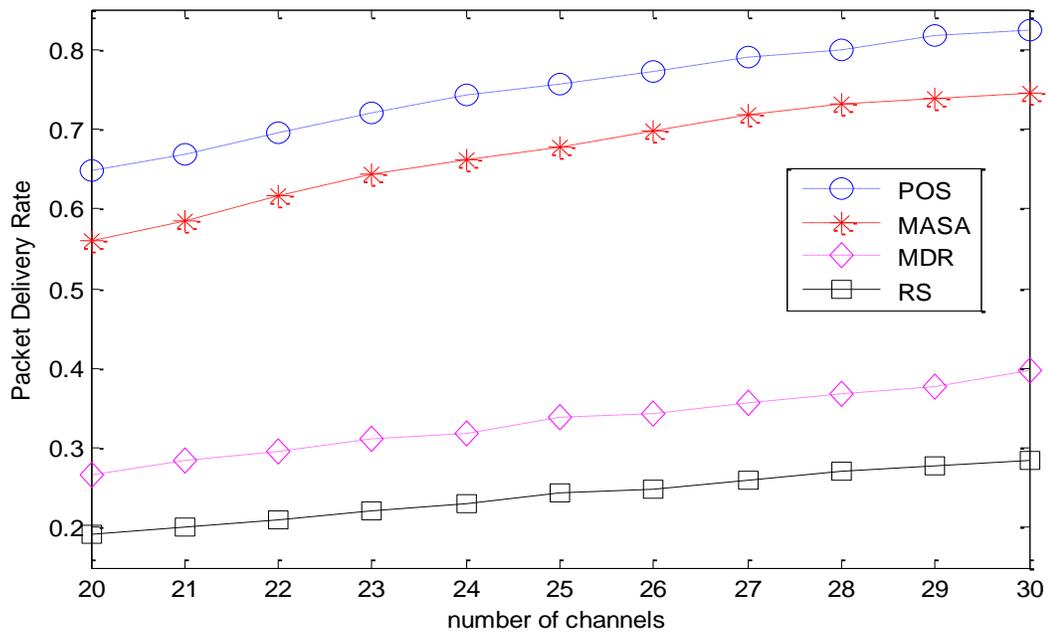

**Fig. 3.32 PDR vs. number of primary channels under $P_I = 0.9$.**

($N = 20$, $N_r = 11$, $P_t = 0.1W$, $BW = 1MHz$, $D = 4KB$)

**3.3.1.4 Throughput and Packet Delivery Rate Performance versus The Transmission Power**

The transmission power is an effective parameter on network performance as shown in Figures 3.33 and 3.34. Increasing the transmission power will increase the data rate and packet delivery rate due to the improvement in the received SINR. However, one watt is too high for such applications because this will increase the interference significantly. These figures show that the proposed scheme outperforms MASA, MDR, and RS schemes in term of throughput by up to 27.4%, 52.5%, and 124.9%, respectively. Also, the proposed scheme outperforms MASA, MDR, and RS schemes in term of packet delivery rate by up to 15.9%, 142.7%, and 240%, respectively.

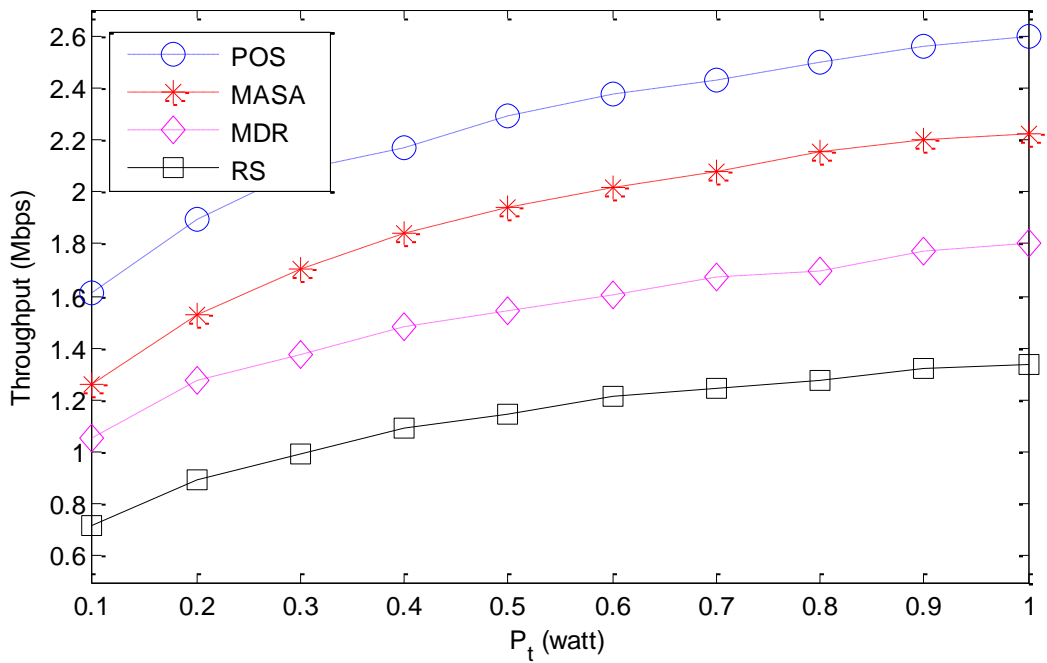

**Fig. 3.33 Throughput vs. the transmission power under $P_I = 0.9$.**

(N = 20, Nr = 11, M = 20, BW = 1MHz, D = 4KB)

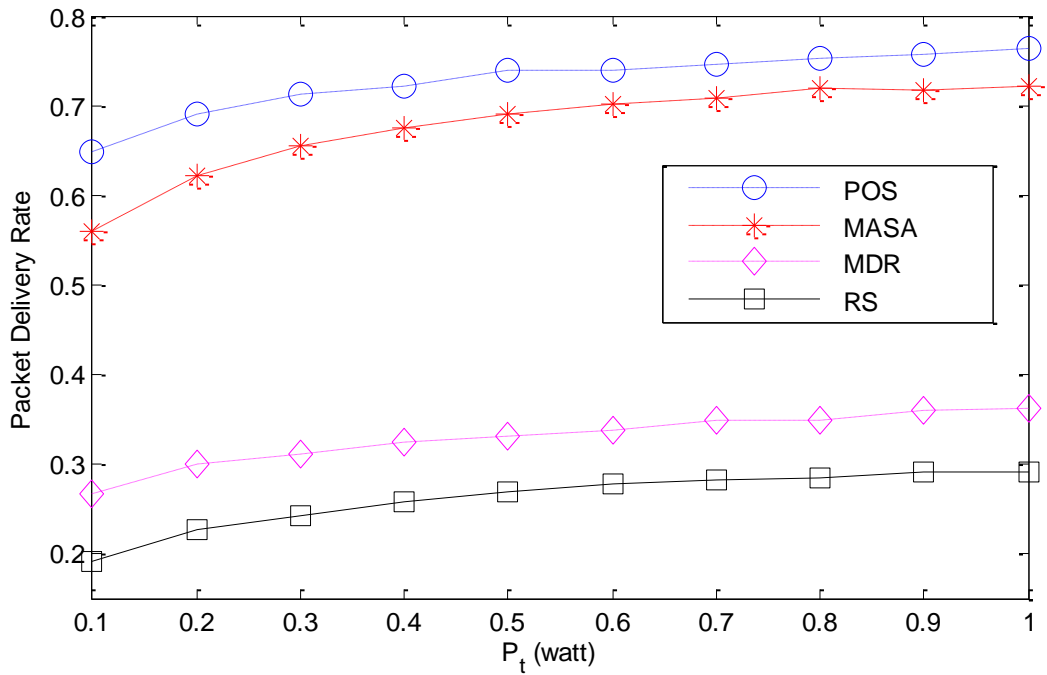

**Fig. 3.34 PDR vs. the transmission power under $P_I = 0.9$.**

(N = 20, Nr = 11, M = 20, BW = 1MHz, D = 4KB)

### 3.3.2 Performance Evaluation under Moderate Traffic Load ($P_I = 0.5$)

As noted before the traffic load is an effective network parameter that cannot be controlled by SUs. The traffic load affects the number of available channels for the CRN. In the rest of this section, we will present simulation results under moderate traffic load (i.e., $P_I = 0.5$).

#### 3.3.2.1 Throughput and Packet Delivery Rate Performance versus Channel Bandwidth

As observed previously, network performance is improved when increasing the channels bandwidth. In addition, the improvement gain from using the proposed scheme is larger when increasing channel bandwidth as shown in Figures 3.35 and 3.36. The proposed scheme outperforms MASA, MDR, and RS schemes in term of throughput by up to 34.6%, 48.5%, and 108.5%, respectively. Also, the proposed scheme outperforms MASA, MDR, and RS schemes in term of packet delivery rate (PDR) by up to 16%, 135%, and 221.8%, respectively.

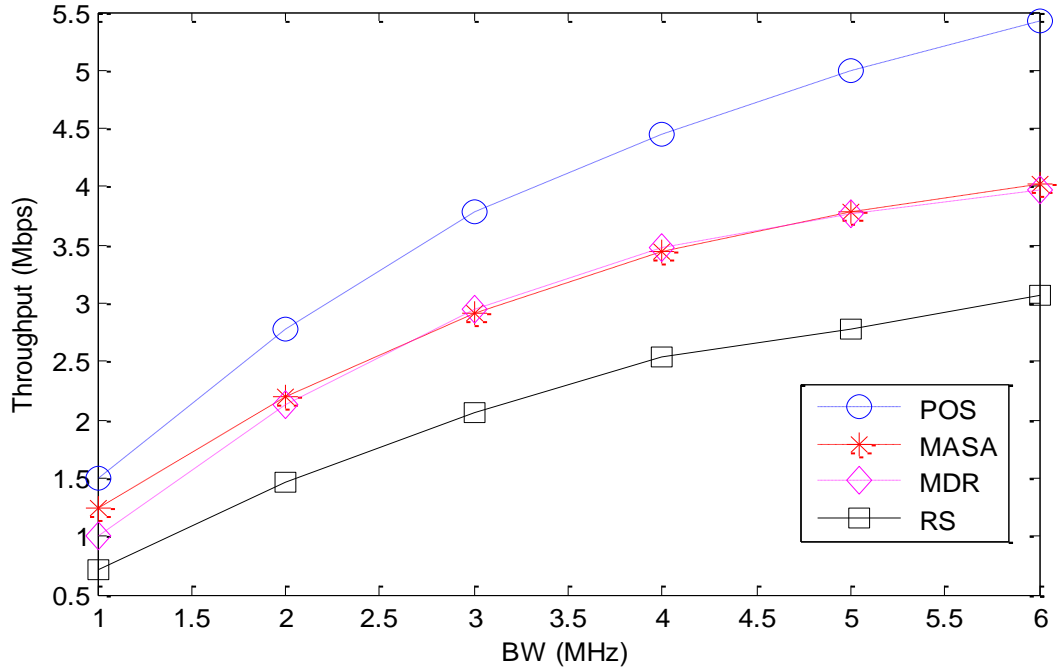

**Fig. 3.35 Throughput vs. channel bandwidth under $P_I = 0.5$.**

(N = 20, Nr = 11, M = 20, $P_t$ = 0.1W, D = 4KB)

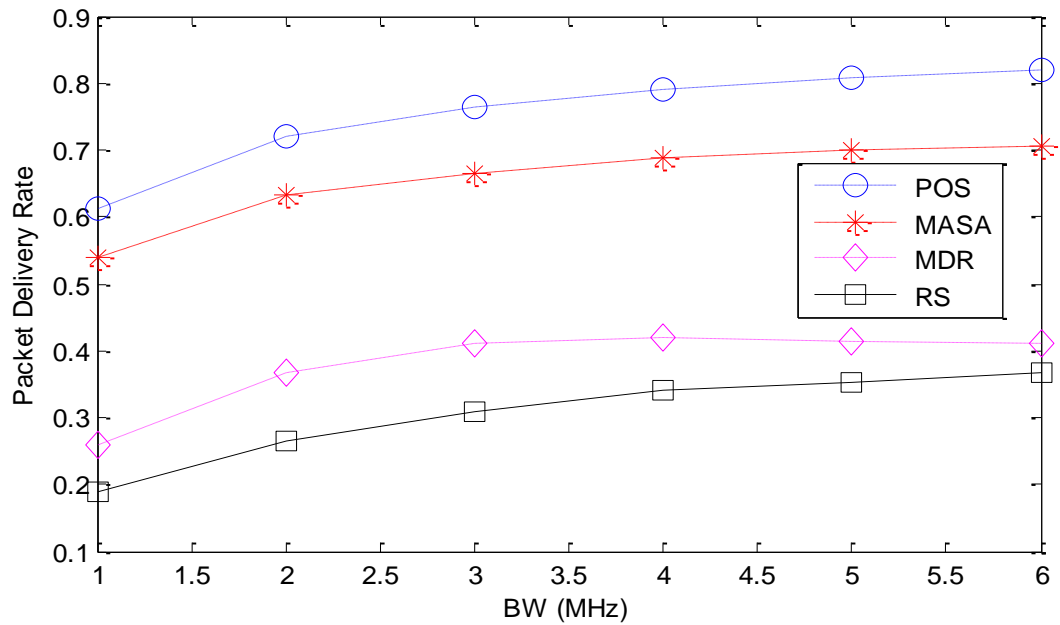

**Fig. 3.36 PDR vs. channel bandwidth under $P_I = 0.5$.**

(N = 20, Nr = 11, M = 20, $P_t$ = 0.1W, D = 4KB)

### 3.3.2.2 Throughput and Packet Delivery Rate Performance versus The Packet Size

We previously noticed the negative impact of increasing packet size on the network performance. Figures 3.37 and 3.38 show the network performance versus the packet size evaluated using the MST at moderate traffic load. The proposed scheme outperforms MASA, MDR, and RS schemes in term of throughput by up to 22.5%, 77.8%, and 141.6%, respectively. Also, the proposed scheme outperforms MASA, MDR, and RS schemes in term of packet delivery rate by up to 18.6%, 189.6%, and 295%, respectively.

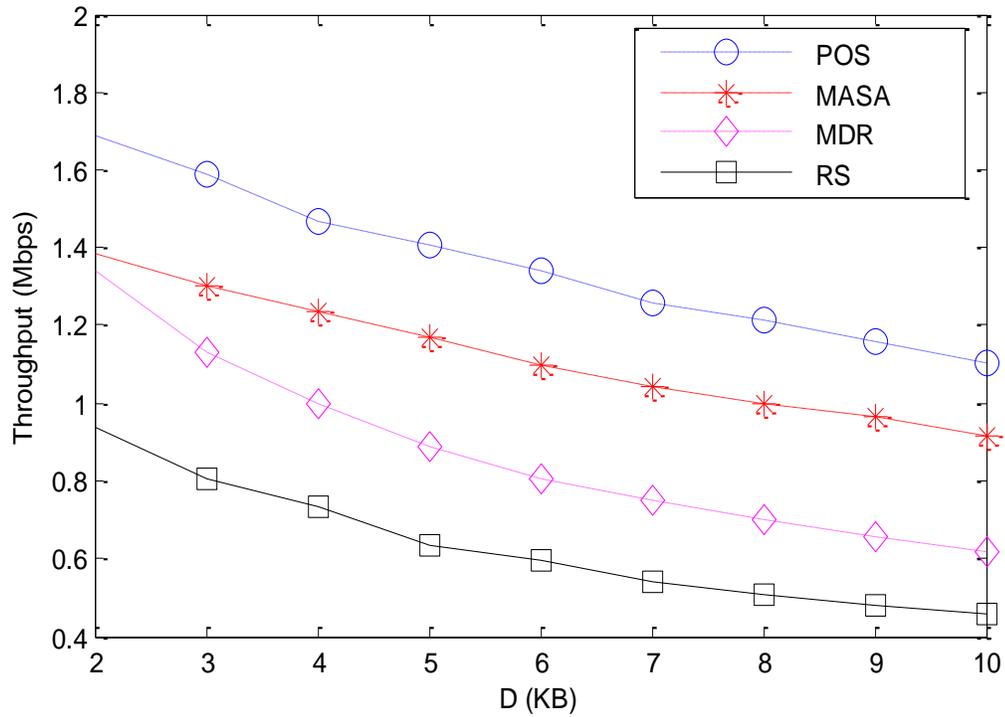

**Fig. 3.37 Throughput vs. the packet size under $P_I = 0.5$.**

(N = 20, Nr = 11, M = 20, $P_t$ = 0.1W, BW = 1MHz)

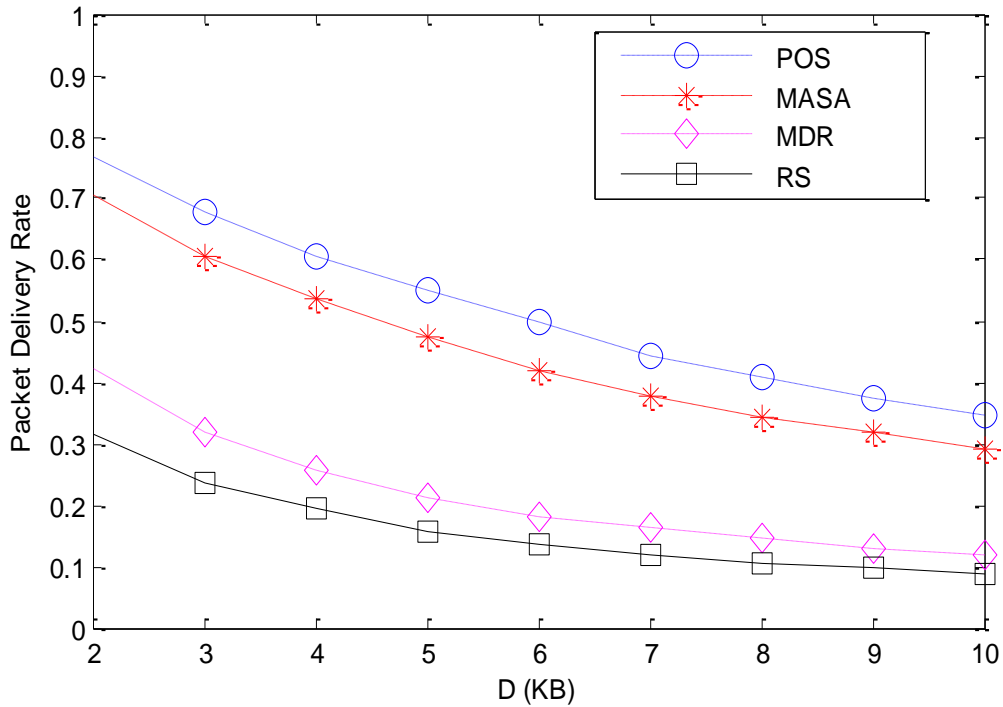

**Fig. 3.38 PDR vs. the packet size under $P_I = 0.5$.**

($N = 20$, $Nr = 11$, $M = 20$, $P_t = 0.1W$, $BW = 1MHz$)

### 3.3.2.3 Throughput and Packet Delivery Rate Performance versus Number of Primary Channels

As mentioned before, increasing the number of primary channels (M) will enhance network performance (Figures 3.39 and 3.40). For idle probability equals to 0.5 that means the number of available channels for the CRN on average is 10-15. The proposed scheme outperforms MASA, MDR, and RS schemes in term of throughput by up to 25%, 49%, and 108.5%, respectively. Also, the proposed scheme outperforms MASA, MDR, and RS schemes in term of packet delivery rate by up to 13.2%, 135.7%, and 221.8%, respectively.

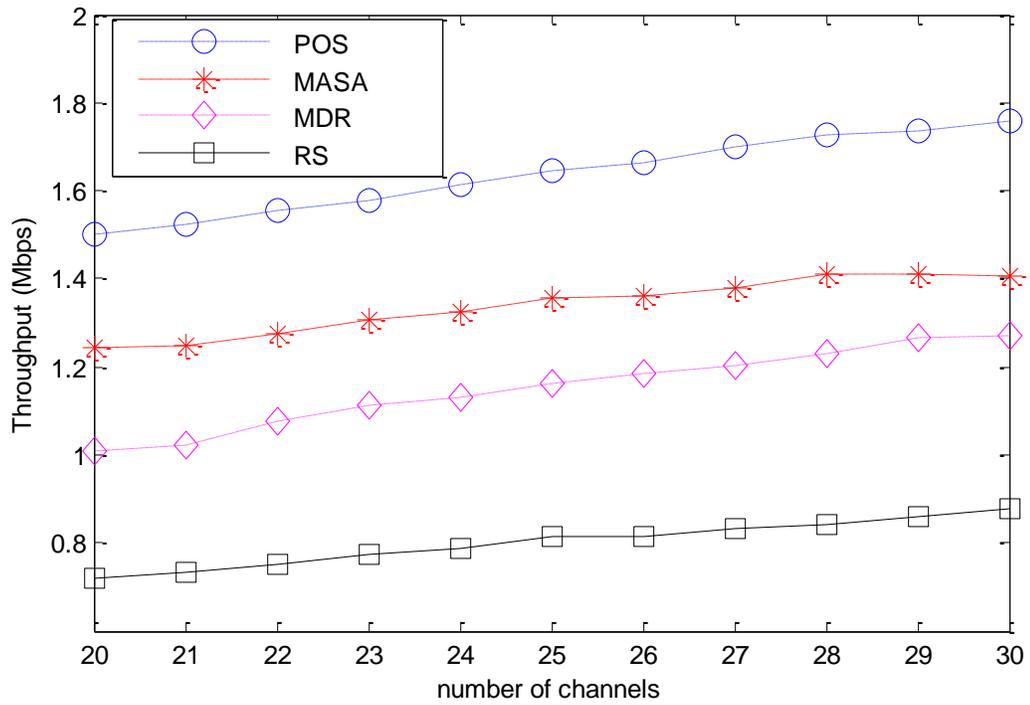

**Fig. 3.39 Throughput vs. number of primary channels under $P_I = 0.5$.**

($N = 20$, $N_r = 11$, $P_t = 0.1W$, $BW = 1MHz$, $D = 4KB$)

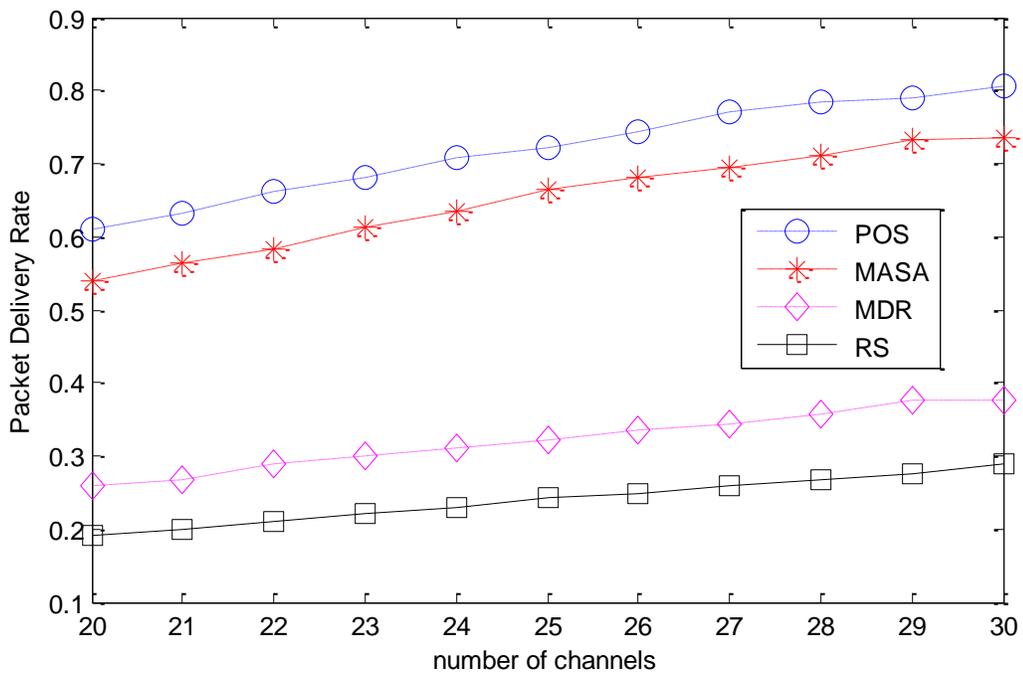

**Fig. 3.40 PDR vs. number of primary channels under $P_I = 0.5$.**

($N = 20$, $N_r = 11$, $P_t = 0.1W$, $BW = 1MHz$, $D = 4KB$)

**3.3.2.4 Throughput and Packet Delivery Rate Performance versus The Transmission Power**

Figures 3.41 and 3.42 demonstrate the effect of the transmission power on network performance. It is shown below that increasing the transmission power improves networks performance in terms of throughput and packet delivery rate. However, one watt is too high for such applications because this will increase the interference significantly. The proposed scheme outperforms MASA, MDR, and RS schemes in term of throughput by up to 20.5%, 48.5%, and 108.5%, respectively. Also, the proposed scheme outperforms MASA, MDR, and RS schemes in term of packet delivery rate by up to 13%, 135%, and 221.8%, respectively.

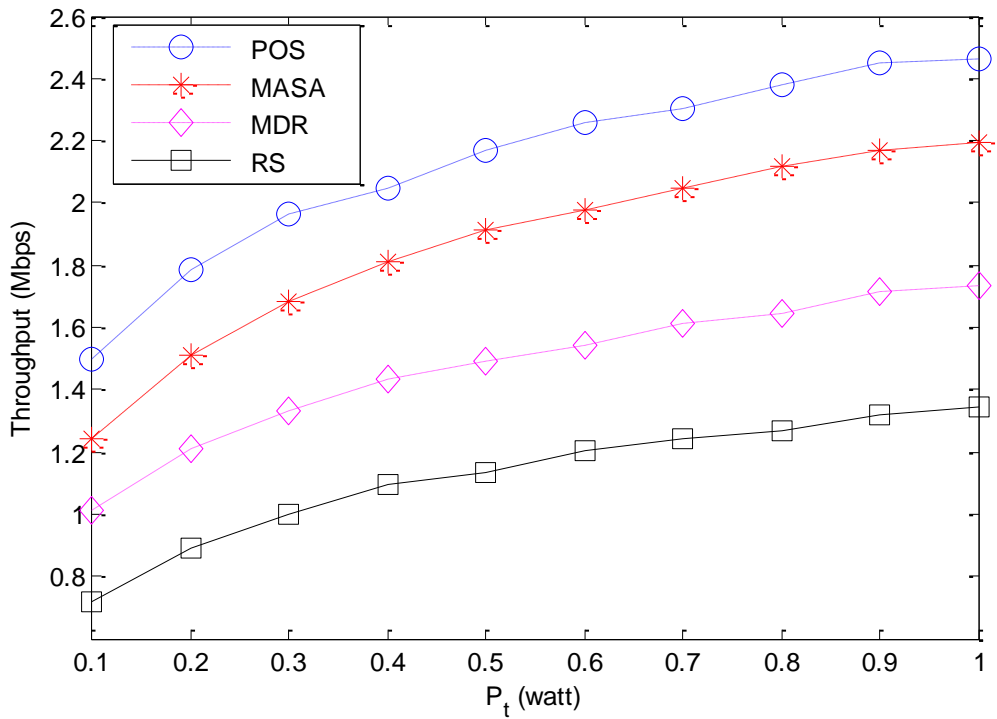

**Fig. 3.41 Throughput vs. the transmission power under $P_I = 0.5$.**

(N = 20, Nr = 11, M = 20, BW = 1MHz, D = 4KB)

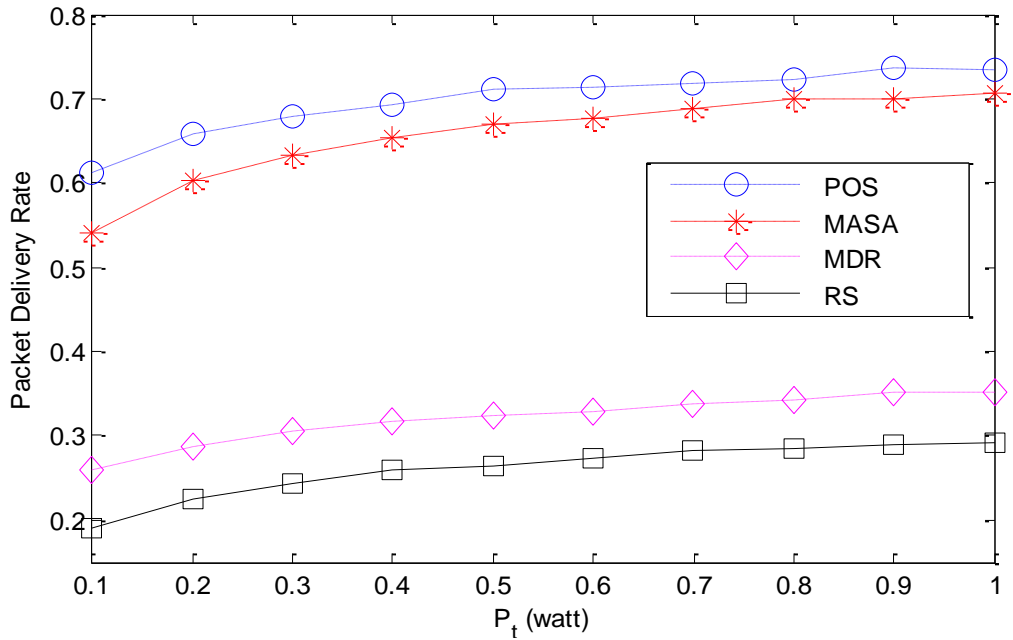

**Fig. 3.42 PDR vs. the transmission power under $P_I$ = 0.5.**

(N = 20, Nr = 11, M = 20, BW = 1MHz, D = 4KB)

### 3.3.3 Performance Evaluation under High Traffic Load ($P_I$ = 0.1)

High PU activity reduces the number of available channels to the CRN and degrades its performance. In this Section, we will study the effect of varying parameters on the CRN performance. It is worth mentioning that the performance of the POS is comparable to the performance of the MASA scheme and outperforms MDR and RS schemes, as we will notice in this section.

**3.3.3.1 Throughput and Packet Delivery Rate Performance versus Channel Bandwidth**

As observed before, increasing the channel bandwidth improves network performance. According to Figures 3.43 and 3.44, the proposed scheme outperforms MASA, MDR, and RS schemes in term of throughput by up to 8%, 20%, and 34%, respectively. Also, the POS scheme outperforms other schemes in term of packet delivery rate by up to 6%, 44%, and 56.5%, respectively.

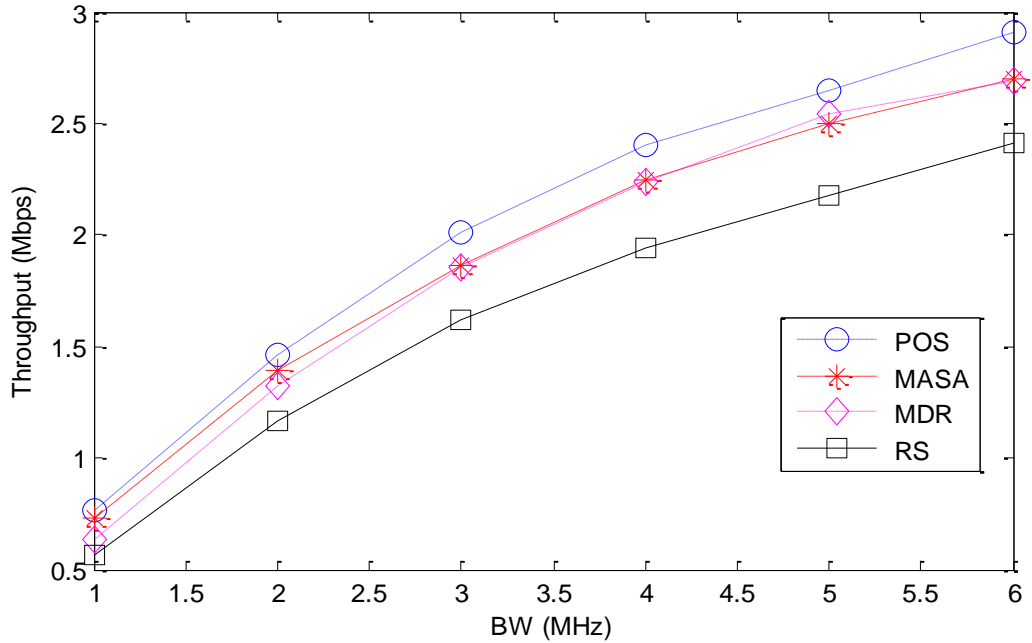

**Fig. 3.43 Throughput vs. channel bandwidth under $P_I = 0.1$.**

(N = 20, Nr = 11, M = 20, $P_t$ = 0.1W, D = 4KB)

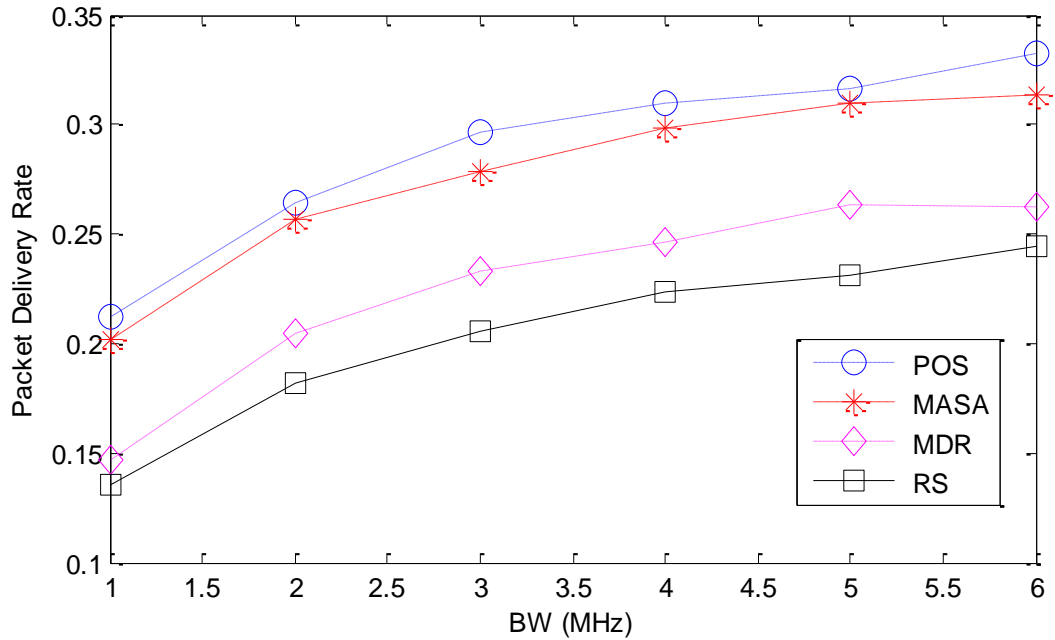

**Fig. 3.44 PDR vs. channel bandwidth under $P_I = 0.1$.**

(N = 20, Nr = 11, M = 20, $P_t$ = 0.1W, D = 4KB)

### 3.3.3.2 Throughput and Packet Delivery Rate Performance versus The packet size

The CRN performance is inversely proportional to the packet size. Figures 3.45 and 3.46 show the performance of the proposed scheme compared to other schemes. The POS and MASA schemes have comparable performance to each other and outperform the other schemes. The proposed scheme outperforms MASA, MDR, and RS schemes in term of throughput by up to 8%, 27.8%, and 46%, respectively. Also, the proposed scheme outperforms MASA, MDR, and RS schemes in term of packet delivery rate (PDR) by up to 6%, 51%, and 71.8%, respectively.

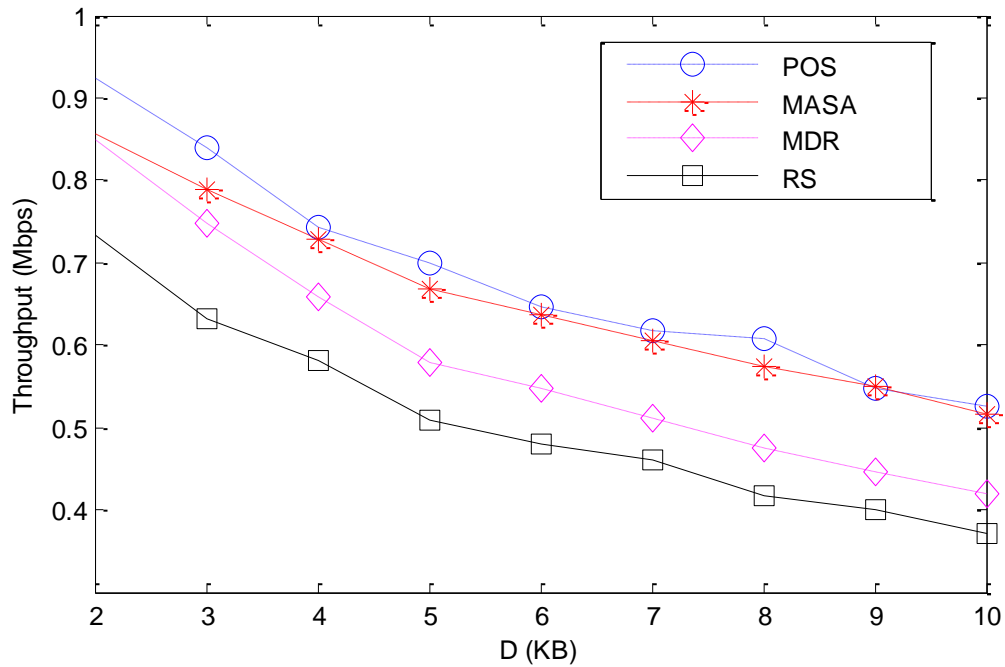

**Fig. 3.45 Throughput vs. the packet size under $P_I = 0.1$.**

(N = 20, Nr = 11, M = 20, $P_t$ = 0.1W, BW = 1MHz)

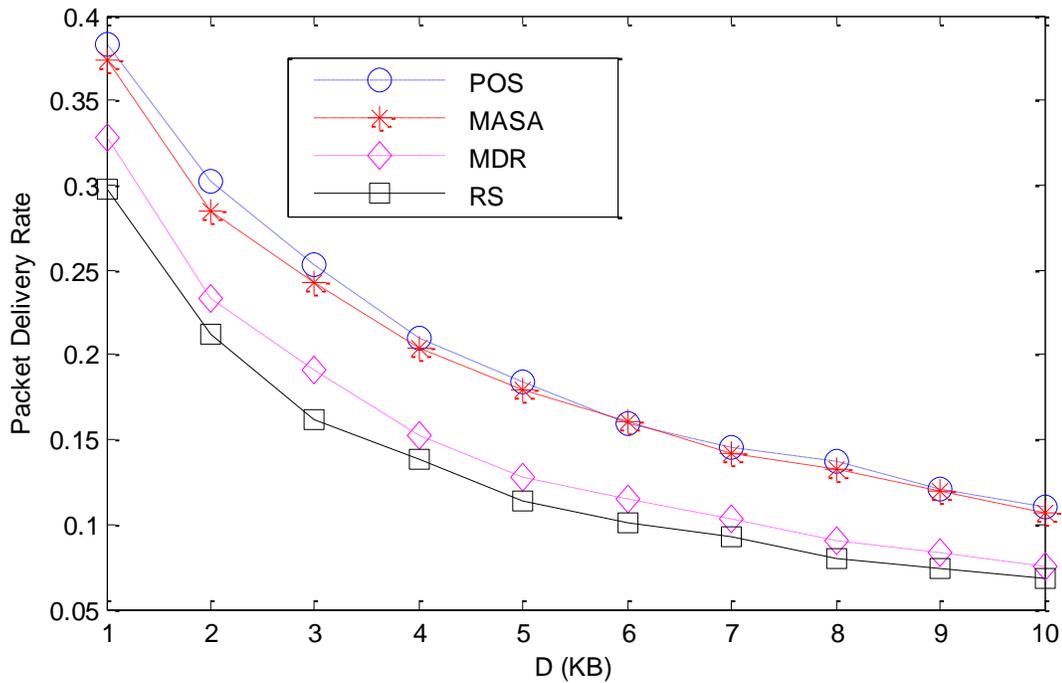

**Fig. 3.46 PDR vs. the packet size under $P_I = 0.1$.**

(N = 20, Nr = 11, M = 20, $P_t$ = 0.1W, BW = 1MHz)

### 3.3.3.3 Throughput and Packet Delivery Rate Performance versus Number of Primary Channels

As shown previously, increasing the number of primary (M) channels improves the network performance. In the case of high PU activity (idle probability is equals to 0.1) the number of available channels for the CRN on average is 2-3. Figures 3.47 and 3.48 show the simulation results that compare the performance of the proposed scheme to that of the different schemes in terms of network throughput and packet delivery rate. The proposed scheme outperforms MASA, MDR, and RS schemes in term of throughput by up to 7.3%, 20.5%, and 45%, respectively. Also, the proposed scheme outperforms MASA, MDR, and RS schemes in term of packet delivery rate by up to 5%, 60%, and 85%, respectively.

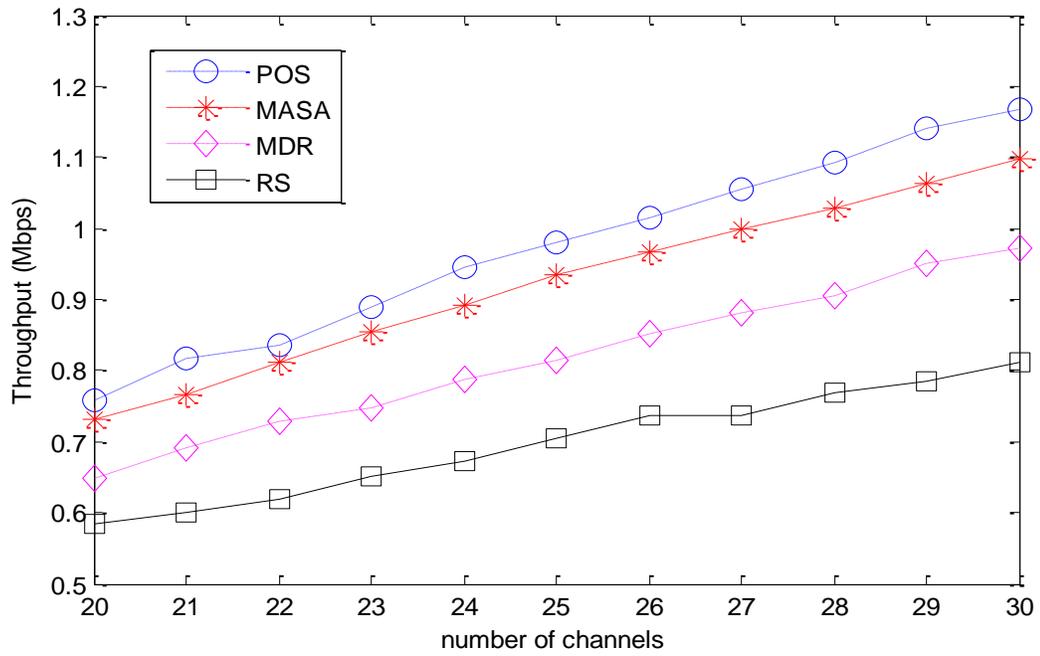

**Fig. 3.47 Throughput vs. number of primary channels under $P_I = 0.1$.**

($N = 20$, $Nr = 11$, $P_t = 0.1W$, $BW = 1MHz$, $D = 4KB$)

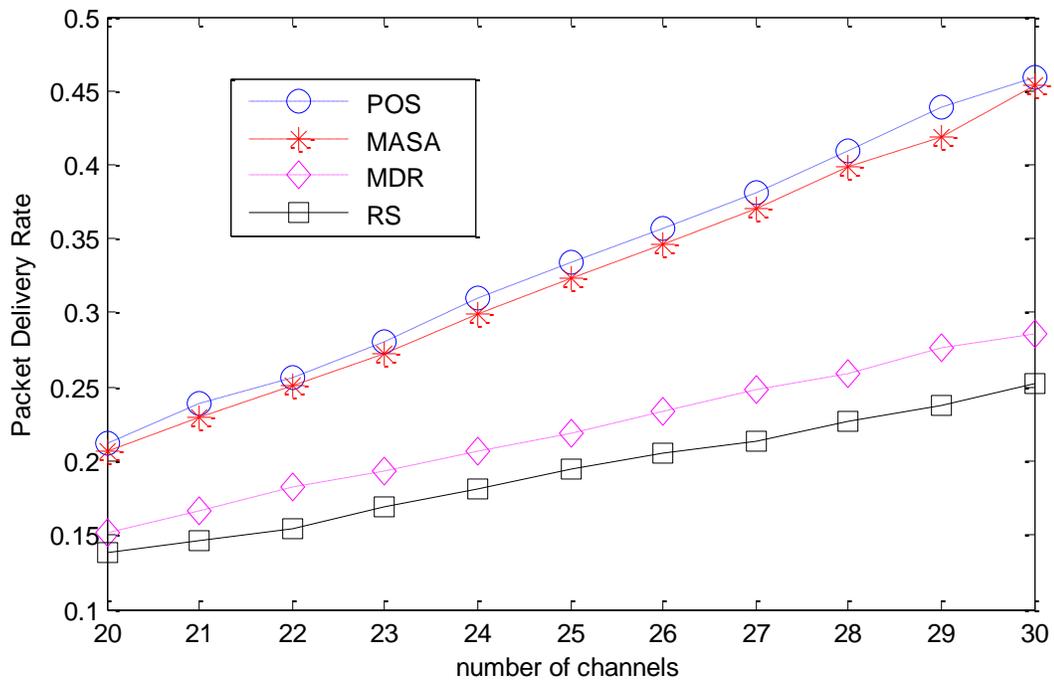

**Fig. 3.48 PDR vs. number of primary channels under $P_I = 0.1$.**

($N = 20$, $Nr = 11$, $P_t = 0.1W$, $BW = 1MHz$, $D = 4KB$)

**3.3.3.4 Throughput and Packet Delivery Rate Performance versus The Transmission Power**

As mentioned before, increasing transmission power improves CRN performance (see Figures 3.49 and 3.50). However, one watt is too high for such applications because this will increase the interference significantly. The proposed scheme outperforms MASA, MDR, and RS schemes in term of throughput by up to 4.9%, 16.7%, and 33%, respectively. Also, the proposed scheme outperforms MASA, MDR, and RS schemes in term of packet delivery rate by up to 5%, 40.5%, and 56%, respectively.

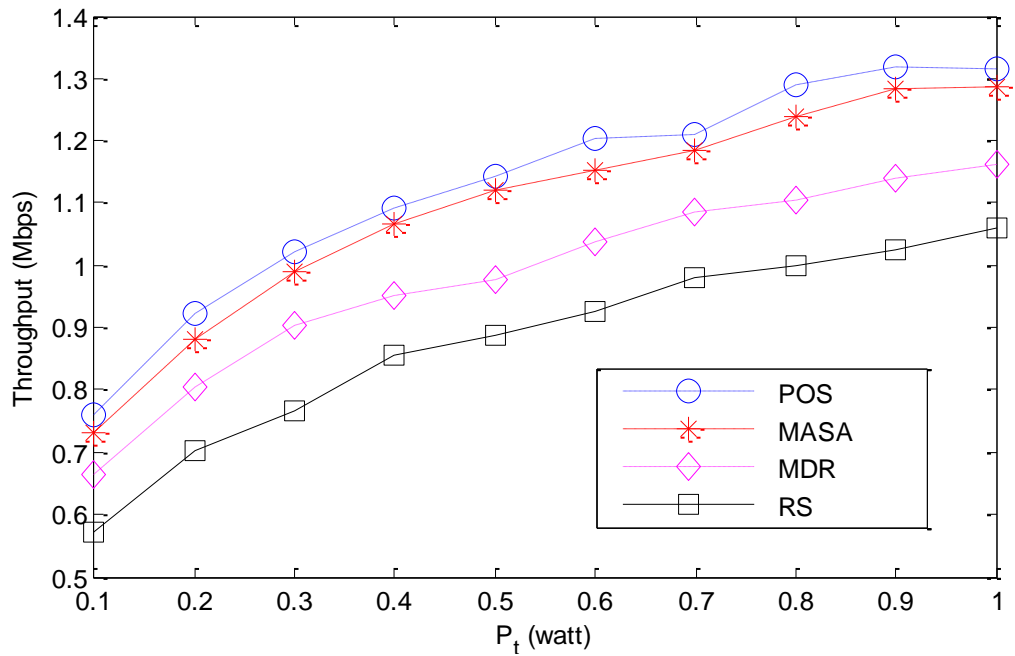

**Fig. 3.49 Throughput vs. the transmission power under $P_I = 0.1$.**

(N = 20, Nr = 11, M = 20, BW = 1MHz, D = 4KB)

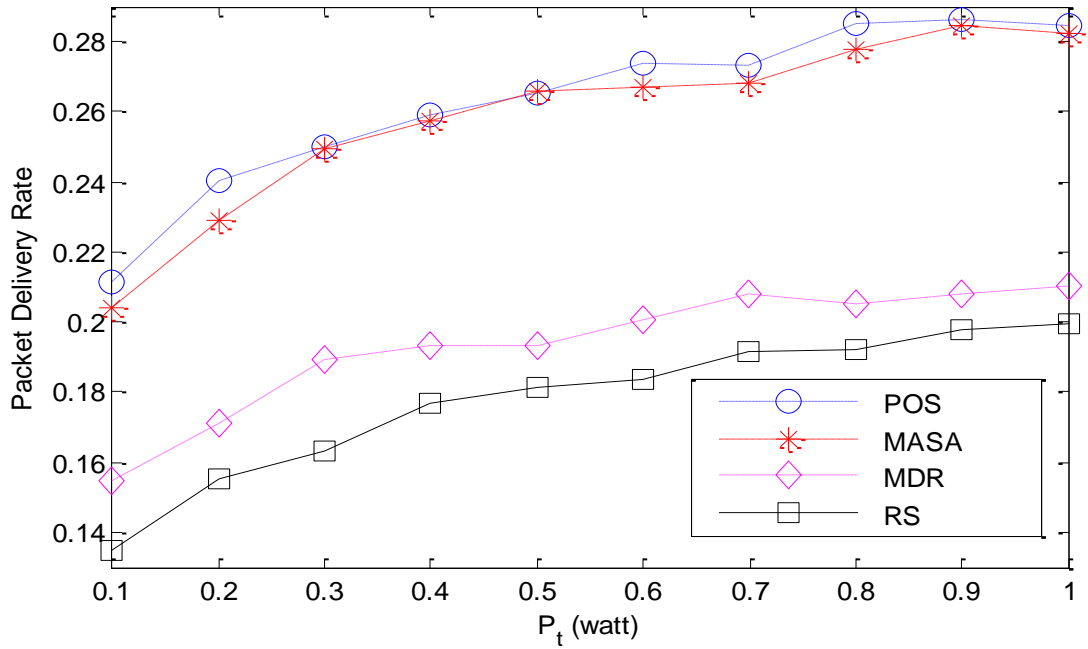

**Fig. 3.50 PDR vs. the transmission power under $P_I = 0.1$.**

(N = 20, Nr = 11, M = 20, BW = 1MHz, D = 4KB)

### 3.3.4 Impact of PUs Traffic Load

Figures 3.51 and 3.52 show the performance evaluation of the throughput and packet delivery rate versus the idle probability ($P_I$). As the idle probability increases (i.e. PU activity decreases) the CRN performance improves. The proposed scheme outperforms MASA, MDR, and RS schemes in term of throughput by up to 28.4%, 53.4%, and 123%, respectively. Also, the proposed scheme outperforms MASA, MDR, and RS schemes in term of packet delivery rate (PDR) by up to 16.5%, 140%, and 236%, respectively.

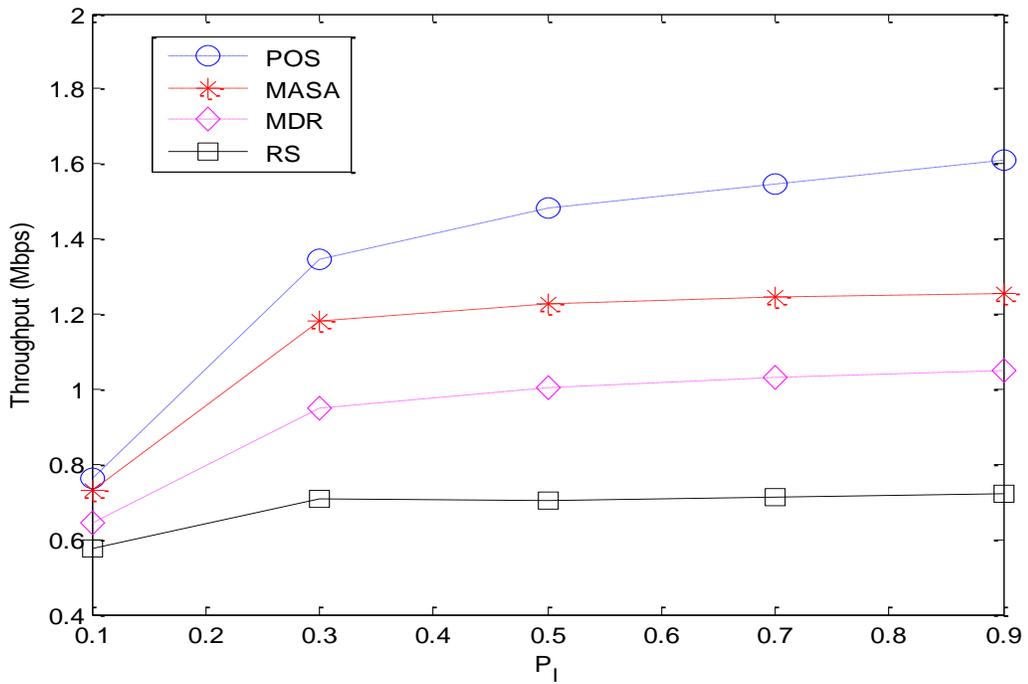

**Fig. 3.51 Throughput vs. idle probability.**

(N = 20, Nr = 11, M = 20, $P_t$ = 0.1W, BW = 1MHz, D = 4KB)

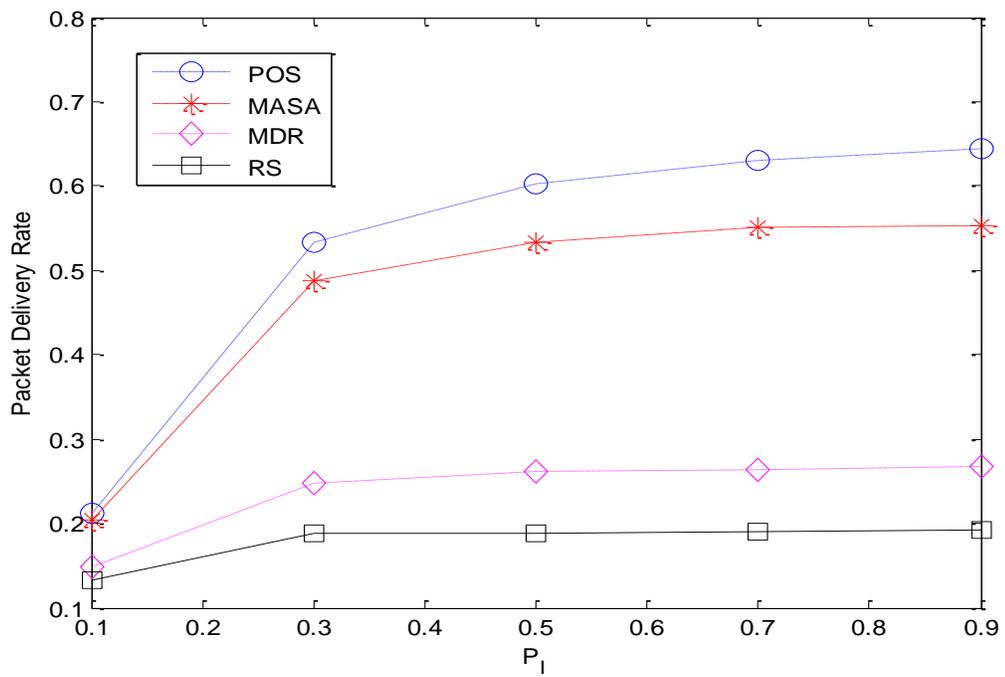

**Fig. 3.52 PDR vs. idle probability.**

(N = 20, Nr = 11, M = 20, $P_t$ = 0.1W, BW = 1MHz, D = 4KB)

## 3.4 Performance Comparison between SPT and MST

In this Section, we present and discuss a brief comparison between SPT and MST schemes in terms of network throughput and packet delivery rate. We consider the total number of nodes (N) is 40 and the total number of destinations (Nr) is 16.

### 3.4.1 Throughput and Packet Delivery Rate Performance versus Channel Bandwidth

When the channel bandwidth is small, the performance of the SPT is better than the performance of the MST. This can be explained based on the fact that the SPT is better when the spectrum is very limited (see Figures 3.53 and 3.54). We note here that SPT show better performance when BW ≤ 1.5 MHz for moderate and low PU activity, whereas it provides better performance for high PU activity. On the other hand, MST shows a better performance when more spectrum is available for the CRN.

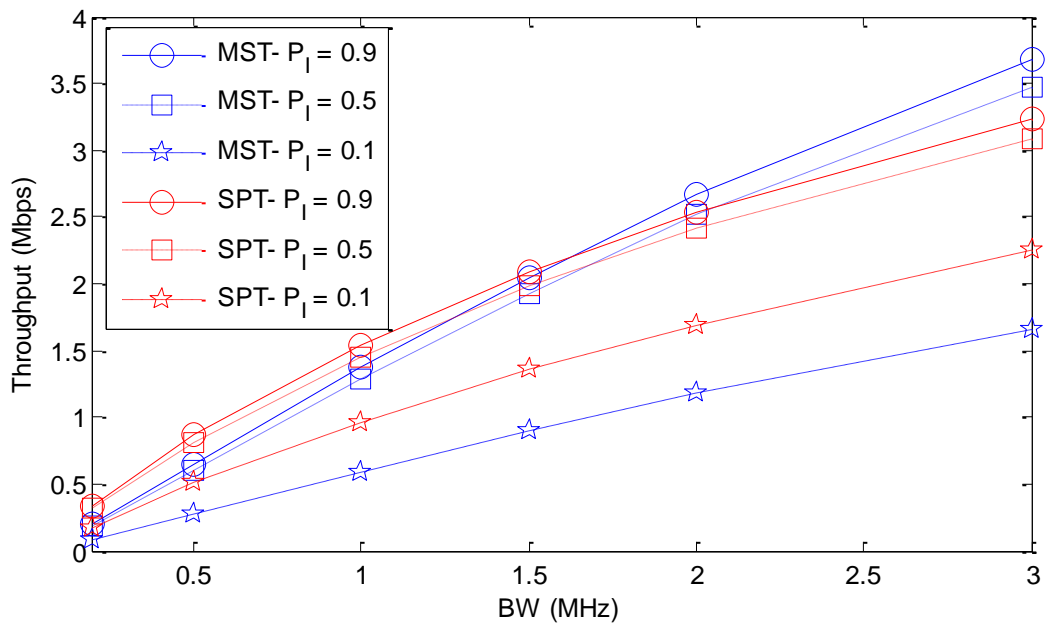

**Fig. 3.53 Throughput vs. channel bandwidth.**

(N = 40, Nr = 16, M = 20, $P_t$ = 0.1W, D = 4KB)

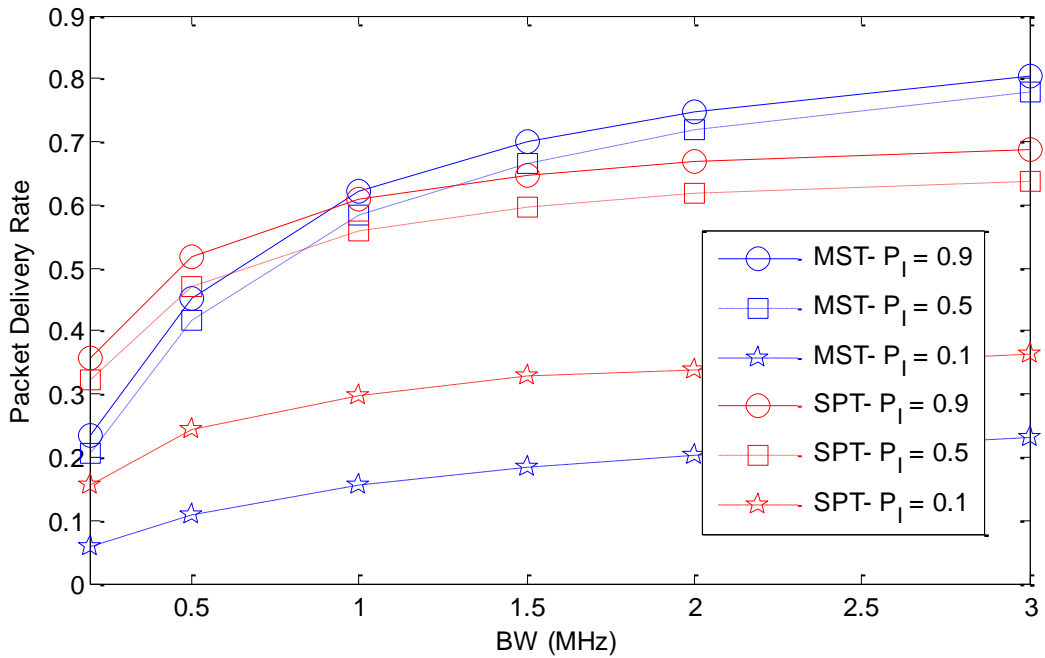

**Fig. 3.54 PDR vs. channel bandwidth.**

(N = 40, Nr = 16, M = 20, $P_t$ = 0.1W, D = 4KB)

### 3.4.2 Throughput and Packet Delivery Rate Performance versus The Packet Size

The throughput of the SPT is greater than for MST that is because a packet requires more hops to be received in the MST than SPT. In addition, the SPT chooses the minimum distances between the source and each node in the network. This reduces the required transmission time and increases throughput (see Figures 3.55 and 3.56). However, increasing the packet size has negative impact on network performance. SPT outperforms MST in term of throughput by up to 13.3%, 15%, and 67.6% under $P_I$ = 0.9, 0.5, and 0.1, respectively. When the idle probability equals to 0.1, SPT achieves an improvement gain of up to 98% in term of PDR compared to MST.

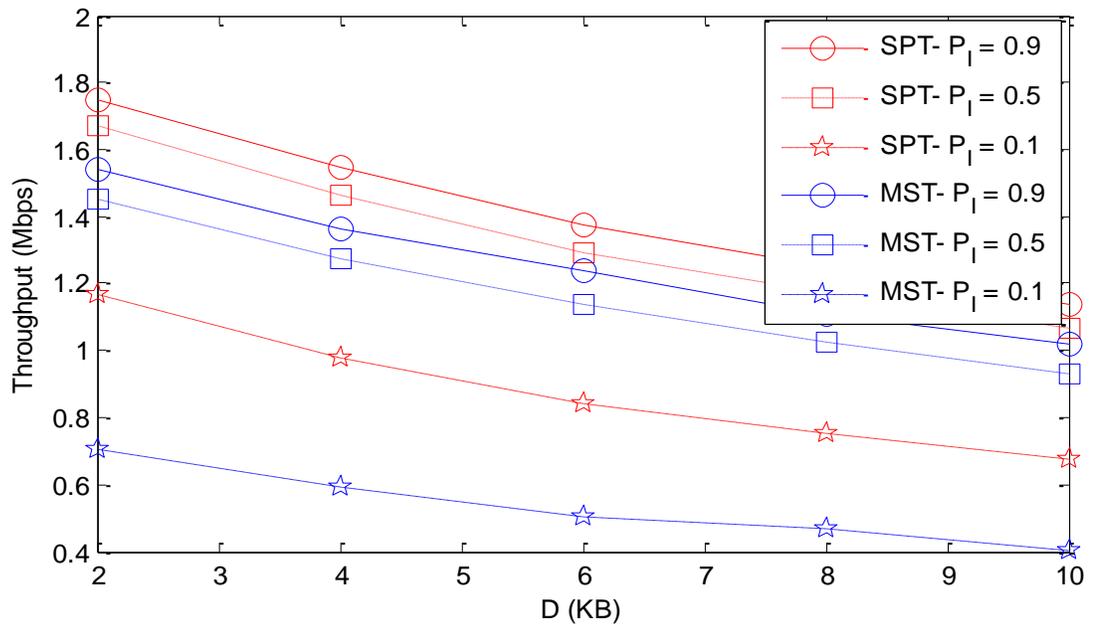

**Fig. 3.55 Throughput vs. the packet size.**

(N = 40, Nr = 16, M = 20, $P_t$ = 0.1W, BW = 1MHz)

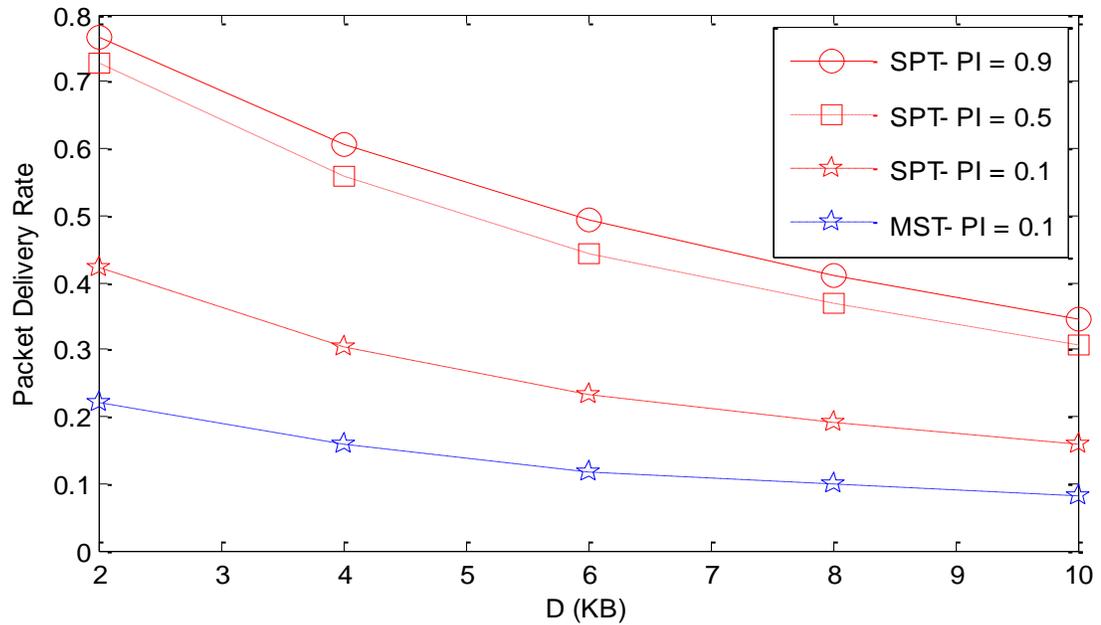

**Fig. 3.56 PDR vs. the packet size.**

(N = 40, Nr = 16, M = 20, $P_t$ = 0.1W, BW = 1MHz)

### 3.4.3 Throughput and Packet Delivery Rate Performance versus The Transmission Power

The SPT scheme indicates a better performance than MST in terms of throughput and PDR as the transmission power increases, see Figures 3.57 and 3.58. However, one watt is too high for such applications because this will increase the interference significantly. We note here that the performance improvement gained from SPT is also increased as the transmission power increases. This improvement can be explained by increasing the SINR and increasing the achieved data rate as a result of increasing the transmission power. The SPT outperforms MST in term of throughput by up to 74%, 75.6%, and 143.7% under $P_I$ = 0.9, 0.5, and 0.1, respectively. Also, The SPT outperforms MST in term of packet delivery rate by up to 152% under $P_I$ = 0.1.

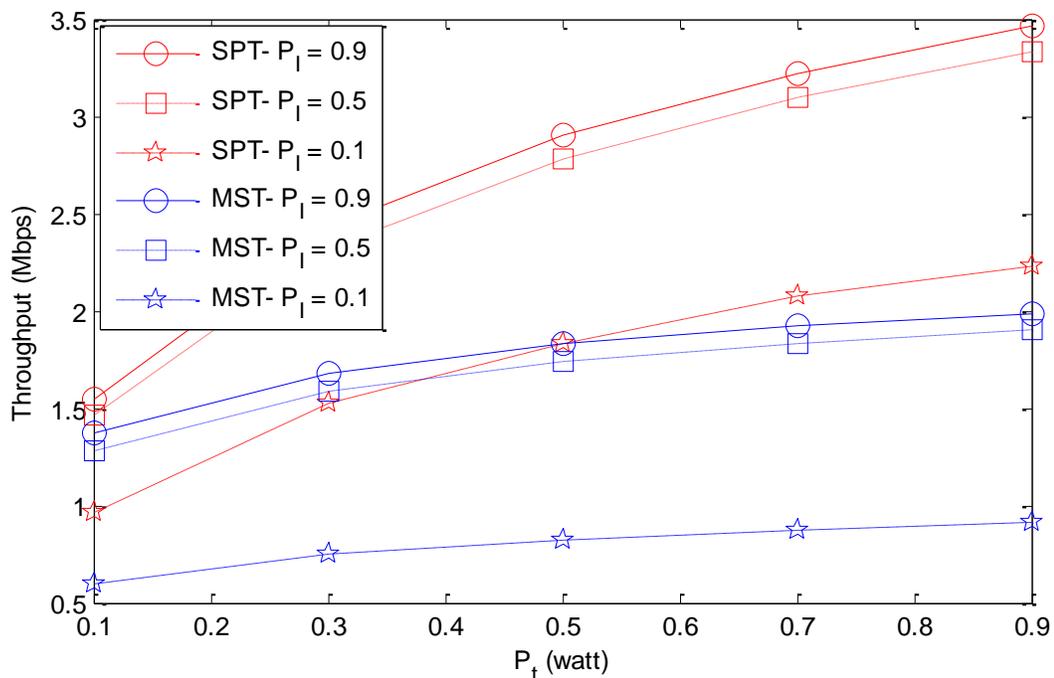

**Fig. 3.57 Throughput vs. the transmission power.**

(N = 40, Nr = 16, M = 20, BW = 1MHz, D = 4KB)

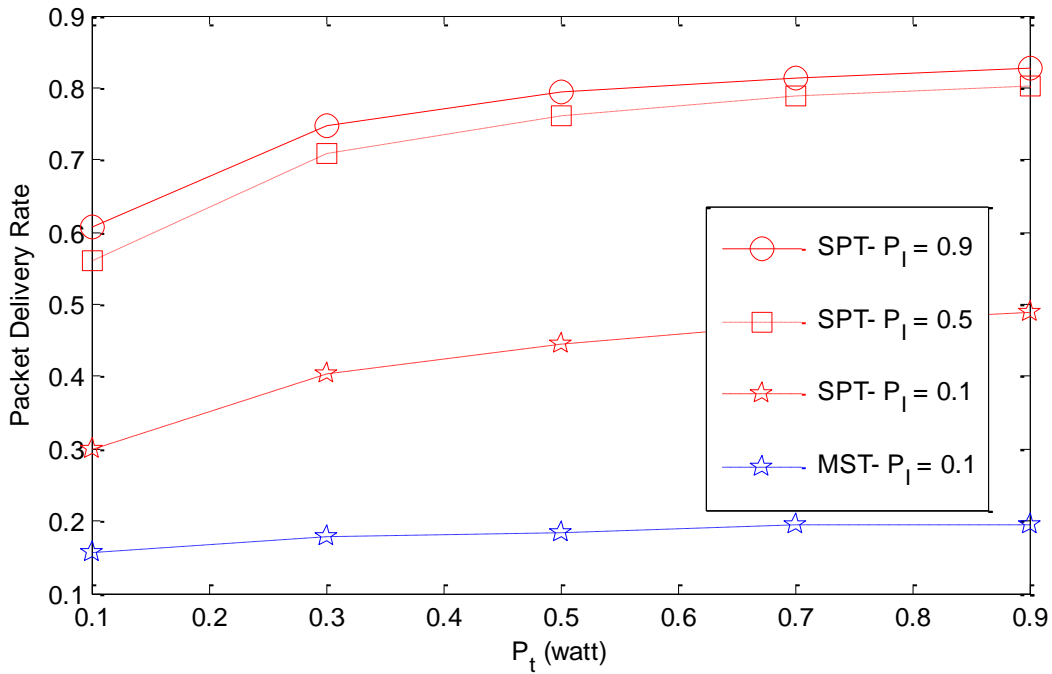

**Fig. 3.58 PDR vs. the transmission power.**

(N = 40, Nr = 16, M = 20, BW = 1MHz, D = 4KB)

### 3.4.4 Throughput and Packet Delivery Rate Performance versus Idle Probability

As shown in Figures 3.59 and 3.60, the throughput of the SPT is greater than that of the MST for all idle probability values (especially when high PU activity), while the PDR of the SPT is less than of the MST when $P_I \geq 0.3$. Lower PDR is not necessarily means lower throughput; may high transmission rates with lower PDR (e.g., in SPT) provides higher throughput than high PDR transmission with low data rate (e.g., MST). This is because of the fact that the transmission time is an impact factor in determining the overall throughput. SPT outperforms MST in terms of throughput and packet delivery rate by up to 62.5% and 91%, respectively. Note that the MST outperforms SPT in term of PDR only when $P_I \geq 0.3$ by up to 4.45%.

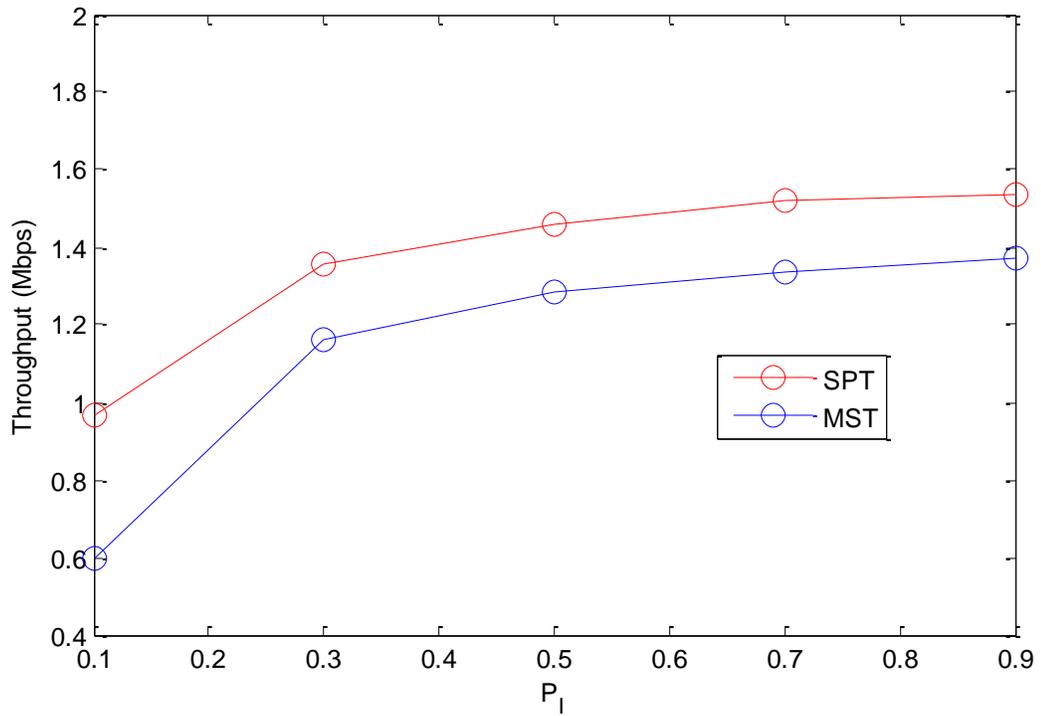

**Fig. 3.59 Throughput vs. idle probability.**

(N = 40, Nr = 16, M = 20, $P_t$ = 0.1W, BW = 1MHz, D = 4KB)

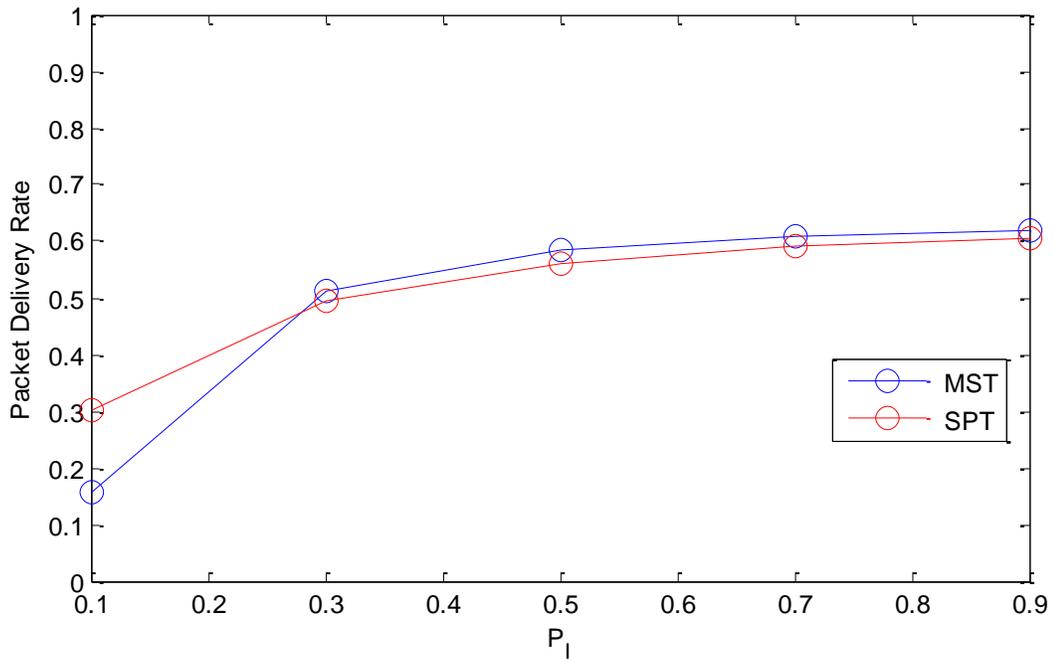

**Fig. 3.60 PDR vs. idle probability.**

(N = 40, Nr = 16, M = 20, $P_t$ = 0.1W, BW = 1MHz, D = 4KB)

## 3.4.5 Throughput and Packet Delivery Rate Performance versus Number of Primary Channels

Increasing the number of primary channels (M) in the network increases the number of available channels for SUs. When spectrum is very limited, the channel assignment is an issue in order to effectively utilize the available spectrum. Moreover, increasing the number of channels has an impact on network performance and routing decisions as well (see Figures 3.61 and 3.62). SPT outperforms MST in term of throughput by up to 26%, 85.6%, and 162% under $P_I$ = 0.9, 0.5, and 0.1, respectively. In term of PDR, SPT outperforms MST by up to 439% only under $P_I$ = 0.1.

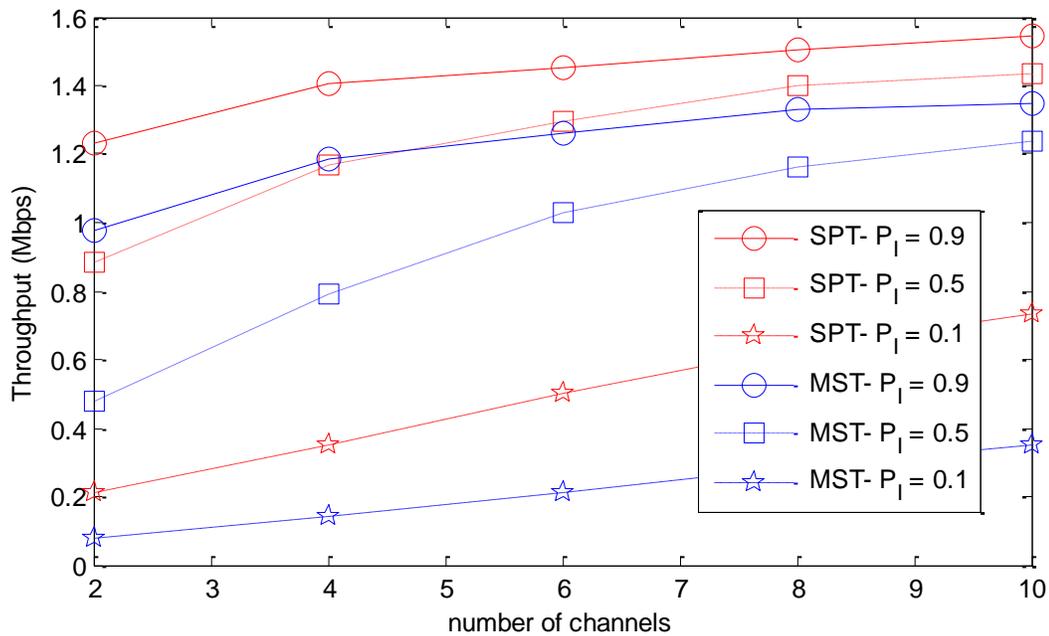

**Fig. 3.61 Throughput vs. number of primary channels.**

(N = 40, Nr = 16, $P_t$ = 0.1W, BW = 1MHz, D = 4KB)

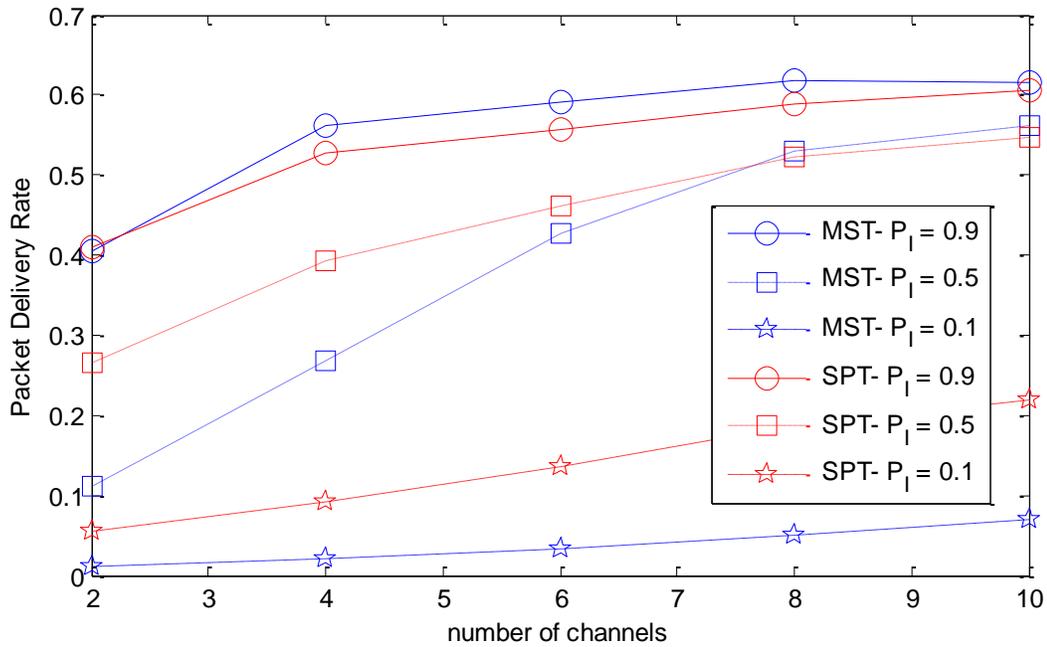

**Fig. 3.62 PDR vs. number of primary channels.**

(N = 40, Nr = 16, $P_t$ = 0.1W, BW = 1MHz, D = 4KB)

### 3.4.6 Throughput and Packet Delivery Rate Performance versus Number of Destinations

Changing the number of destinations (Nr) has an impact on the overall network performance and routing decision as well. We note here that, increasing the number of destinations in MST doesn't have effect on network performance because already the source needs multiple hops to reach each destination. On the other hand, increasing the number of destinations has negative effects on the SPT performance because of the increasing in the branching points on the multicast tree which negatively affected on the channel assignment decision. As stated before, the SPT provides better performance when the spectrum is limited than MST. Specifically, the difference between the performance of the SPT and MST scheme is maxima when spectrum is very limited (see Figures 3.63 and 3.64). SPT outperforms MST in term of throughput by up to 33.9%, 35%, and 76% under $P_I = 0.9$, 0.5, and 0.1, respectively. Also, SPT outperforms MST in term of PDR by up to 105% only under $P_I = 0.1$.

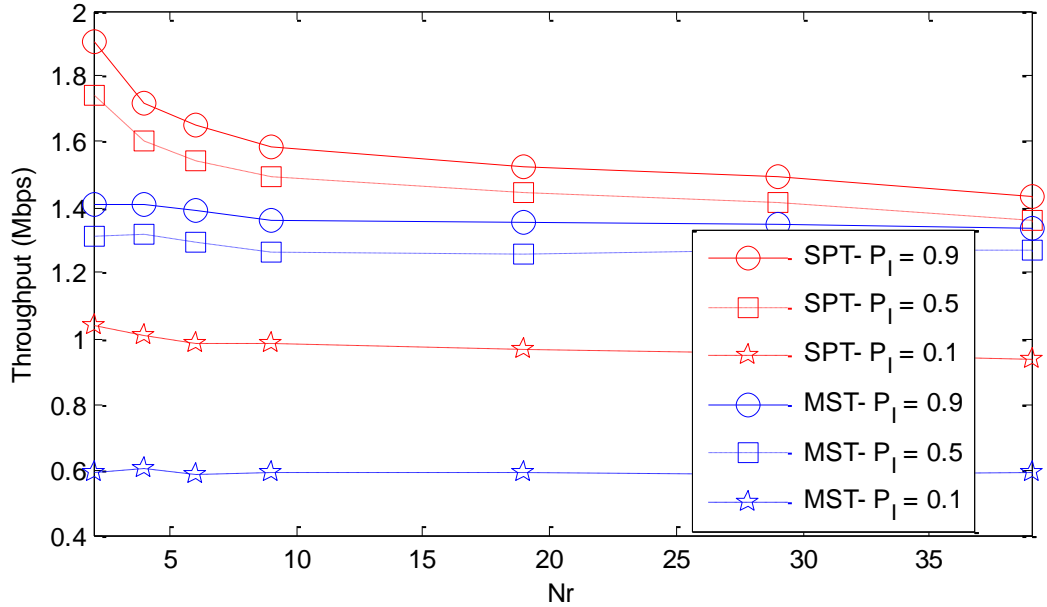

**Fig. 3.63 Throughput vs. number of destinations.**

(N = 40, M = 20, $P_t$ = 0.1W, BW = 1MHz, D = 4KB)

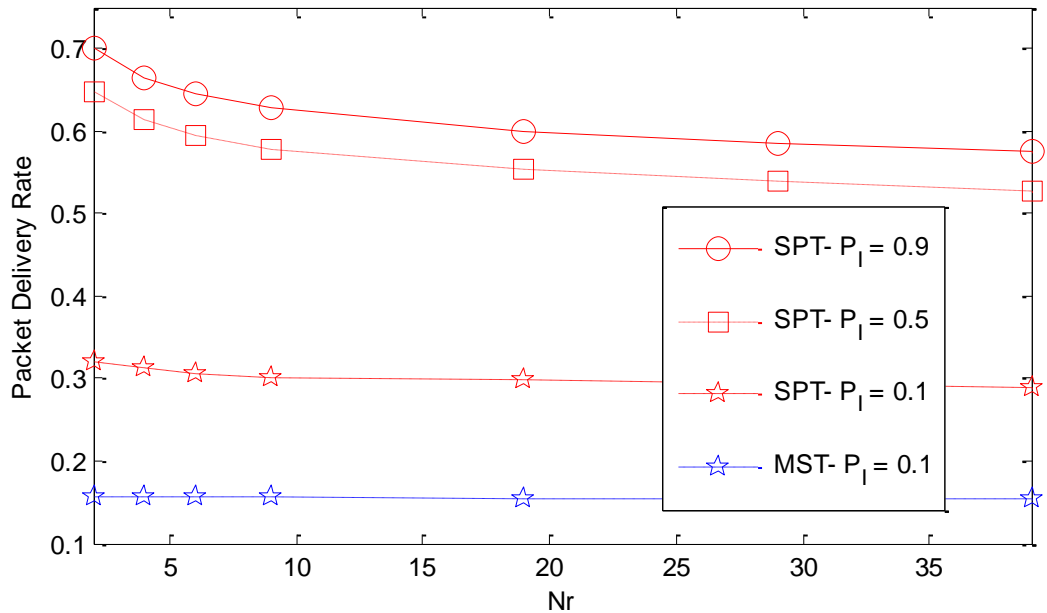

**Fig. 3.64 PDR vs. number of destinations.**

(N = 40, M = 20, $P_t$ = 0.1W, BW = 1MHz, D = 4KB)

### 3.4.7 Throughput and Packet Delivery Rate Performance versus Number of Nodes

As shown in Figures 3.65 and 3.66 below, increasing the number of nodes (N) is resulting in degradation performance for MST while the performance for SPT is unaffected. Again that is because number of hops that are needed to reach each destination in both trees. SPT outperforms MST under $P_I$ = 0.9, 0.5, and 0.1 in term of throughput by up to 61%, 62%, and 168%, respectively. SPT outperforms MST in term of PDR by up to 245% under $P_I$ = 0.1.

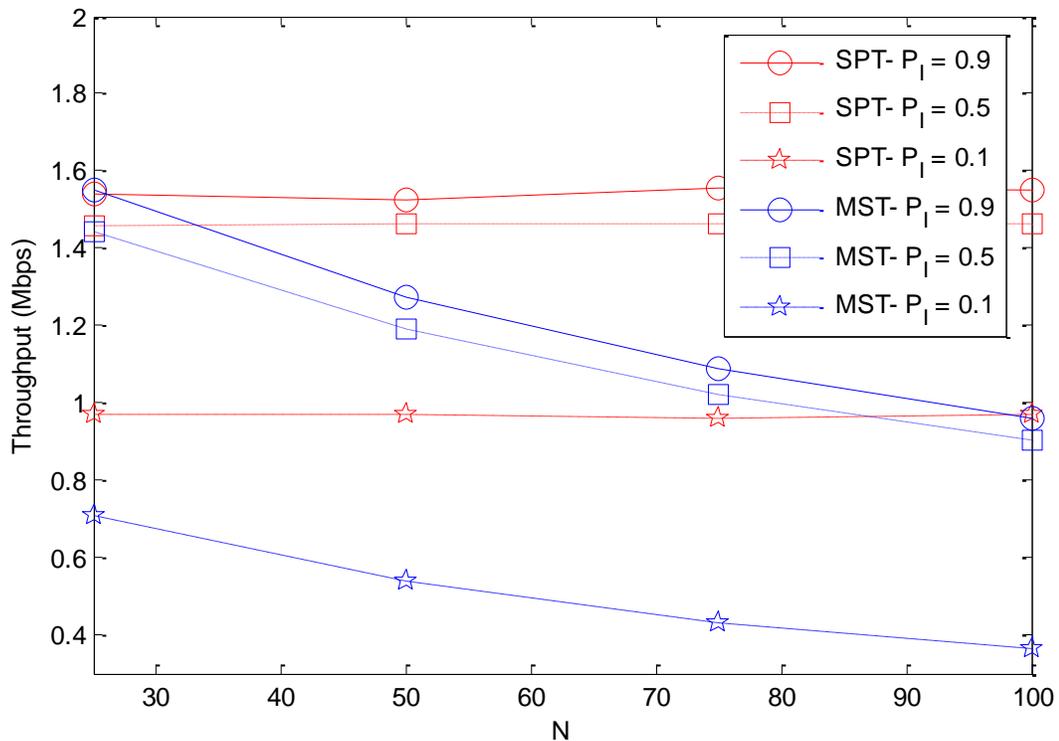

**Fig. 3.65 Throughput vs. number of nodes.**

($N_r$ = 16, M = 20, $P_t$ = 0.1W, BW = 1MHz, D = 4KB)

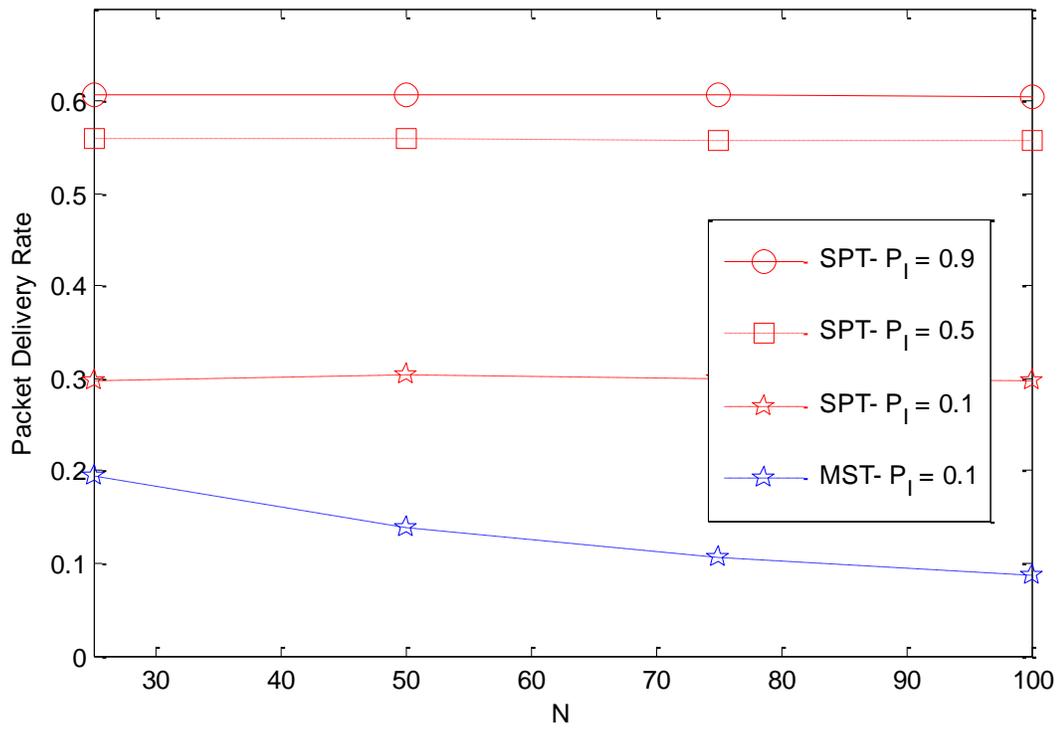

**Fig. 3.66 PDR vs. number of nodes.**

($N_r = 16$, $M = 20$, $P_t = 0.1W$, $BW = 1MHz$, $D = 4KB$)

# Chapter Four: Conclusions and Future Work

## 4.1 Conclusions

Routing and channel assignment design are challenging problems in multi-hop mobile Ad Hoc CRNs. Many attempts have been made to design efficient routing protocols, but none of them considers cross layer multi-hop routing protocol for mobile Ad Hoc CRNs. In this thesis, we proposed multi-hop multicast routing protocol for mobile Ad Hoc CRNs using the shortest path tree (SPT) and minimum spanning tree (MST). The channel assignment schemes used in the proposed protocol is based on the probability of success (POS). The proposed protocol transformed the network topology into tree (i.e., MST or SPT) and applied the POS-based channel assignment scheme to enhance network performance. The proposed protocol deals with each generated tree as a multi-layer transmission for routing purposes. We compared the performance of the proposed scheme to other variants and proved that the POS-based channel assignment showed the best performance under different network parameters.

When applying SPT, the proposed scheme (i.e., POS) outperforms MASA, MDR, and RS schemes in term of the packet delivery rate (PDR) by up to 9.3%, 110%, and 162%, respectively. Also, POS scheme outperforms MASA, MDR, and RS schemes in term of throughput by up to 11.28%, 69.5%, and 87%, respectively.

When applying MST, the proposed scheme outperforms MASA, MDR, and RS schemes in term of the packet delivery rate by up to 42.6%, 52%, and 122.6%, respectively. Also, POS scheme outperforms MASA, MDR, and RS schemes in term of throughput by up to 18%, 142%, and 236%, respectively.

SPT outperforms MST in term of throughput under all network conditions. The simulation results showed that the SPT achieves the maximum improvement gain compared to MST under high PU activity.

## 4.2 Future Work

In this thesis, we proposed multilayer routing protocol with single session that uses the POS-based channel assignment scheme. Other future work could consider multi source instead of single multicasting source, or use multisession scenario instead of single session and compare their performances under different PU activity and other network parameters. Recall that the proposed multicasting routing protocols (i.e., SPT and MST) in this thesis are distance-based, new protocols can be investigated by using the expected transmission count (ETX) to construct SPT and MST trees [40]. Also, the Steiner tree could be used to construct the multicast tree.

# References

[1] J. Mitola and J. Maguire, G.Q., "Cognitive radio: making software radios more personal," *IEEE Personal Communications*, pp. 13–18, Vol. 6, No. 4, Aug 1999.

[2] J. Mitola, "Cognitive radio for flexible mobile multimedia communications," *Proceedings of IEEE International Workshop on Mobile Multimedia Communications*, pp. 3–10, 1999.

[3] H. Bany Salameh, "Rate-maximization channel assignment scheme for cognitive radio networks," in *IEEE Global Telecommunications Conference (GLOBECOM 2010)*, pp. 1–5, Dec 2010.

[4] H. Bany Salameh, M. Krunz, and D. Manzi, "Spectrum bonding and aggregation with guard-band awareness in cognitive radio networks," *IEEE Transactions on Mobile Computing*, pp. 523–532, Vol. 13, No. 3, March 2014.

[5] H. Bany Salameh and O. S. Badarneh, "Opportunistic medium access control for maximizing packet delivery rate in dynamic access networks," *Journal of Network and Computer Applications*, pp. 523–532, Vol. 36, No. 1, Jan. 2013.

[6] J.-P. Sheu and I.-L. Lao, "Cooperative routing protocol in cognitive radio ad-hoc networks," *Proceedings of IEEE Wireless Communications and Networking Conference (WCNC)*, pp. 2916–2921, April 2012.

[7] O. S. Badarneh and H. Bany Salameh, "Probabilistic quality aware routing in cognitive radio networks under dynamically varying spectrum opportunities," *Comput. Electr. Eng.*, pp. 1731–1744, Vol. 38, No. 6, Nov. 2012.

[8] R. Engelman, K. Abrokwah, G. Dillon, G. Foster, G. Godfrey, T. Hanbury, C. Lagerwerff, W. Leighton, M. Marcus, R. Noel, J. Payton, J. Tomchin, J. Williams, and A. Yang, "Report

of the Spectrum Efficiency Working Group," US Federal Communications Commission Spectrum Policy Task Force, Tech. Rep., 2002.

[9] I. Akyildiz, W. Lee, M. Vuran, "NeXt generation/dynamic spectrum access/cognitive radio wireless networks: A survey", Comp. Net., pp. 2127-2159, Vol. 50, No. 13, May 2006.

[10] R. Vaishampayan and J. Garcia-Luna-Aceves, "Efficient and robust multicast routing in mobile ad hoc networks," in *Proceedings of IEEE International Conference on Mobile Ad-hoc and Sensor Systems*, pp. 304–313, Oct 2004.

[11] J. Garcia-Luna-Aceves and E. Madruga, "The core-assisted mesh protocol," *IEEE Journal on Selected Areas in Communications*, pp. 1380–1394, Vol. 17, No. 8, Aug 1999.

[12] S.-J. Lee, M. Gerla, and C.-C. Chiang, "On-demand multicast routing protocol," in *Proceedings of IEEE Wireless Communications and Networking Conference (WCNC)*, pp. 1298–1302 Vol.3, 1999.

[13] E. M. Royer and C. E. Perkins, "Multicast operation of the ad-hoc on-demand distance vector routing protocol," in *Proceedings of the 5$^{th}$ Annual ACM/IEEE International Conference on Mobile Computing and Networking*, ser. MobiCom '99. pp. 207–218, New York, NY, USA: ACM, 1999.

[14] Y.-B. Ko, S.-J. Lee, and K.-Y. Lee, "A multicast protocol for physically hierarchical ad hoc networks," in *Proceedings of The 57th IEEE Semiannual Vehicular Technology Conference (VTC)*, pp. 1238–1242, Vol. 2, April 2003.

[15] P.-Y. Chen, S.-M. Cheng, W.-C. Ao, and P.-Y. Chen, "Multi-path routing with end-to-end statistical qos provisioning in underlay cognitive radio networks," *Proceedings of IEEE Conference on Computer Communications Workshops (Infocom Wkshps)*, pp. 7–12, April 2011.

[16] G. Lei, W. Wang, T. Peng, and W. Wang, "Routing metrics in cognitive radio networks," *Proceedings of 4th IEEE International Conference on Circuits and Systems for Communications*, pp. 265–269, May 2008.

[17] I. Pefkianakis, S. Wong, and S. Lu, "Samer: Spectrum aware mesh routing in cognitive radio networks," *Proceedings of 3rd IEEE Symposium on New Frontiers in Dynamic Spectrum Access Networks (DySPAN)*, pp. 1–5, Oct 2008.

[18] G. Cheng, W. Liu, Y. Li, and W. Cheng, "Joint on-demand routing and spectrum assignment in cognitive radio networks," *Proceedings of IEEE International Conference on Communications (ICC)*, pp. 6499–6503, June 2007.

[19] Y. Liu and D. Grace, "Improving capacity for wireless ad hoc communications using cognitive routing," in *Proceedings of 3rd International Conference on Cognitive Radio Oriented Wireless Networks and Communications (CrownCom)*, pp. 1–6, May 2008.

[20] H. Khalife, S. Ahuja, N. Malouch, and M. Krunz, "Probabilistic path selection in opportunistic cognitive radio networks," in *Proceedings of IEEE Global Telecommunications Conference (GLOBECOM)*, pp. 1–5, Nov 2008.

[21] L. Yun, Q. Fengxie, L. Zhanjun, and Z. Hongcheng, "Cognitive radio routing algorithm based on the smallest transmission delay," in *Proceedings of 2nd International Conference on Future Computer and Communication (ICFCC)*, pp. V2–306–V2–310, Vol. 2, May 2010.

[22] A. Cacciapuoti, C. Calcagno, M. Caleffi, and L. Paura, "Caodv: Routing in mobile ad-hoc cognitive radio networks," in *Proceedings of Wireless Days (WD), 2010 IFIP*, Oct 2010, pp. 1–5.

[23] A. C. Talay and D. T. Altilar, "Ropcorn: Routing protocol for cognitive radio ad hoc networks," in *Proceedings of International Conference on Ultra Modern Telecommunications (ICUMT)*, 2009, pp. 1–6.

[24] D. Hu, S. Mao, and J. H. Reed, "On Video Multicast in Cognitive Radio Networks", INFOCOM, IEEE, pp. 2222-2230, Rio de Janeiro, 19-25 April 2009.

[25] J. Jin, H. Xu, B. Li, "Multicast Scheduling with Cooperation and Network Coding in Cognitive Radio Networks", INFOCOM, Proceedings IEEE, pp. 1-9, San Diego, CA, 14-19 March 2010.

[26] W. Ren, X. Xiao, Q. Zhao, "Minimum-energy Multicast Tree in Cognitive Radio Networks", Conference Record of the Forty-Third Asilomar Conference on Signals, Systems and Computers, pp. 312-316, Pacific Grove, CA , 1-4 Nov. 2009.

[27] L. Ding, T. Melodia, S. Batalama, J. Matyjas, and M. Medley "Crosslayer Routing and Dynamic Spectrum Allocation in Cognitive Radio Ad Hoc Networks," *IEEE Transactions on Vehicular Technology*, pp. 1969-1979, Vol. 59, No. 4, May 2010.

[28] L. Geng, Y.- C. Liang, and F. Chin "Network Coding for Wireless Ad Hoc Cognitive Radio Networks", *The 18th Annual IEEE International Symposium on Personal, Indoor and Mobile Radio Communications (PIMRC)*, pp. 1-5, Athens, 3-7 Sept. 2007.

[29] D. Hu, et al. "Scalable Video Multicast in Cognitive Radio Networks", *IEEE Journal On Selected Areas In Communications,* pp. 334-344, Vol. 28, No. 3, April 2010.

[30] Y. Huang, et al. "Robust Multicast Beamforming for Spectrum Sharing-Based Cognitive Radios", *IEEE Transactions on Signal Processing,* pp. 527-533, Vol. 60, No. 1, Jan. 2012.

[31] W. Ren, Q. Zhao, and A. Swami, "Power Control in Cognitive Radio Networks: How to Cross A Multi-lane Highway", *IEEE Journal on Selected Areas in Communications (JSAC): Special Issue on Stochastic Geometry and Random Graphs for Wireless Networks*, pp. 1283–1296, Vol. 27, No. 7, Sept. 2009.

[32] G. Zeng, B. Wang, Y. Ding, L. Xiao, and M. Mutka, "Multicast Algorithms for Multi-channel Wireless Mesh Networks", *IEEE International Conference on Network Protocols (ICNP),* pp. 1–10, Beijing, 16-19 Oct. 2007.

[33] H.L. Nguyen and U.T. Nguyen, "Channel Assignment for Multicast in Multi-channel Multi-radio Wireless Mesh Networks", *Wireless Communications and Mobile Computing,* pp. 557–571, Vol. 9, No. 4, April 2009.

[34] Gao, Cunhao, et al. "Multicast Communications in Multi-hop Cognitive Radio Networks", *IEEE Journal on Selected Areas in Communications,* pp. 784-793, Vol. 29, No. 4, April 2011.

[35] O. S. Bdarneh and H. Bany Salameh "Opportunistic Routing in Cognitive Radio Network: Exploiting Spectrum Availability and Rich Channel Diversity", *IEEE GlobeCom*, pp. 1-5, Houston, TX, USA, 5-9 Dec. 2011.

[36] Y. Shi, Y. Hou, "A Distributed Optimization Algorithm for Multi-hop Cognitive Radio Networks", *The 27th IEEE Conference on Computer Communications, INFOCOM*, Phoenix, AZ, 13-18 April 2008.

[37] M. Xie, W. Zhang, K.-K. Wong, "A Geometric Approach to Improve Spectrum Efficiency for Cognitive Relay Networks", *IEEE Transactions on Wireless communications,* pp. 268-281, Vol. 9, No. 1, Jan. 2010.

[38] H. M. Almasaeid, T. H. Jawadwala, and A. E. Kamal, "On-Demand Multicast Routing in Cognitive Radio Mesh Networks", *Global Telecommunications Conference, GlobeCom,* pp. 1-5, Miami, FL, 6-10 Dec. 2010.

[39] Y. Mhaidat, et. al., "A Cross-Layer Video Multicasting Routing Protocol for Cognitive Radio Networks", seventh international workshop on selected topics in mobile and wireless comuting, 2014.

[40] D. De Couto et. al., "A HighThroughput Path Metric for MultiHop Wireless Routing," *MobiCom '03,* San Diego, California, USA, September 14–19, 2003.

[41] O. S. Badarneh and M. Kadoch, "Multicast Routing Protocols in Mobile Ad Hoc Networks: A Comparative Survey and Taxonomy", *Hindawi Publishing Corporation, EURASIP Journal on Wireless Communications and Networking*, 42 Pages, Vol. 2009, No. 26, 2009.

[42] J. Villalon, et. al., "Multiservice unicast/multicast communications over IEEE 802.11e networks," *Telecommunications systems, Springer*, pp. 59-72, Vol. 43, no. 1-2, 2009.

[43] C. T.M. "Comments on broadcast channels," *IEEE Transaction on Information Theory*, pp. 2524-2530, Vol. 44, no. 6, Oct 1998.

[44] B. Ye Wu and K. Chao, "Spanning Trees and Optimization Problems", *CRC Press,* 2004.

[45] V.K. Balakrishnan, "*Schaum's Outline of Theory and Problems of* Graph Theory", *McGraw-Hill*, 1997.

[46] MathWorks. [Online]. http://www.mathworks.com.

[47] H. Bany Salameh, M. Krunz, "Channel access protocols for multihop opportunistic networks: Challenges and recent developments," IEEE Network, vol. 23, no. 4, pp. 14–19, Jul.–Aug. 2009, Special Issue on Networking over Multi-Hop Cognitive Networks.

[48] H. A. Bany Salameh, "Throughput-oriented Channel Assignment for Opportunistic Spectrum Access Networks", *Mathematical and Computer Modelling*, pp. 2108-2118, Vol. 53, No. 11, 2011.

# الملخص باللغة العربية

**العنوان:** طريقه متعددة الطبقات لاجل مسار متعدد الارسال في الشبكات الراديوية الذكية ذات الخطوات المتعددة

**الاسم:** مصطفى مهدي علي (2013976007)

**المشرف:** د. هيثم احمد بني سلامة

توجيه الارسال المتعدد (Multicast) يعد واحدا من اهم العمليات في الاتصالات وذلك لكثرة التطبيقات عليه والتي منها المؤتمرات الفيديويه ونشر البيانات والاغاثة في الكوارث الطبيعية وغيرها من التطبيقات. يمكن تطبيق توجيه الارسال المتعدد بدون اي صعوبات في حال الوصول الى الوسائط متاحا من قبل اي مرسل في اي وقت, لكن هذا غير وارد في الشبكات الراديوية الادراكية (CRN) بسبب عدم تجانس القنوات الراديوية عبر الشبكة. في هذه الاطروحة, سنطرح فكرة المسار الاقصر من المرسل الى المستلم (SPT) وكذلك فكرة ربط الشبكه باقل مجموع مسافلت بين المستخدمين (MST) في الشبكات الراديوية الادراكية ذات القفزات المتعددة التي لاتحتوي على خادم. الخوارزمية المقترحة لتوجية الارسال المتعدد توظف احتمالية نجاح الارسال (POS) في عملية تعيين القنوات في جلسه واحده (single session). اختيار المسار الاقصر بين المصدر وكل مستقبل مستقبل اظهر اداءا افضل للشبكة من اخيار اقل مجموع مسافات بين المستخدمين , حيث كان افضل اداء عند استخدام مقياس احتمالية نجاح الارسال لتعيين القنوات افضل من بقية الخوارزميات في الابحاث السابقة. زودنا في هذه الاطروحة نتائج تجارب المحاكاة والتي اثبتت صحة النتائج الانفة الذكر بالاضافة الى تمحيص اداء وفعالية الخوارزميات المقترحه ضمن ظروف ومعايير مختلفه للشبكة ومقارنة النتائج مع نتائج الابحاث السابقة بدلالة معدل سرعة تدفق البيانات ونسبة توصيل المعلومة.